\documentclass[10pt,onecolumn]{IEEEtran}
\usepackage[mathscr]{eucal}
\usepackage[cmex10]{amsmath}
\usepackage{epsfig,epsf,psfrag}
\usepackage{amssymb,amsmath,amsthm,amsfonts,latexsym}
\usepackage{amsmath,graphicx,bm,xcolor,url,overpic}
\usepackage{array}%array and tabular environments
\usepackage{verbatim}
\usepackage{bm}
\usepackage{algorithm}
\usepackage{verbatim}
\usepackage{textcomp}
\usepackage{mathrsfs}
\usepackage{epstopdf}

\newcommand{\openone}{\leavevmode\hbox{\small1\normalsize\kern-.33em1}}

\catcode`~=11 \def\UrlSpecials{\do\~{\kern -.15em\lower .7ex\hbox{~}\kern .04em}} \catcode`~=13 

\allowdisplaybreaks[4]

\newcommand{\nn}{\nonumber}

\newcommand{\calA}{\mathcal{A}}
\newcommand{\calB}{\mathcal{B}}
\newcommand{\calC}{\mathcal{C}}
\newcommand{\calD}{\mathcal{D}}

\newcommand{\calF}{\mathcal{F}}

\newcommand{\calI}{\mathcal{I}}

\newcommand{\calM}{\mathcal{M}}

\newcommand{\calP}{\mathcal{P}}

\newcommand{\calR}{\mathcal{R}}

\newcommand{\calV}{\mathcal{V}}

\newcommand{\calX}{\mathcal{X}}
\newcommand{\calY}{\mathcal{Y}}

\newcommand{\by}{\mathbf{y}}
\newcommand{\bY}{\mathbf{Y}}

\newcommand{\rmseq}{\mathrm{seq}}
\newcommand{\rmtp}{\mathrm{tp}}

\newcommand{\rmA}{\mathrm{A}}

\newcommand{\rmG}{\mathrm{G}}

\newcommand{\rmH}{\mathrm{H}}

\newcommand{\rmN}{\mathrm{N}}

\newcommand{\rmP}{\mathrm{P}}

\newcommand{\rmr}{\mathrm{r}}

\newcommand{\bbE}{\mathsf{E}}

\newcommand{\bbN}{\mathbb{N}}

\newcommand{\bbP}{\mathbb{P}}

\newcommand{\bbR}{\mathbb{R}}

\DeclareMathAlphabet{\mathbsf}{OT1}{cmss}{bx}{n}
\DeclareMathAlphabet{\mathssf}{OT1}{cmss}{m}{sl}% slanted sans serif

\DeclareSymbolFont{bsfletters}{OT1}{cmss}{bx}{n}  
\DeclareSymbolFont{ssfletters}{OT1}{cmss}{m}{n}
\DeclareMathSymbol{\bsfGamma}{0}{bsfletters}{'000}
\DeclareMathSymbol{\ssfGamma}{0}{ssfletters}{'000}
\DeclareMathSymbol{\bsfDelta}{0}{bsfletters}{'001}
\DeclareMathSymbol{\ssfDelta}{0}{ssfletters}{'001}
\DeclareMathSymbol{\bsfTheta}{0}{bsfletters}{'002}
\DeclareMathSymbol{\ssfTheta}{0}{ssfletters}{'002}
\DeclareMathSymbol{\bsfLambda}{0}{bsfletters}{'003}
\DeclareMathSymbol{\ssfLambda}{0}{ssfletters}{'003}
\DeclareMathSymbol{\bsfXi}{0}{bsfletters}{'004}
\DeclareMathSymbol{\ssfXi}{0}{ssfletters}{'004}
\DeclareMathSymbol{\bsfPi}{0}{bsfletters}{'005}
\DeclareMathSymbol{\ssfPi}{0}{ssfletters}{'005}
\DeclareMathSymbol{\bsfSigma}{0}{bsfletters}{'006}
\DeclareMathSymbol{\ssfSigma}{0}{ssfletters}{'006}
\DeclareMathSymbol{\bsfUpsilon}{0}{bsfletters}{'007}
\DeclareMathSymbol{\ssfUpsilon}{0}{ssfletters}{'007}
\DeclareMathSymbol{\bsfPhi}{0}{bsfletters}{'010}
\DeclareMathSymbol{\ssfPhi}{0}{ssfletters}{'010}
\DeclareMathSymbol{\bsfPsi}{0}{bsfletters}{'011}
\DeclareMathSymbol{\ssfPsi}{0}{ssfletters}{'011}
\DeclareMathSymbol{\bsfOmega}{0}{bsfletters}{'012}
\DeclareMathSymbol{\ssfOmega}{0}{ssfletters}{'012}

\newcommand{\hats}{\hat{s}}
\newcommand{\hatS}{\hat{S}}

\newcommand{\tilt}{\tilde{t}}

\newcommand{\tilx}{\tilde{x}}

\newcommand{\tily}{\tilde{y}}

\DeclareMathOperator*{\argmin}{arg\,min}

\newtheorem{theorem}{Theorem} 
\newtheorem{lemma}{Lemma}

\usepackage{graphicx,amssymb,amstext,amsmath}
\usepackage{makecell}
\usepackage{amsthm}
\usepackage{color}
\usepackage{subfigure}
\usepackage{algpseudocode}
\usepackage{multirow}
\usepackage[switch,pagewise]{lineno}
\usepackage{enumerate}
\usepackage[colorlinks = true,
linkcolor = blue,
urlcolor  = blue,
citecolor = red,
anchorcolor = green,]{hyperref}
\usepackage{color}

\definecolor{Dyellow}{RGB}{254,152,0}
\definecolor{Dgreen}{RGB}{0,176,80}
\begin{document}
\title{Exponentially Consistent Outlier Hypothesis Testing for Continuous Sequences}

\author{\IEEEauthorblockN{Lina Zhu and Lin Zhou}\\
\thanks{L. Zhu is with Space Information Research Institute, Hangzhou Dianzi University, Hangzhou, Zhejiang, China, 310018 (Email: zhulina@hdu.edu.cn). L. Zhou is with the School of Cyber Science and Technology, Beihang University, Beijing 100191, China (Email: lzhou@buaa.edu.cn).}
%\thanks{This work was supported in part by the National Natural Science Foundation of China under Grants 62341105, 62201022 and U22B2008 and in part by the Beijing Natural Science Foundation under Grant 4232007.}
}

\maketitle
\flushbottom

\begin{abstract}
In outlier hypothesis testing, one aims to detect outlying sequences among a given set of sequences, where most sequences are generated i.i.d. from a nominal distribution while outlying sequences (outliers) are generated i.i.d. from a different anomalous distribution. Most existing studies focus on discrete-valued sequences, where each data sample takes values in a finite set. To account for practical scenarios where data sequences usually take real values, we study outlier hypothesis testing for continuous sequences when both the nominal and anomalous distributions are \emph{unknown}. Specifically, we propose distribution free tests and prove that the probabilities of misclassification error, false reject and false alarm decay exponentially fast for three different test designs: fixed-length test, sequential test, and two-phase test. In a fixed-length test, one fixes the sample size of each observed sequence; in a sequential test, one takes a sample sequentially from each sequence per unit time until a reliable decision can be made; in a two-phase test, one adapts the sample size from two different fixed values. Remarkably, the two-phase test achieves a good balance between test design complexity and theoretical performance. We first consider the case of at most one outlier, and then generalize our results to the case with multiple outliers where the number of outliers is unknown.
\end{abstract}


\section{Introduction}
\label{sec:intro}

In outlier hypothesis testing (OHT), one is given a set of $M$ observed sequences. Most of the sequences are generated i.i.d. from a nominal probability density function (pdf) $f_\rmN$ while the rest of the sequences named outliers are generated i.i.d. from an anomalous pdf $f_\rmA$. The task of OHT is to identify the set of outliers. A misclassification error event occurs if a wrong set of sequences are claimed to be outliers; a false alarm event occurs if a set of sequences are claimed to be outliers while there is no outlier in observed sequences; a false reject event occurs if no sequence is claimed to be outliers while there are outliers in the observed sequences. Motivated by practical applications of anomalous detection including analyzing anomalous traffic pattern to identify sensitive data from a hacked computer \cite{kumar2005parallel} and extracting anomalous MRI images to alarm malignant tumors \cite{spence2001detection}, we study the tradeoff among the probabilities of misclassification error, false alarm and false reject of outlier hypothesis testing for continuous sequences when the number of outliers is unknown.

When the generating distributions are known, outlier hypothesis testing generalizes binary hypothesis testing. In binary hypothesis testing, one is given a test sequence $Y^n=(Y_1,\ldots,Y_n)$ and two \emph{known} distributions $f_1$ and $f_2$. The task is to decide whether $Y^n$ is generated i.i.d. from $f_1$ or $f_2$. There are two types of error events: type-I error and type-II error events. Specifically, a type-I error event occurs if $Y^n$ is claimed to be generated from $f_2$ while $Y^n$ is actually generated from $f_1$. Analogously, a type-II error event occurs if $Y^n$ is claimed to be generated from $f_1$ while $Y^n$ is actually generated from $f_2$. Typically, there are two types of tests: the fixed-length and the sequential tests. In a fixed-length  test, one is given a sequence of a fixed length $n$. In a sequential test, one takes each sample per unit time until a reliable decision can be made and thus the length of the observed sequence is a random variable. For a fixed-length test, the tradeoff between the type-I and type-II error probabilities for optimal tests is characterized by the Chernoff-Stein lemma~\cite{Chernoff1952AMO} in the Neyman-Pearson setting and by Blahut~\cite{Blahut1974HypothesisTA} for the Bayesian setting. Specifically, the Chernoff-Stein lemma  states that for the optimal test, the type-II error probability decays exponentially fast with respect to the sample length $n$ at the speed of the Kullback-Leibler (KL) divergence~\cite{kullback1951information} of two distributions $D(f_1\|f_2)$ when the type-I error probability is upper bounded by a constant. Balahut characterized the tradeoff between the decay rates (also known as error exponents) of the type-I and type-II error probabilities by showing that when type-I error exponent is $\lambda$, the maximal type-II error exponent is given by a function $E(\lambda)$ of $\lambda$ such that $\max_{\lambda\geq 0}E(\lambda)=E(0)=D(f_1\|f_2)$.

For sequential tests, the optimal performance of binary hypothesis testing was derived by Wald \cite{wald1948optimum} in the Bayesian setting, who showed that the error exponent pair $(D(f_2\|f_1),D(f_1\|f_2))$ can be achieved simultaneously. In other words, an optimal sequential test resolves the tradeoff between the type-I and type-II error exponents for fixed-length tests and allows both type-I and type-II error probabilities to decay exponentially fast at maximal possible rates. Thus, an optimal sequential test has significantly better performance than an optimal fixed-length test. However, the superior performance of an optimal sequential test is achieved at the cost of high design complexity. This is because in a sequential test, after collecting each data sample, one needs to determine whether to continue collecting additional samples or to make a decision. In contrast, in a fixed-length test, one makes a decision directly when a fixed number of samples are collected.

One might wonder whether there exists a test that can balance the design complexity and performance between the fixed-length and the sequential tests for binary hypothesis testing. To answer this question, Lalitha and Javidi~\cite{AFL} proposed a two-phase test and demonstrated that the test could achieve error exponents close to the sequential test with design complexity similar to a fixed-length test. Specifically, the test consists of two phases: the first phase is a fixed-length test taking $n$ samples, the second phase is another fixed-length test taking additional $(K-1)n$ samples for some real number $K\geq 1$ and the second phase proceeds only if the first phase outputs a reject option.

For the more practical case where the generating distributions are \emph{unknown}, outlier hypothesis testing is closely related with statistical classification initiated by Gutman~\cite{Asymptotically_optimal_classification}. In the binary case, one is given a testing sequence $Y^n$ and two training sequences $(X_1^n,X_2^n)$, where for each $i\in\{1,2\}$, $X_i^n$ is generated i.i.d. from an unknown distribution $f_i$. The task is to determine whether $Y^n$ is generated i.i.d. from an unknown distribution $f_1$ or another unknown distribution $f_2$. Note that binary classification is more practical than binary hypothesis testing since in practical applications, the exact distributions are usually unavailable. Instead, training samples are provided. The performance of fixed-length tests was characterized by Gutman~\cite{Asymptotically_optimal_classification} and Zhou, Tan, Motani~\cite{second} while the performance of sequential tests was characterized by Haghifam, Tan and Khsiti \cite{Sequential_classification} and Hsu, Li and Wang \cite{hsu2022universal}. The two-phase tests were proposed and analyzed by Diao, Zhou and Bai \cite{bai2022achievable,diao2023achievable}.

For outlier hypothesis testing with \emph{discrete-valued} sequences, the performance of fixed-length tests was characterized by Li and Veeravalli~\cite{Universal_Outlier_Hypothesis} and by Zhou, Wei and Hero~\cite{Joint}. The performance of sequential tests was characterized by Li and Veeravalli~\cite{Universal_sequential_Hypothesis}. However, the corresponding results for {continuous} sequences are limited. The only known result is by Zou \emph{et al.}~\cite{MMD} for fixed-length tests. Specifically, the authors of \cite{MMD} designed a testing using the maximum mean discrepancy (MMD) metric with a reproducing kernel Hilbert space~\cite{Hilbert}, and showed that the test is exponentially consistent under mild conditions in the sense that all error probabilities decay exponentially fast for any unknown nominal and anomalous distributions.

However, the results of Zou \emph{et al.}~\cite{MMD} have several limitations. Firstly, the results were only established for fixed-length tests while both sequential and two-phase tests were not studied. Secondly, for fixed-length tests, Zou \emph{et al.}~\cite{MMD} proposed to check whether each sequence is an outlier individually instead of considering the more powerful joint test that identifies the set of outliers simultaneously. To solve the above two problems, we propose distribution free tests and prove that the probabilities of misclassification error, false reject and false alarm decay exponentially fast for three different test designs: fixed-length test, sequential test, and two-phase test. Our main contributions are summarized as follows.

\subsection{Main contributions}
We study outlier hypothesis testing for continuous observed sequences and propose exponentially consistent tests when both generating distributions are unknown. We first consider the case where there is at most one outlier. In this setting, the null hypothesis indicates that there is no outlier and each non-null hypothesis specifies a possible outlier. Depending on the way where the samples are obtained, we propose three test designs: the fixed-length test, the sequential test and the two-phase test. In a fixed-length test, one fixes the sample size of each observed sequence; in a sequential test, one takes one sample sequentially from each sequence per unit time until a reliable decision can be made; in a two-phase test, one adapts the sample size from two different fixed values. For all three tests, we prove that the probabilities of misclassification, false reject and false alarm decay exponentially fast. Furthermore, we show that the sequential test achieves larger exponents than the fixed-length test. Finally, analytically and numerically, we show that by changing design parameters, our two-phase test either reduces to a fixed-length test or approximately achieves the performance of a sequential test and thus strikes a good balance between test design complexity and outlier detection performance.

We next consider the more practical case where the number of outliers could be more than one and is \emph{unknown}. To deal with the unknown number of outliers, for all three types of tests, we first estimate the number of outliers and then identify the set of outliers if the estimated number is positive. For each test, we analyze its achievable performance and characterize the exponential decay rates for the probabilities of misclassification, false reject and false alarm. In this case, a misclassification event occurs if the number of outliers is estimated positive and incorrectly or if the number of outliers is estimated correctly but the set of outliers is identified incorrectly. A false reject event occurs if the number of outliers is estimated as zero when there are outliers while a false alarm event occurs if the number of outliers is estimated positive when there is no outlier. Compared with the case of at most one outlier, the analysis of the misclassification probability is further complicated since the error event concerning estimating the number of outliers requires additional efforts. Analogous to the case of at most one outlier, we show that the two-phase test bridges over the fixed-length test and the sequential test by having performance close to the sequential test and having design complexity propositional to the fixed-length test. Furthermore, we show that there is a penalty of not knowing the number of outliers, analytically and numerically.

\subsection{Other Related works}
We recall other non-exhaustive studies on binary hypothesis testing, binary classification and outlier hypothesis testing. Sason~\cite{Moderate_Deviations} studied moderate-deviations of binary hypothesis testing. Zeitouni, Ziv and Merhav~\cite{generalized_likelihoodratio_test} demonstrated the asymptotic optimality of the generalized likelihood ratio
test (GLRT) for composite binary hypothesis testing. Merhav and Ziv \cite{Merhav1991ABA} derived the Bayesian error exponent for binary classification. Hsu and Wang~\cite{Hsu2020OnBS} considered binary classification with mismatched empirically observed statistics, where the training sequences are generated from distributions that are perturbed versions of the true generating distribution under each hypothesis. For outlier hypothesis testing of discrete data samples, Bu, Zou and Veeravalli proposed a linear-complexity test using the clustering idea and showed that the test is exponentially consistent~\cite{bu2019linear}.

\subsection{Organization of the Rest of the Paper}
The rest of the paper is organized as follows. In Section \ref{Problem_formulation}, we set up the notation, formulate the problem of outlier hypothesis testing and recall the definition of the MMD metric. In Section \ref{Main}, we present theoretical results when there is at most one outlier for three test designs: fixed-length, sequential and two-phase tests. In Section \ref{Main_unS}, we generalize our results to the case of multiple outliers, where the number of outliers is unknown. Numerical examples are provided in Section \ref{simulation} to illustrate our theoretical benchmarks. In Section \ref{sec:conc}, we conclude the paper and discuss future research directions. All proofs are deferred to appendices for smooth presentation of the paper.

\section{Problem Formulation}
\label{Problem_formulation}

\subsection*{Notation}
We use $\calR$, $\calR_+$ and $\bbN$ to denote the set of real numbers, non-negative real numbers and natural numbers, respectively. All logarithms are base $e$. All sets are
denoted in calligraphic font (e.g., $\calX$). Random variables and their realizations are in upper case (e.g., $X$) and lower case (e.g., $x$), respectively. Given any event $\calA$, we use $\Pr\{\calA\}$ to denote the probability that $\calA$ occurs. We use $Y^n = (Y_1,\ldots , Y_n)$ to denote a random vector of length $n\in\bbN$. The set of all probability density functions (pdf) defined on $\calR$ is denoted as $\calF(\calR)$. For any integers $(a,b)\in\bbN^2$, we use $[a:b]$ to denote the set of integers between $a$ and $b$ and we use $[a]$ to denote $[1:a]$.

\subsection{Case of at Most One Outlier}
Fix integers $(M,n)\in\bbN^2$. Let $\bY^n=\{Y_1^n,Y_2^n,...,Y_M^n\}\in (\calR^n)^M$ be a set of $M$ sequences, where for each $i\in [M]$, $Y_i^n$ is generated i.i.d. from either a nominal distribution $f_\rmN\in \calF(\calR)$ or an anomalous distribution $f_\rmA\in \calF(\calR)$. Any sequence that is generated i.i.d. from $f_\rmA$ is an outlier. Motivated by practical applications, we assume that $(f_\rmN,f_\rmA)$ are both \emph{unknown}. Let $\calM:=[M]$ denote the set of all integers from $1$ to $M$. Assume that there is at most one outlier among all $M$ sequences $\bY^n$.

The task is to design a test $\Phi=(\tau, \phi_\tau)$, which includes a potentially random stopping time $\tau\in\bbN$ and a decision rule $\phi_\tau$, to decide among the following $M+1$ hypotheses:
\begin{itemize}
\item $\rmH_i$,~$i\in \calM$: the $i$-th sequence is the outlier.
\item $\rmH_\rmr$: there is no outlier.
\end{itemize}
The random stopping time $\tau$ is with respect to the filtration $\{\calF_n\}_{n\in\bbN}$, where $\calF_n$ is generated by $\sigma$-algebra $\sigma\{Y_1^n,\ldots Y_M^n\}$ for each $n\in\bbN$.

Under any pair of nominal and anomalous distributions $(f_\rmN,f_\rmA)\in \calF(\calR)^2$, to evaluate the performance of a test $\Phi=(\tau, \phi_\tau)$, for each $i\in\calM$, we consider the following misclassification error and false reject probabilities under hypothesis $\rmH_i$:
\begin{align}\label{Error_define}
\beta_i(\phi_\tau|f_\rmN,f_\rmA)& := \bbP_i\{\phi_\tau(\bY^\tau)\notin\{\rmH_i,\rmH_\rmr\}\},\\
\zeta_i(\phi_\tau|f_\rmN,f_\rmA)& := \bbP_i\{\phi_\tau(\bY^\tau)=\rmH_\rmr\},
\end{align}
where under the distribution $\bbP_i$, $Y_i^n$ is generated i.i.d. from the unknown anomalous distribution $f_\rmA$ and $Y_j^n$ is generated i.i.d. from the unknown nominal distribution $f_N$ when $j\neq i$. Note that $\beta_i(\phi_\tau|f_\rmN,f_\rmA)$ bounds the probability of the misclassification error event where the test incorrectly claims a wrong sequence $Y_j^n$ with $j\neq i$ as the outlier, while $\zeta_i(\phi_\tau|f_\rmN,f_\rmA)$ bounds the probability of the false reject event where the test incorrectly claims that there is no outlier while $Y_i^n$ is the outlier.

In addition, we also need the following false alarm probability under the null hypothesis:
\begin{align}
\rmP_{\rm{FA}}(\phi_\tau|f_\rmN,f_\rmA)& := \bbP_\rmr\{\phi_\tau(\bY^\tau)\neq \rmH_\rmr\},\label{FA_s1}
\end{align}
where under the distribution $\bbP_\rmr$, all sequences are generated i.i.d. from the unknown nominal distribution $f_N$. Note that $\rmP_{\mathrm{FA}}(\phi_\tau|f_\rmN,f_\rmA)$ bounds the probability of the false alarm event where the test incorrectly claims there is an outlier when all the sequences are nominal samples.

In the first part of this paper, we propose a fixed-length test, a sequential test and a two-phase test for the case of at most one outlier. For each test, we show that all three error probabilities decay exponentially fast. In particular, the sequential test has the largest exponential decay rates while the fixed length test enjoys the simplest design. Our two-phase test, which combines two fixed-length tests with different sample sizes, strikes a good balance between design complexity and performance.

\subsection{Case of Multiple Outliers}
A more practical case is where there are multiple outliers but the number of outliers is unknown. Since at most less than half of all sequences can be outliers, the number of outliers is naturally upper bounded by $T:=\lceil \frac{M}{2}\rceil-1\geq1$. In this case, our task is to estimate the number of outliers and identify the set of outliers if the estimated number is positive.

Let $\calB\subseteq \calM$ be a subset of $\calM=[M]$, and let $\bY_\calB:=\{Y_i^n\}_{i\in\calB}$ be the sequences in $\bY^n$ with indices specified by the set $\calB$. Fix any positive integer $t\in\bbN$ and let $\calC_t:=\{\calB\subseteq\calM:|\calB|=t\}$ be the set of all subsets of $\calM$ of size $t$. Furthermore, let $\calC:=\bigcup_{t\in[T]}\calC_t$. The number of hypotheses thus increases from $M+1$ for the case of at most one outlier to $|\calC|+1=\sum_{t\in[T]}|\calC_t|+1$ for the case of at most $T$ outliers. The test design and performance metrics are similar as the case with at most one outlier except that the non-null hypotheses change from $\{\rmH_i\}_{i\in\calM}$ to $\{\rmH_\calB\}_{\calB\in\calC}$. Specifically, our task is to design a test $\Phi=(\tau, \phi_\tau)$ to decide among the following $|\calC|+1$ hypotheses:
\begin{itemize}
\item $\rmH_\calB$,~$\calB\in \calC$: the indices of the outliers are specified by $\calB$.
\item $\rmH_\rmr$: there is no outlier.
\end{itemize}
Under any pair of nominal and anomalous distributions $(f_\rmN,f_\rmA)\in \calF(\calR)^2$, for each $\calB\in\calC$,  we consider the following misclassification error and false reject probabilities under the non-null hypothesis $\rmH_\calB$:
\begin{align}
\beta_\calB(\phi_\tau|f_\rmN,f_\rmA)& := \bbP_\calB\{\phi_\tau(\bY^\tau)\notin\{\rmH_\calB,\rmH_\rmr\}\},\label{Error_define2}\\  \zeta_\calB(\phi_\tau|f_\rmN,f_\rmA)& := \bbP_\calB\{\phi_\tau(\bY^\tau)=\rmH_\rmr\},\label{Error_define3}
\end{align}
where under the distribution $\bbP_\calB$, $Y_j^n$ is generated i.i.d. from the unknown anomalous distribution $f_\rmA$ if $j\in\calB$ and $Y_j^n$ is generated i.i.d. from the unknown nominal distribution $f_\rmN$ if $j\notin\calB$. The false alarm probability $\rmP_{\rm{FA}}(\phi_\tau|f_\rmN,f_\rmA)$ under the null hypothesis is exactly the same as \eqref{FA_s1}.

Analogously to \eqref{Error_define2} and \eqref{Error_define3}, $\beta_\calB(\phi_\tau|f_\rmN,f_\rmA)$ bounds the probability of the misclassification error event where the test incorrectly claims a wrong set of sequences as outliers, while $\zeta_\calB(\phi_\tau|f_\rmN,f_\rmA)$ bounds the probability of the false reject event where the test claims there is no outlier when there are outliers.

\subsection{MMD Metric}
Consistent with \cite{MMD}, we adopt the MMD metric \cite{gretton2012kernel} to design our tests. Given any two distributions $f_1$ and $f_2$, the MMD distance  between $f_1$ and $f_2$ is defined as
\begin{align}
\mathrm{MMD}^2(f_1,f_2)
&:=\bbE_{f_1f_1}[k(X,X')]-2\bbE_{f_1f_2}[k(X,Y)]+\bbE_{f_2f_2}[k(Y,Y')],
\end{align}
where $k(\cdot,\cdot)$ is a kernel function associated with the Reproducing Kernel Hilbert Spaces~\cite{sun2023kernel} and $(X,X',Y,Y')\sim f_1f_1f_2f_2$. Note that $\mathrm{MMD}^2(f_1,f_2)=0$ if $f_1=f_2$. A usually adopted kernel function is the following Gaussian kernel function
\begin{align}
k(x,y):=\exp\left\{-\frac{(x-y)^2}{2\sigma_0^2}\right\}\label{Gaussiankernel},
\end{align}
where $(x,y)\in\calR^2$ and $\sigma_0\in\bbR_+$ is an positive real number. In this paper, we adopt the Gaussian kernel function in numerical examples to illustrate our theoretical results.

Given two sequences $x^{n_1}=[x_1,x_2,...,x_{n_1}]$ and $y^{n_2}=[y_1,y_2,...,y_{n_2}]$ sampled i.i.d. from distributions $f_1$ and $f_2$, respectively, the MMD between the two sequences is defined as
\begin{align}
\mathrm{MMD}^2(x^{n_1},y^{n_2})
\nn&:=\frac{1}{n_1(n_1-1)}\sum_{i,j\in[n_1],i\neq j}k(x_i,x_j)+\frac{1}{n_2(n_2-1)}\sum_{i,j\in[n_2],i\neq j}k(y_i,y_j)\\*
&\qquad-\frac{2}{n_1n_2}\sum_{i\in[n_1],j\in[n_2]}k(x_i,y_j)\label{MMDcompute}.
\end{align}
It was shown in~\cite[Lemma 6]{gretton2012kernel} that $\mathrm{MMD}^2(x^{n_1},y^{n_2})$ is an unbiased estimator of $\mathrm{MMD}^2(f_1, f_2)$, i.e., $\lim\limits_{n_1,n_2\rightarrow\infty}\mathrm{MMD}^2(x^{n_1},y^{n_2})=\mathrm{MMD}^2(f_1, f_2)$.

\begin{figure}[tb]
\centering
\includegraphics[height=0.4\textwidth]{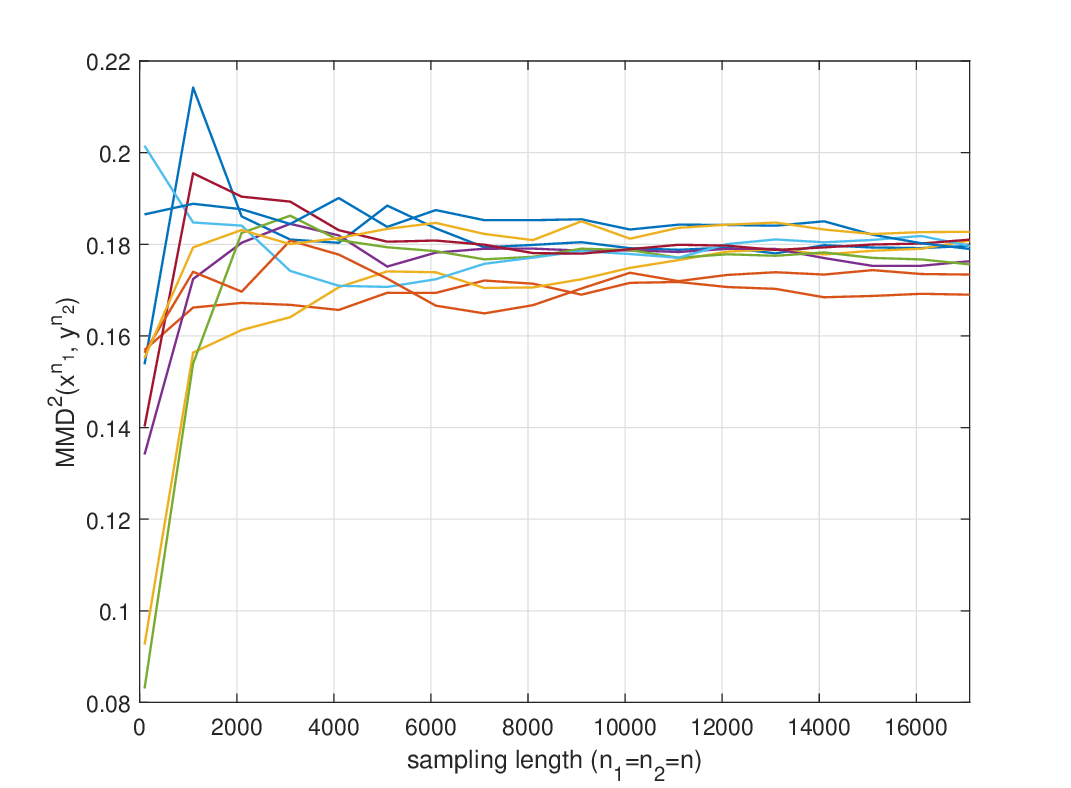}
\caption{The values of $\mathrm{MMD}^2(x^{n_1},y^{n_2})$ when $(x^{n_1},y^{n_2})$ are generated i.i.d. from two Gaussian distributions with mean values $(0,1)$ and the same variance $1$  when the Gaussian kernel is used and $n_1=n_2=n$.}
\label{MMDtest}
\end{figure}

To illustrate the MMD metric, in Fig. \ref{MMDtest},  we plot $10$ realizations of $\mathrm{MMD}^2(x^{n_1},y^{n_2})$ as a function of $n_1$ and $n_2$ when $f_1=\mathcal{N}(0,1)$, $f_2= \mathcal{N}(1,1)$, where $n_1=n_2=n$ increases from $100$ to $n=18000$ with the interval of 100. The results show that $\mathrm{MMD}^2(x^{n_1},y^{n_2})$ converges to its expected value $\mathrm{MMD}^2(f_1,f_2)$ as the sample size $n$ tends to infinity.

\begin{figure}[tb]
\centering
\includegraphics[height=0.4\textwidth]{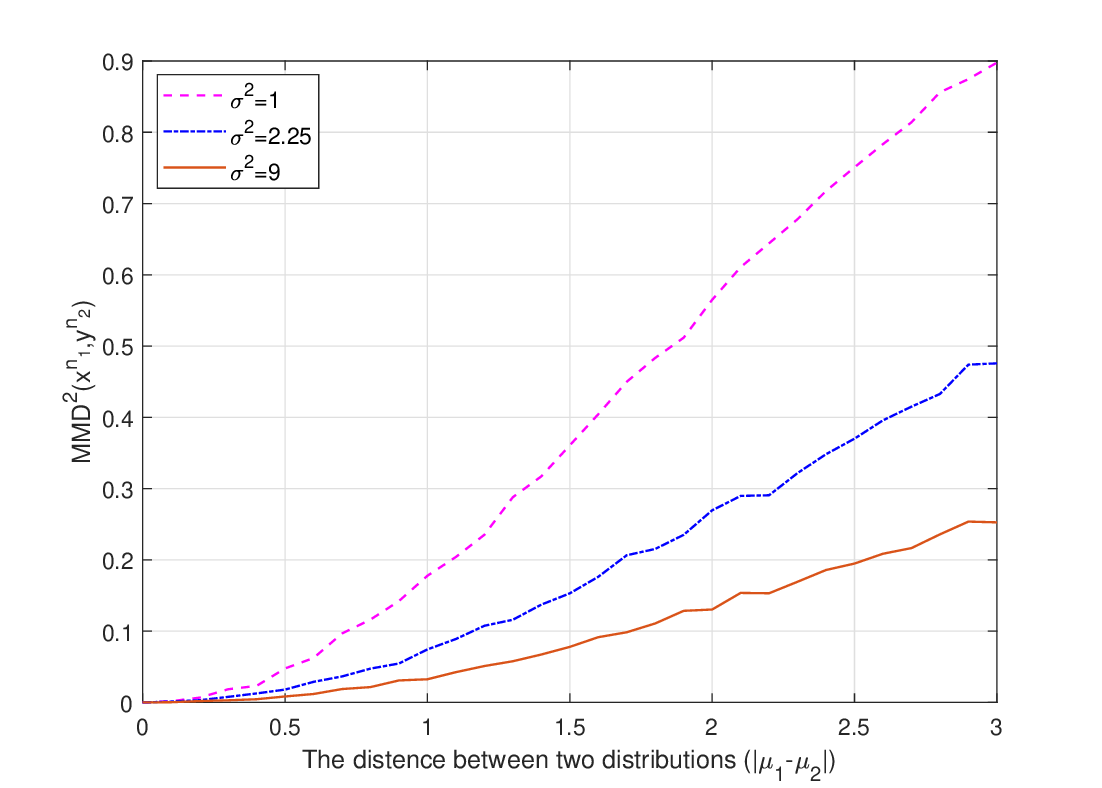}
\caption{The values of $\mathrm{MMD}^2(x^{n_1},y^{n_2})$ when $(x^{n_1},y^{n_2})$ are generated i.i.d. from two Gaussian distributions with mean values $(\mu_1,\mu_2)$ when the Gaussian kernel is used and $n_1=n_2=6000$.}
\label{MMDtestUn}
\end{figure}

In Fig. \ref{MMDtestUn}, we plot  $\mathrm{MMD}^2(x^{n_1},y^{n_2})$ when $n_1=n_2=6000$, $f_1=\mathcal{N}(\mu_1,\sigma^2)$, $f_2= \mathcal{N}(\mu_2,\sigma^2)$ as a function for the absolute difference between the mean values $(\mu_1,\mu_2)$ for different variance values of $\sigma^2$. When $\mu_1=\mu_2$, the two distributions $(f_1,f_2)$ are the same and it is observed that $\mathrm{MMD}^2(x^{n_1},y^{n_2})$ is roughly zero. As $|\mu_1-\mu_2|$ grows, the two distributions become more different and the value of $\mathrm{MMD}^2(x^{n_1},y^{n_2})$ increases. The MMD metric is a distance metric for the generating distributions of continuous sequences and can be used similarly as the KL divergence for the discrete sequences in~\cite{second,Universal_Outlier_Hypothesis} to design tests for outlier hypothesis testing with continuous sequences.

% \red{Comment: In the above figure, use $\sigma^2=1$ and so on to denote the variance and make sure to put the legend with the increasing order of variance.}

\section{Results for at Most One Outlier}
\label{Main}

In this section, we present our results for fixed-length, sequential and two-phase tests when there is at most one outlier. For each case, we present the test, explain the asymptotic intuition why the test works, present our theoretical results and discuss the significance of our results.

\subsection{Fixed-length Test}
\label{S-FLMT}

\subsubsection{Test Design and Asymptotic Intuition}
In a fixed-length test, the stopping time $\tau$ is fixed and set to be a positive integer $n\in\bbN$. To present our test, given any observed sequences $\by^n=(y_1^n,\ldots,y_M^n)$, we need the following scoring function for each $i\in[M]$:
\begin{equation}
\label{GB}
\rmG_i(\by^n)=\max_{j\in \calM_i}\mathrm{MMD}^2(y_j^n,\bar{\by}_{i,j}^n),
\end{equation}
where $\calM_i:=\{j\in[M],~j\neq i\}$ denotes all integers in $[M]$ except $i$, and $\bar{\by}_{i,j}^n:=\{y_k^n\}_{k\in[M]:k\neq i,k\neq j}$ collects all observed sequences $\by^n$ except $y_i^n$ and $y_j^n$. Furthermore, define the following two quantities:
\begin{align}
& i^*(\by^n):=\argmin_{i\in [M]}\rmG_i(\by^n),\label{MinG}\\
&h(\by^n):=\min_{i\in [M]:~i\neq i^*(\by^n)}\rmG_i(\by^n),\label{secMinG}
\end{align}
where $i^*(\by^n)$ denotes the index of sequence that has the smallest scoring function and $h(\by^n)$ denotes the second minimal value of all scoring functions.

For any positive real number $\lambda\in\bbR_+$, given observed sequences $\by^n$, our fixed-length test operates as follows:
\begin{equation}\label{FLTest}
\phi_n(\by^n)=
\left\{
\begin{aligned}
& \rmH_i: \text{ if } h(\by^n)>\lambda\mathrm{~and~}i^*(\by^n)=i;\\
& \rmH_\mathrm{r}:  \text{ otherwise}.
\end{aligned}
\right.
\end{equation}
Specifically, the test $\phi_n$ claims that $y_i^n$ is the outlier if the $i$-th scoring function $\rmG_i(\by^n)$ is smallest among all $M$ scoring functions and the second minimal scoring function $h(\by^n)=\min\limits_{k\in \calM_i}\rmG_k(\by^n)$ is above the threshold $\lambda$ while the test claims that there is no outlier if the second minimal scoring function $h(\by^n)$ is no greater than the threshold $\lambda$.

We now explain the asymptotic intuition why the test in \eqref{FLTest} works. First consider the case that the $i$-th sequence $y_i^n$ is the outlier for some $i\in[M]$. It follows that $\rmG_i(\by^n)\rightarrow 0$ as the sample size $n\rightarrow\infty$. On the other hand, for any $k\neq i$, the scoring function satisfies $\rmG_k(\by^n)\geq \mathrm{MMD}^2(y_i^n,\bar{\by}_{k,i}^n)\to \mathrm{MMD}^2(f_\rmN,f_\rmA)>0$ as $n\to\infty$. Thus,  $\min\limits_{k\in \calM_i}\rmG_k(\by^n)$ converges to a value greater than $ \mathrm{MMD}^2(f_\rmN,f_\rmA)$. Therefore, if the threshold satisfies $0<\lambda<\mathrm{MMD}^2(f_\rmN,f_\rmA)$, as the sample size $n$ increases, the outlier $y_i^n$ can always be found correctly. Next consider the case that there is no outlier. In this case, all sequences are generated i.i.d. from the unknown nominal distribution $f_\rmN$. Analogously, it follows that $\rmG_i(\by^n)\rightarrow 0$ for all $i\in[M]$. Thus, as long as $\lambda>0$, the correct decision $\rmH_\rmr$ would be decided by the fixed-length test $\phi_n$ when the sample size is large enough. In summary, when $0<\lambda<\mathrm{MMD}^2(f_\rmN,f_\rmA)$, no error event would be made asymptotically. In the next subsection, we show that all three kinds of error probabilities decay exponentially fast.

\subsubsection{Theoretical Results and Discussions}
Consider any kernel function $k(x,y)$ such that the maximum value is finite, i.e., $K_0:=\max_{(x,y)\in\calX\times\calY}k(x,y)<\infty$. One valid kernel function is the Gaussian kernel function in \eqref{Gaussiankernel}, which satisfies the above constraint with $K_0=1$. The following theorem characterizes the exponential decay rates of the probabilities of misclassification, false reject, and false alarm for the test in \eqref{FLTest} when the number of outliers is at most one.

\begin{theorem}\label{FLMT}
Under any pair of nominal and anomalous distributions $(f_\rmN,f_\rmA)$, for any positive real number $\lambda\in\bbR_+$, the test in \eqref{FLTest} ensures that
\begin{enumerate}
\item for each $i\in[M]$, the misclassification and false reject probabilities satisfy
\begin{align}
\liminf_{n\to\infty}-\frac{1}{n}\log \beta_i(\phi_n|f_\rmN,f_\rmA)
&\geq \frac{\lambda^2}{ 32K_0^2\left(1+\frac{1}{M-2}\right)},\\
\liminf_{n\to\infty}-\frac{1}{n}\log \zeta_i(\phi_n|f_\rmN,f_\rmA)
&\geq \frac{\left(\mathrm{MMD}^2(f_\rmN,f_\rmA)-\lambda\right)^2}{32K_0^2\left(1+\frac{1}{M-2}\right)}\mathbb{I}(\lambda<\mathrm{MMD}^2(f_\rmN,f_\rmA)),
\end{align}
where $\mathbb{I}(\cdot)$ is the indicator function.
\item the false alarm probability satisfies
\begin{align}
\liminf_{n\to\infty}-\frac{1}{n}\log \rmP_{\rm{FA}}(\phi_n|f_\rmN,f_\rmA)&\geq \frac{\lambda^2}{ 32K_0^2\left(1+\frac{1}{M-2}\right)}.
\end{align}
\end{enumerate}
\end{theorem}
The proof of Theorem \ref{FLMT} is provided in Appendix \ref{proof_of_FLMT}. The key point is to use the statistical properties of the scoring function $\rmG_i(\bY^n)$ for each $i\in\calM$ and apply the McDiarmid's inequality~\cite{mcdiarmid1989method}, which is recalled in Lemma \ref{McDiarmid}.

Analogously to \cite[Theorem 3]{Joint} for the discrete case, Theorem \ref{FLMT} implies that the threshold $\lambda$ tradeoffs the homogeneous misclassification and false alarm exponent $\frac{\lambda^2}{ 32K_0^2\left(1+\frac{1}{M-2}\right)}$ and the false reject exponent $\frac{\left(\mathrm{MMD}^2(f_\rmN,f_\rmA)-\lambda\right)^2}{32K_0^2\left(1+\frac{1}{M-2}\right)}$. Specifically, if $\lambda$ increases, the false reject exponent decreases while the homogeneous misclassification and false alarm exponent increases.

Theorem \ref{FLMT} implies that all three kinds of error probabilities decay exponentially with respect to the sample size $n$ if the threshold $\lambda$ satisfies that $0<\lambda<\mathrm{MMD}^2(f_\rmN,f_\rmA)$. Since the nominal and anomalous distributions $(f_\rmN,f_\rmA)$ are both unknown, choosing a threshold $\lambda$ determines the set of nominal and anomalous distributions for which all three kinds of error probabilities decay exponentially fast, which is given by $\calF(\lambda):=\big\{(f_\rmN,f_\rmA)\in\calF(\calR)^2:~\mathrm{MMD}^2[f_\rmN,f_\rmA]>\lambda\big\}$.

Furthermore, it follows from Theorem \ref{FLMT} that as the number of observed sequences $M$ increases, all three exponent rates increase. This result is consistent with the intuition and the corresponding achievability result~\cite[Theorem 3]{Joint} for the discrete case. This is because with more observed sequences, it is easier to estimate the nominal distribution and identify the outlier.

Finally, we remark that our fixed-length test is not comparable to the test in~\cite[Eq. (6)]{MMD} for a known number of outliers or \cite[Eq. (6)]{MMD} for unknown number of outliers. This is because for both tests in \cite{MMD} focus on the non-null hypothesis and the false alarm probability was not studied.

\subsection{Sequential test}\label{ST}

\subsubsection{Test Design and Asymptotic Intuition}

In this section, we present a sequential test with a random stopping time $\tau$ and the decision rule $\phi_\tau$. Let $N\in\bbN$ be a fixed integer. Given two positive real numbers $(\lambda_1,\lambda_2)\in\bbR_+^2$, for any observed sequences $\by^n=\{y_1^n,\ldots,y_M^n\}$, the stopping time $\tau$ of our sequential test is defined as follows:
\begin{align}
\tau
&=\inf\{n\in\bbN:~n\geq N-1,~\mathrm{and~}h(\by^n)>\lambda_1,\text{ or }h(\by^n)<\lambda_2\},\label{Taulength2}
\end{align}
where $N$ is a design parameter of the sequential test to avoid stopping too early. At the stopping time $\tau$, we run the fixed-length test in \eqref{FLTest} with $(n,\lambda)$ replaced by $(\tau,\lambda_1)$.
Note that if $\lambda_1\leq \lambda_2$, the stopping time $\tau$ always equals to $N-1$ and the sequential test reduces to the fixed-length test in \eqref{FLTest}. Thus, we require that $\lambda_1>\lambda_2$ to ensure the test has random stopping time and thus superior performance as demonstrated in Theorem \ref{SJMT}.

We next discuss the asymptotic intuition why the sequential test works. First consider the case that $y_i^n$ is the outlier for some $i\in[M]$. In this case, it follows that $\rmG_i(\by^n)\to 0$ and $\rmG_k(\by^n)\to \mathrm{MMD}^2(f_\rmN,f_\rmA)$ for any $k\neq i$. It follows that $h(\by^n)\to \mathrm{MMD}^2(f_\rmN,f_\rmA)$. Thus, when $N$ is sufficiently large, if $0<\lambda_2<\lambda_1<\mathrm{MMD}^2(f_\rmN,f_\rmA)$, the sequential test could make a correct decision. On the other hand, if there is no outlier, it follows that $\rmG_k(\by^n)\to 0$ for all $k\in[M]$ and thus $h(\by^n)\to 0$. When $N$ is sufficiently large and  $\lambda_2>0$, the null hypothesis could be correctly output. In summary, as long as the thresholds satisfy $0<\lambda_2<\lambda_1<\mathrm{MMD}^2(f_\rmN,f_\rmA)$, when $N$ is sufficiently large, the sequential test makes no error asymptotically. In the following theorem, we explicitly bound the average stopping time $\bbE[\tau]$ under each hypothesis and show that all three kinds of error probabilities decay exponentially fast with the \emph{same} exponent rate with respect to the average stopping time if $0<\lambda_2<\lambda_1<\mathrm{MMD}^2(f_\rmN,f_\rmA)$.

\subsubsection{Theoretical Results and Discussions}
\begin{theorem}\label{SJMT}
Under any pair of nominal and anomalous distributions $(f_\rmN,f_\rmA)$, for any positive real numbers $(\lambda_1,\lambda_2)\in\bbR_+^2$ such that $\lambda_1>\lambda_2$, our sequential test ensures that
\begin{enumerate}
\item  when $N$ is sufficiently large, the average stopping time satisfies
\begin{align}
\max_{i\in[M]}\bbE_{\bbP_i}[\tau]
&\le
\left\{
\begin{array}{ll}
N&\mathrm{if~}\lambda_1<\mathrm{MMD}^2(f_\rmN,f_\rmA),\\\infty&\mathrm{otherwise.}
\end{array}
\right.\\
\bbE_{\bbP_\rmr}[\tau]&\leq N.
\end{align}
\item for each $i\in[M]$, the misclassification and false reject error probabilities satisfy that if $\lambda_1<\mathrm{MMD}^2(f_\rmN,f_\rmA)$,
\begin{align}
\liminf_{N\rightarrow\infty}-\frac{1}{\bbE_{\bbP_i}[\tau]}\log \beta_i^{\rmseq}(\phi_\tau|f_\rmN,f_\rmA|)
&\geq \frac{\lambda_1^2}{32K_0^2\left(1+\frac{1}{M-2}\right)},\\
\liminf_{N\to\infty}-\frac{1}{\bbE_{\bbP_i}[\tau]}\log \zeta_i^{\rmseq}(\phi_\tau|f_\rmN,f_\rmA)
&\geq \frac{(\mathrm{MMD}^2(f_\rmN,f_\rmA)-\lambda_2)^2}{32K_0^2\left(1+\frac{1}{M-2}\right)},
\end{align}
\item the false alarm probability satisfies
\begin{align}
\liminf_{N\to\infty}-\frac{1}{\bbE_{\bbP_\rmr}[\tau]}\log \rmP^{\rmseq}_{\rm{FA}}(\phi_\tau|f_\rmN,f_\rmA)
\geq \frac{\lambda_1^2}{  32K_0^2\left(1+\frac{1}{M-2}\right)}.
\end{align}
\end{enumerate}
\end{theorem}

The proof of Theorem \ref{SJMT} is provided in Appendix \ref{proof_of_SJMT}. To prove Theorem \ref{SJMT}, we first upper bound the average stopping time under each non-null hypothesis and the null hypothesis. Subsequently, we use the similar idea to prove Theorem \ref{FLMT} to bound all three kinds of error probabilities,  including calculating the expected value of the scoring function and applying the McDiarmid's inequality~\cite{mcdiarmid1989method}.

It follows from Theorems \ref{FLMT} and \ref{SJMT} that both the sequential test and the fixed-length test achieve the same misclassification and false alarm exponent while the sequential test achieves a larger false reject exponent. Furthermore, the sequential test resolves the tradeoff between the false reject exponent and the homogeneous misclassification and false alarm exponent. The superior performance of the sequential test results from the freedom to stop at any possible time and the uses of two different thresholds $(\lambda_1,\lambda_2)$.

If one considers the Bayesian error probability criterion by calculating a weighted sum of three error probabilities, the error exponent is then the smallest exponent rate among exponent rates of three error probabilities. For the fixed-length test, the achievable Bayesian exponent is $\max_{\lambda\in(0,\mathrm{MMD}^2(f_\rmN,f_\rmA))}\min\bigg\{\frac{\lambda^2}{32K_0^2(1+\frac{1}{M-2})},\frac{(\mathrm{MMD}^2(f_\rmN,f_\rmA)-\lambda)^2}{32K_0^2(1+\frac{1}{M-2})}\bigg\}$, which equals to $\frac{(\mathrm{MMD}^2(f_\rmN,f_\rmA))^2}{32K_0^2(1+\frac{1}{M-2})}$ when $\lambda=\frac{1}{2}\mathrm{MMD}^2(f_\rmN,f_\rmA)$. In contrast, for the sequential test, the Bayesian exponent equals to
\begin{align}
\max_{(\lambda_1,\lambda_2)\in(0,\mathrm{MMD}^2(f_\rmN,f_\rmA)):\lambda_1>\lambda_2}
\min\left\{\frac{\lambda_1^2}{32K_0^2\left(1+\frac{1}{M-2}\right)},\frac{(\mathrm{MMD}^2(f_\rmN,f_\rmA)-\lambda_2)^2}{  32K_0^2\left(1+\frac{1}{M-2}\right)}\right\},
\end{align}
which is greater than $\frac{\left(\mathrm{MMD}^2(f_\rmN,f_\rmA)-\varepsilon\right)^2}{32K_0^2\left(1+\frac{1}{M-2}\right)}$ for any $\varepsilon\in(0,\frac{1}{2}\mathrm{MMD}^2(f_\rmN,f_\rmA))$ by setting $\lambda_2=\varepsilon$ and $\lambda_1=\mathrm{MMD}^2(f_\rmN,f_\rmA)-\varepsilon$. Thus, the Bayesian exponent of the sequential test is larger than the Bayesian exponent of the fixed-length test.

The above analyses show that the sequential test achieves significantly better performance than the fixed-length test. This mainly results from the freedom of choosing a random stopping time. Specifically, if it is challenging to identify the outlier among a tuple of sequences, one can choose to collect more samples until a reliable decision could be made. However, such a test design is complicated since one needs to check the stopping criterion after collecting each data sample. One might wonder whether we can achieve the superior performance of the sequential test with a simple test design similar to a fixed-length test. In the next subsection, we answer this question affirmatively by proposing a test with two possible stopping times and prove that the test balances the performance of a fixed-length or a sequential test by choosing different test parameters.

\subsection{Two-phase test}
\label{S-AFLJMT}

\subsubsection{Test Design and Asymptotic Intuition}
Fix two integers $(K,n)\in\bbN^2$. Similarly to the sequential test, our two-phase test has a random stopping time $\tau$. However, the difference is that $\tau$ can only take two possible values, either $n$ or $Kn$. This is the reason why the test is named the two-phase test. Fix three positive real numbers $(\lambda_1,\lambda_2,\lambda_3)\in\bbR_+^3$. For any observed sequences $\by^{Kn}=\{y_1^{Kn},\ldots,y_M^{Kn}\}$, the random stopping time $\tau$ satisfies
\begin{align}\label{Taulength}
\tau:=\left\{
\begin{aligned}
n :& \text{ if }h(\by^n)>\lambda_1 \text{ or } h(\by^n)<\lambda_2;\\
Kn :& \text{ otherwise}.
\end{aligned}
\right.
\end{align}
At the stopping time $\tau$, our two-phase test operates as follows. When $\tau=n$, $\phi_\tau(\by^\tau)$ applies the fixed-length test $\phi_n(\by^n)$ in \eqref{FLTest} with $\lambda$ replaced by $\lambda_1$; when $\tau=Kn$, $\phi_\tau(\by^\tau)$ applies the fixed-length test $\phi_n(\by^n)$ in \eqref{FLTest} with $n$ replaced by $Kn$ and $\lambda$ replaced by $\lambda_3$.

In summary, our two-phase test operates as follows. In the first phase, the test takes $n$ samples to perform a fixed-length test with a reject option. The reject decision is output when $\lambda_2<\min\limits_{k\in \calM_i}\rmG_k(\by^n)<\lambda_1$, which means that no hypothesis could be reliably decided and thus more samples are required. Once a reject decision is made in the first phase, our test proceeds in the second phase, where $(K-1)n$ additional samples are collected and a fixed-length test without a
rejection is used to make a final decision. Note that the thresholds $(\lambda_1,\lambda_2)$ play a similar role in the two-phase test as in the sequential test. The asymptotic intuition why the two-phase test works is highly similar to the sequential test and thus omitted.

\subsubsection{Theoretical Results and Discussions}
The performance of our two-phase test is characterized in the following theorem.
\begin{theorem}\label{AFLJMT}
Under any pair of nominal and anomalous distributions $(f_\rmN,f_\rmA)$, for any positive real numbers $(\lambda_1, \lambda_2, \lambda_3)\in\bbR_+^3$ such that $\lambda_2<\lambda_1$, our two-phase test ensures that
\begin{enumerate}
\item  when $n$ is sufficiently large, the average stopping time satisfies
\begin{align}
\max_{i\in[M]}\bbE_{\bbP_i}[\tau]&\leq
\left\{
\begin{array}{ll}
n+1, &\mathrm{if~}\lambda_1<\mathrm{MMD}^2(f_\rmN,f_\rmA),\\
Kn&\mathrm{otherwise.}
\end{array}
\right.\\
\bbE_{\bbP_\rmr}[\tau]&\leq n+1.
\end{align}
\item for each $i\in[M]$, the misclassification and false reject error probabilities satisfy that if $\lambda_1<\mathrm{MMD}^2(f_\rmN,f_\rmA)$,
\begin{align}
\liminf_{n\to\infty}-\frac{1}{\bbE_{\bbP_i}[\tau]}\log \beta^{\mathrm{tp}}_i(\phi_{\tau}|f_\rmN,f_\rmA)
&\geq \min\left\{\frac{\lambda_1^2}{ 32K_0^2\left(1+\frac{1}{M-2}\right)},\frac{K\lambda_3^2}{ 32K_0^2\left(1+\frac{1}{M-2}\right)}\right\},\\
\liminf_{n\to\infty}-\frac{1}{\bbE_{\bbP_i}[\tau]}\log \zeta^{\mathrm{tp}}_i(\phi_\tau|f_\rmN,f_\rmA)
&\geq \min\Bigg\{ \frac{\left(\mathrm{MMD}^2(f_\rmN,f_\rmA)-\lambda_2\right)^2}{32K_0^2\left(1+\frac{1}{M-2}\right)},\\*
&\qquad\qquad\frac{K\left(\mathrm{MMD}^2(f_\rmN,f_\rmA)-\lambda_3\right)^2}{32K_0^2\left(1+\frac{1}{M-2}\right)}\mathbb{I}(\lambda_3<\mathrm{MMD}^2(f_\rmN,f_\rmA))\Bigg\}.
\end{align}
Otherwise, if $\lambda_1\geq \mathrm{MMD}^2(f_\rmN,f_\rmA)$, the exponents of misclassification and false reject error probabilities are the same as those of fixed-length test with $\lambda_3$ playing the role of $\lambda$.
\item the false alarm probability satisfies
\begin{align}
\liminf_{n\to\infty}-\frac{1}{\bbE_{\bbP_\rmr}[\tau]}\log \rmP_\mathrm{FA}^{\mathrm{tp}}(\phi_n|f_\rmN,f_\rmA)
&\geq \min\left\{\frac{\lambda_1^2}{ 32K_0^2\left(1+\frac{1}{M-2}\right)},\frac{K\lambda_3^2}{ 32K_0^2\left(1+\frac{1}{M-2}\right)}\right\}.
\end{align}
\end{enumerate}
\end{theorem}

The proof of Theorem \ref{AFLJMT} is provided in Appendix \ref{proof_of_AFLJMT}. To prove Theorem \ref{AFLJMT}, we first upper bound the average stopping time $\bbE[\tau]$ under each non-null hypothesis and the null hypothesis, using the definition of the stopping time $\tau$ in \eqref{Taulength} and the equality $\bbE[X]=\sum_{a\in\bbN}\Pr\{X\geq a\}$ for any positive integer random variable $X$. Subsequently, we upper bound three kinds of error probabilities at the stopping time by generalizing the proof techniques for Theorem \ref{FLMT}.

Comparing Theorem \ref{AFLJMT} with Theorems \ref{FLMT} and \ref{SJMT}, we conclude that the two-phase test bridges over the fixed-length test and the sequential test.
Specifically, if we set $\lambda_1=\lambda_2=\lambda_3=\lambda$ and $K=1$, the two-phase test reduces to a fixed-length test. On the other hand, if $K$ is large enough, it follows from Theorem \ref{AFLJMT} that the achievable error exponents of the two-phase test are exactly the same as the sequential test. Analogous to the discussion of the Bayesian exponent below Theorem \ref{SJMT}, we list in Table. \ref{Table} the best achievable Bayesian exponent of the fixed-length test, the sequential test and the two-phase test, where $\varepsilon\in(0,\mathrm{MMD}^2(f_\rmN,f_\rmA))$ is arbitrary. As observed, by changing the design parameters, the two-phase test bridges over the fixed-length and the sequential test.
\begin{table}[!h]
\centering
\caption{The best achievable Bayesian exponent of the fixed-length test, the sequential test and the two-phase test.}
\begin{tabular}{ c| c| c| c}
\hline
Test & Fixed-length &Sequential & Two-phase\\
\hline
\makecell{The best achievable\\ Bayesian exponent} & $\frac{(\mathrm{MMD}^2(f_\rmN,f_\rmA))^2}{32K_0^2(1+\frac{1}{M-2})}$
& $\frac{(\mathrm{MMD}^2(f_\rmN,f_\rmA)-\varepsilon)^2}{32K_0^2(1+\frac{1}{M-2})}$
& $ \min\left\{\frac{\left(\mathrm{MMD}^2(f_\rmN,f_\rmA)-\varepsilon\right)^2}{32K_0^2\left(1+\frac{1}{M-2}\right)},\frac{K(\mathrm{MMD}^2(f_\rmN,f_\rmA))^2}{32K_0^2\left(1+\frac{1}{M-2}\right)} \right\}$\\
\hline
\end{tabular}
\label{Table}
\end{table}

\section{Results for Multiple Outliers}
\label{Main_unS}
In this section, we present our results when there might exist multiple outliers and the number of outliers is unknown.

\subsection{Fixed-length Test}\label{FLMT_unS}
Recall that the unknown number of outliers $s$ satisfies $0\leq s\leq T=\lceil \frac{M}{2}\rceil-1$, $\calC_t=\{\calB\subseteq\calM:~|\calB|=t\}$ for each $t\in[T]$ collects all subsets of $\calM$ of size $t$ and $\calC=\bigcup_{t\in[T]}\calC_t$. To present our test, we generalize the scoring function in \eqref{GB} as follows. Given any sequences $\by^n$, for each $\calB\in\calC$, define the following scoring function:
\begin{equation}\label{GB2}
\rmG_\calB(\by^n)=\max_{j\in \calM_\calB}\mathrm{MMD}^2(y_j^n,\bar{\by}_{\calB,j}^n),
\end{equation}
where $\calM_\calB:=\{j\in[M],~j\notin\calB\}$ denotes all indices in $[M]$ that do not belong to the set $\calB$, and $\bar{\by}_{\calB,j}^n:=\{y_i^n\}_{i\in\calM_\calB:~i\neq j}$ collects all sequences $\by_{\calM_\calB}^n$ except $y_j^n$. Furthermore, for each $t\in[T]$, define the following two quantities:
\begin{align}
& \calI_t^*(\by^n):=\argmin_{\calB\in \calC_{t}}\rmG_\calB(\by^n),\label{MinG2}\\
&h_t(\by^n):=\min_{\calD\in \calC_{t}:~\calD\neq\calI_t^*(\by^n)}\rmG_\calD(\by^n),\label{secMinG2}
\end{align}
where $\calI_t^*(\by^n)$ denotes the set that minimizes the scoring function with size $t$ and $h_t(\by^n)$ denotes the second minimal value of the scoring function.

\subsubsection{Test Design and Asymptotic Intuition}
Since the number of outliers $s$ is unknown, our task is two fold: estimate the number of outliers as $\hats$ and identify the set of outliers. Using the scoring function, given observed sequences $\by^n$, our test estimates the number of outliers as follows:
\begin{align}
\hats:=
\left\{
\begin{array}{ll}
t\in[T]&\mathrm{if~}h_t(\by^n)>\lambda\mathrm{~and~}\forall~\tilt\in[t+1:T],~h_{\tilt}(\by^n)< \lambda,\\
0&\mathrm{otherwise}.
\end{array}
\right.
\label{sdetect}
\end{align}
If $\hats=0$, a null hypothesis $\rmH_\rmr$ is decided. Otherwise, our test further identifies the set of outliers using the following minimal scoring function test:
\begin{equation}
\label{FLTest_un}
\phi_n(\by^n)=\calI_{\hats}^*(\by^n).
\end{equation}

In summary, the test $\phi_n$ claims that there is no outlier if the second minimal scoring function $h_t(\by^n)$ is less than the threshold $\lambda$ for each $t\in[T]$. Otherwise, the test $\phi_n$ estimates the number of outliers as the largest number in $[T]$ such that the second minimal scoring function $h_t(\by^n)$ is greater than the threshold $\lambda$. Subsequently, given a positive estimated number $\hats$, the test $\phi_n$ identifies $\by_\calB^n$ as the indices of outliers if the scoring function $\rmG_\calB(\by^n)$ is smallest among all $|\calC_{\hats}|$ scoring functions.

We now explain the asymptotic intuition why the above test works. First consider $s>0$ and let $\calB\in\calC_s$ denote the indices set of outliers. For each $t>s$, the set $\calC_t^\calB:=\{\calD\in\calC_t:~\calB\subseteq\calD\}$ has size of at least $M-s>2$. Given any set $\calD\in\calC_t^\calB$, the scoring function satisfies $\rmG_\calD(\by^n)\rightarrow 0$ as $n\rightarrow\infty$ since each sequence $y_j^n$ with $j\in\calM_\calD$ is generated from the same nominal distribution. Thus, for any $t>s$, the second minimal scoring function satisfies $\liminf_{n\to\infty}h_t(\by^n)\to 0$. When $t=s$, the scoring function $\rmG_{\calB}(\by^n)\to 0$. For any $\calD\in\calC_s$ such that $\calD\neq\calB$,  the scoring function satisfies that as $n\to\infty$,
\begin{align}
\rmG_{\calD}(\by^n)\geq\max_{j\in \calM_\calD\cap\calB} \mathrm{MMD}^2(y_j^n,\bar{\by}_{\calD,j}^n)\to \left(1-\frac{s}{M-s-1}\right)^2\mathrm{MMD}^2(f_\rmN,f_\rmA).
\end{align}
Thus, as $n\to\infty$, $h_s(\by^n)$ converges to a value greater than $\Big(1-\frac{s}{M-s-1}\Big)^2\mathrm{MMD}^2(f_\rmN,f_\rmA)$. Therefore, the correct number of outliers could be estimated if $0<\lambda<\Big(1-\frac{s}{M-s-1}\Big)^2\mathrm{MMD}^2(f_\rmN,f_\rmA)$. The intuition why the test can identify the correct set of outliers is very similar to the case of at most one outlier and thus omitted. Next consider the case of $s=0$. In this case, it follows that for any set $\calD\in\calC$, the scoring function satisfies $\rmG_\calD(\by^n)\rightarrow 0$ as $n\rightarrow\infty$. Thus, if $\lambda>0$, the correct decision of null hypothesis can be made.

\subsubsection{Theoretical Results and Discussions}

For ease of notation, given any non-negative real number $y\in\bbR_+$, define the following two exponent functions:
\begin{align}
g_1(y|f_\rmN,f_\rmA)
&:=\frac{y^2}{32K_0^2\left(1+\frac{1}{M-T-1}\right)}.\label{nota1}\\
g_2(y|f_\rmN,f_\rmA)
&:=\frac{\left(\left(1-\frac{s}{M-s-1}\right)^2\mathrm{MMD}^2(f_\rmN,f_\rmA)-y\right)^2}{32K_0^2\left(1+\frac{1}{M-s-1}\right)}.\label{nota2}
\end{align}

The following theorem characterizes the exponential decay rates of the probabilities of misclassification, false reject, and false alarm for the test in \eqref{FLTest_un} when the number of outliers is at most $T$.

\begin{theorem}\label{FLMT_Un}
Under any pair of nominal and anomalous distributions $(f_\rmN,f_\rmA)$, given any positive real number $\lambda\in\bbR_+$,
the test in \eqref{FLTest_un} ensures that
\begin{enumerate}
\item when $s>1$ and $\lambda<\left(1-\frac{s}{M-s-1}\right)^2\mathrm{MMD}^2(f_\rmN,f_\rmA)$, for each $\calB\in\calC_s$, the misclassification and false reject probabilities satisfy
\begin{align}
\liminf_{n\to\infty}-\frac{1}{n}\log \beta_\calB(\phi_n|f_\rmN,f_\rmA)
&\geq
\min\left\{g_1(\lambda|f_\rmN,f_\rmA),g_2(\lambda|f_\rmN,f_\rmA)\right\},\\
\liminf_{n\to\infty}-\frac{1}{n}\log \zeta_\calB(\phi_n|f_\rmN,f_\rmA)
&\geq
g_2(\lambda|f_\rmN,f_\rmA)\label{misexponent:uk},
\end{align}
\item when $s=1$ and $\lambda<\mathrm{MMD}^2(f_\rmN,f_\rmA)$, for each $\calB\in\calC_s$, the misclassification and false reject probabilities satisfy
\begin{align}
\liminf_{n\to\infty}-\frac{1}{n}\log \beta_\calB(\phi_n|f_\rmN,f_\rmA)
&\geq g_1(\lambda|f_\rmN,f_\rmA),\\
\liminf_{n\to\infty}-\frac{1}{n}\log \zeta_\calB(\phi_n|f_\rmN,f_\rmA)
&\geq\frac{\left(\mathrm{MMD}^2(f_\rmN,f_\rmA)-\lambda\right)^2}{ 32K_0^2\left(1+\frac{1}{M-2}\right)}.
\end{align}
\item the false alarm probability satisfies
\begin{align}
\liminf_{n\to\infty}-\frac{1}{n}\log \rmP_{\rm{FA}}(\phi_n|f_\rmN,f_\rmA)&\geq g_1(\lambda|f_\rmN,f_\rmA).
\end{align}
\end{enumerate}
\end{theorem}

The proof of Theorem \ref{FLMT_Un} is provided in Appendix \ref{proof_of_FLMT_Un}, which generalizes the proof of Theorem \ref{FLMT} by considering multiple outliers with a more complicated analyses for misclassification probabilities, where additional analyses for the probability of estimating the number of outliers incorrectly are required. In particular, under each non-null hypothesis, a misclassification event occurs if the number of outliers is estimated incorrectly or if a wrong set of sequences is claimed as outliers. The event of estimating the number of outliers incorrectly can be further decomposed to the event that the estimated number $\hatS>s$ and the other event that $0<\hatS<s$, where the latter event occurs if $s>1$. The exponents for the probability of the two events lead to two exponent terms inside the minimization. Furthermore, the exponent for the probability of finding a wrong set of sequences as outliers equals to $\frac{\lambda^2}{ 32K_0^2\left(1+\frac{1}{M-s-1}\right)}$, which is lower bounded by $\frac{\lambda^2}{ 32K_0^2\left(1+\frac{1}{M-T-1}\right)}$ since $s\leq T$.

Theorem \ref{FLMT_Un} implies that when $s>1$, under each non-null hypothesis $\rmH_\calB$, the misclassification exponent is lower bounded by the minimal value of the exponent lower bounds for false reject and false alarm probabilities.  This is because when the number of outliers is unknown, the misclassification error event could occur under both the cases of correct and incorrect estimate of $s$.
We show in Appendix \ref{proof_of_FLMT_Un} that the analyses of the false reject and false alarm probabilities share similarly to the analyses of the probabilities of the events $0<\hatS<s$ and $\hatS>s$, respectively.

Similar to the case of at most one outlier, one can also consider the Bayesian error probability criterion for the fixed-length test with multiple outliers when the number of outliers is unknown. Specifically, for $s>1$, the achievable Bayesian exponent is lower bounded by
\begin{align}
\max_{\lambda\in\left(0,\left(1-\frac{s}{M-s-1}\right)^2\mathrm{MMD}^2(f_\rmN,f_\rmA)\right)}\min\bigg\{g_1(\lambda|f_\rmN,f_\rmA),g_2(\lambda|f_\rmN,f_\rmA)\bigg\}
&\geq\frac{\left(1-\frac{s}{M-s-1}\right)^4(\mathrm{MMD}^2(f_\rmN,f_\rmA))^2}{32K_0^2(1+\frac{1}{M-T-1})},
\end{align}
where the inequality is achieved when $\lambda=\frac{1}{2}\left(1-\frac{s}{M-s-1}\right)^2\mathrm{MMD}^2(f_\rmN,f_\rmA)$. The Bayesian exponent decreases when $s$ increases. This is because the second term inside the minimization decreases as $s$ increases since both $\Big(\frac{s}{M-s-1},\frac{1}{M-s-1}\Big)$ increase in $s$. Intuitively, this is because when the number of outliers increases, the differences among scoring functions decrease, which makes it more challenging to identify the correct set of outliers. Specifically, for any $\calD$ that collects the indices of both nominal sequences and outliers, the scoring function $\max_{j\in \calM_\calD\cap\calB} \mathrm{MMD}^2(y_j^n,\bar{\by}_{\calD,j}^n)$ decreases when the number $s$ of outliers increases. This is because the mixture distribution of $\bar{\by}_{\calD,j}^n$ will be closer to $f_\rmA$ when $s$ increases.

Furthermore, the case of at most outlier corresponds to $T=1$ and the result in Theorem \ref{FLMT} is recovered from Theorem \ref{FLMT_Un} by setting $T=1$. However, we note that when $T>1$, even if $s=1$, the misclassification exponent under each hypothesis $\rmH_\calB$ with $\calB\in\calC_s$ is smaller compared to the case of $T=1$. This is because when $T>1$, we need to consider the error event that the estimated number of outliers is greater than $s$, which results in the exponent $\frac{\lambda^2}{ 32K_0^2\left(1+\frac{1}{M-T-1}\right)}$.

When the number of outliers $s>0$ is known, there would be no false alarm event and any reject decision is a false reject, which indicates that further investigation is required to identify the correct set of outliers. In this case, we can apply the test in \eqref{FLTest} with $(i^*(\by^n),h(\by^n))$ replaced by $(\calI_s^*(\by^n),h_s(\by^n))$, respectively.  Similarly to the proof of Theorem \ref{FLMT_Un}, for each $\calB\in\calC_s$, the misclassification and false reject probabilities satisfy
\begin{align}
\liminf_{n\to\infty}-\frac{1}{n}\log \beta_\calB(\phi_n|f_\rmN,f_\rmA)
&\geq
\frac{\lambda^2}{ 32K_0^2\left(1+\frac{1}{M-s-1}\right)},\label{misKnowns}\\
\liminf_{n\to\infty}-\frac{1}{n}\log \zeta_\calB(\phi_n|f_\rmN,f_\rmA)
&\geq g_2(\lambda|f_\rmN,f_\rmA).\label{frejectKnowns}
\end{align}
Note that both exponents are no less than the corresponding results of unknown $s$. The above results imply that there is a penalty in the exponents when the number of outliers is unknown. In Fig. \ref{detection_error_multi2}, we provide a numerical example to show the penalty of not knowing the number of outliers.

Finally, we remark that Theorem \ref{FLMT_Un} generalizes \cite[Theorem 6]{second} for discrete sequences to continuous sequences, analogous to how Theorem \ref{FLMT} generalizes \cite[Theorem 3]{second}. In fact, \cite[Theorem 6]{second} holds only if the number of outliers is either zero or known as a positive number $s$. In contrast, Theorem \ref{FLMT_Un} holds when the number of outliers is \emph{unknown}, which can be any number from zero to the maximal allowable number of outliers $T$. To deal with the more complicated case, Theorem \ref{FLMT_Un} further solves the problem of estimating the number of outliers and introduces additional terms inside the misclassification exponent when $s>1$. When specializing Theorem \ref{FLMT_Un} to the setting in \cite[Theorem 6]{second}, we can apply the same test when the number of outliers is known in the last comment, and consider the additional possible event of false alarm. Analogously to the proof of Theorem \ref{FLMT_Un}, one can show that the misclassification and false reject probabilities satisfy \eqref{misKnowns} and \eqref{frejectKnowns} while the false alarm probability satisfies
\begin{align}
\liminf_{n\to\infty}-\frac{1}{n}\log \rmP_{\rm{FA}}(\phi_n|f_\rmN,f_\rmA)&\geq \frac{\lambda^2}{32K_0^2\left(1+\frac{1}{M-s-1}\right)}.
\end{align}

\subsection{Sequential test}\label{ST_un}

\subsubsection{Test Design and Asymptotic Intuition}

In this section, we present a sequential test with a random stopping time $\tau$ and the decision rule $\phi_\tau$. Let $N\in\bbN$ be a fixed integer. Given two positive real numbers $(\lambda_1,\lambda_2)\in\bbR_+^2$, for any observed sequences $\by^n=\{y_1^n,\ldots,y_M^n\}$, the stopping time $\tau$ of our sequential test is defined as follows:
\begin{align}
\tau
\nn=\inf\{n\in\bbN:~n\geq N-1, &\text{ and }\exists~ t\in[T]:~\forall~ \bar{t}\in[t+1:T] ,~h_{\bar{t}}(\by^n)<\lambda_2,~h_t(\by^n)>\lambda_1,\\
&\text{ or }\forall~ t\in[T],~h_t(\by^n)<\lambda_2\},\label{Taulength2_unS}
\end{align}
where $N$ is a design parameter of the sequential test to avoid stopping too early. At the stopping time $\tau$, given sequences $\by^\tau$,
our sequential test estimates the number of outliers as follows:
\begin{align}
\hats:=
\left\{
\begin{array}{ll}
t\in[T]&\mathrm{if~}h_t(\by^\tau)>\lambda_1\mathrm{~and~}\forall~\tilt\in[t+1:T],~h_{\tilt}(\by^\tau)< \lambda_2\\
0&\mathrm{otherwise}
\end{array}
\right.
\label{sdetect_ST}
\end{align}
If $\hats=0$, a null hypothesis $\rmH_\rmr$ is decided. Otherwise, our test further identifies the set of outliers using the minimal scoring function test specified in \eqref{FLTest_un} with $n$ replaced by $\tau$.
Note that if $\lambda_1\leq \lambda_2$, the stopping time $\tau$ always equals to $N-1$ and the sequential test reduces to the fixed-length test in \eqref{FLTest_un}. Thus, we require that $\lambda_1>\lambda_2$ to ensure the test has random stopping time and thus superior performance as demonstrated in Theorem \ref{SJMT_un}.

The asymptotic intuition why the sequential test works follows by combining the corresponding arguments for the sequential test tailored to the case of at most one outlier and the arguments for fixed-length test for an unknown number of outliers.
% We next discuss the asymptotic intuition why the sequential test works. First consider the case that a nonempty set $\calB\in\calC$ denotes the indices of outliers such that the number of outliers is $s=|\calB|$. In this case, as $n\to\infty$, for any $\calD\in$

% it follows that $\rmG_\calD(\by^n)\to 0$ for $\calD\supseteq\calB$ and $\rmG_\calD(\by^n)\to \left(1-\frac{s}{M-s-1}\right)^2\mathrm{MMD}^2(f_\rmN,f_\rmA)$ for any $\calD\in\calC_s$ and $\calD\neq \calB$. Thus, similar as the fixed-length test, when $N$ is sufficiently large, if $0<\lambda_2<\lambda_1<\left(1-\frac{s}{M-s-1}\right)^2\mathrm{MMD}^2(f_\rmN,f_\rmA)$, the sequential test could find correct number of outliers. The intuition why the test can identify the correct set of outliers is very similar to the case of at most one outlier and thus omitted. On the other hand, if there is no outlier, it follows that for any set $\calD\in\calC$, the scoring function satisfies $\rmG_\calD(\by^n)\rightarrow 0$ as $n\rightarrow\infty$. Thus, if $\lambda_2>0$, the null hypothesis could be correctly output.

\subsubsection{Theoretical Results and Discussions}

Recall the definitions of  $g_1(y|f_\rmN,f_\rmA)$ and $g_2(y|f_\rmN,f_\rmA)$ in \eqref{nota1} and \eqref{nota2}, respectively. The following theorem characterizes the exponential decay rates of the probabilities of misclassification, false reject, and false alarm for the sequential test when the number of outliers is unknown.

\begin{theorem}\label{SJMT_un}
Under any pair of nominal and anomalous distributions $(f_\rmN,f_\rmA)$, given any positive real numbers $(\lambda_1,\lambda_2)\in\bbR_+^2$ such that $\lambda_1>\lambda_2$, our sequential test ensures that
\begin{enumerate}
\item  when $N$ is sufficiently large and $s>1$, the average stopping time satisfies
\begin{align}
\max_{\calB\in\calC}\bbE_{\bbP_\calB}[\tau]
&\le
\left\{
\begin{array}{ll}
N&\mathrm{if~}\lambda_1<\left(1-\frac{s}{M-s-1}\right)^2\mathrm{MMD}^2(f_\rmN,f_\rmA),\\\infty&\mathrm{otherwise.}
\end{array}
\right.\\
\bbE_{\bbP_\rmr}[\tau]&\leq N.
\end{align}
If $s=1$, the results above still hold by replacing $\left(1-\frac{s}{M-s-1}\right)^2\mathrm{MMD}^2(f_\rmN,f_\rmA)$ with $\mathrm{MMD}^2(f_\rmN,f_\rmA)$.
\item when $s>1$ and $\lambda_1<\left(1-\frac{s}{M-s-1}\right)^2\mathrm{MMD}^2(f_\rmN,f_\rmA)$, for each $\calB\in\calC_s$, the misclassification and false reject probabilities satisfy
\begin{align}
\liminf_{N\to\infty}-\frac{1}{\bbE_{\bbP_\calB}[\tau]}\log \beta_\calB^{\mathrm{seq}}(\phi_n|f_\rmN,f_\rmA)
&\geq
\min\left\{g_1(\lambda_1|f_\rmN,f_\rmA),g_2(\lambda_2|f_\rmN,f_\rmA)\right\},\\
\liminf_{N\to\infty}-\frac{1}{\bbE_{\bbP_\calB}[\tau]}\log \zeta_\calB^{\mathrm{seq}}(\phi_n|f_\rmN,f_\rmA)
&\geq
g_2(\lambda_2|f_\rmN,f_\rmA)\label{STmisexponent:uk},
\end{align}
\item when $s=1$ and $\lambda_1<\mathrm{MMD}^2(f_\rmN,f_\rmA)$, for each $\calB\in\calC_s$, the misclassification and false reject probabilities satisfy
\begin{align}
\liminf_{N\to\infty}-\frac{1}{\bbE_{\bbP_\calB}[\tau]}\log \beta_\calB^{\mathrm{seq}}(\phi_n|f_\rmN,f_\rmA)
&\geq g_1(\lambda_1|f_\rmN,f_\rmA),\\
\liminf_{N\to\infty}-\frac{1}{\bbE_{\bbP_\calB}[\tau]}\log \zeta_\calB^{\mathrm{seq}}(\phi_n|f_\rmN,f_\rmA)
&\geq\frac{\left(\mathrm{MMD}^2(f_\rmN,f_\rmA)-\lambda_2\right)^2}{ 32K_0^2\left(1+\frac{1}{M-2}\right)}.
\end{align}
\item the false alarm probability satisfies
\begin{align}
\liminf_{N\to\infty}-\frac{1}{\bbE_{\bbP_\rmr}[\tau]}\log \rmP^{\mathrm{seq}}_{\rm{FA}}(\phi_n|f_\rmN,f_\rmA)&\geq g_1(\lambda_1|f_\rmN,f_\rmA).
\end{align}
\end{enumerate}
\end{theorem}
The proof of Theorem \ref{SJMT_un} is similar to the proof of Theorem \ref{SJMT} and provided in Appendix \ref{proof_of_SJMT_un} for completeness.

Analogous to the result for the fixed-length test in Theorem \ref{FLMT_Un}, the misclassification exponent of the sequential test equals to the minimum of the false reject and false alarm exponents. Comparing Theorems \ref{FLMT_Un} and \ref{SJMT_un}, we conclude that the sequential test  resolves the tradeoff between the false reject exponent and the false alarm exponent of the fixed-length test. Analogous to the case of at most one outlier, when the number of outliers is unknown, the superior performance of the sequential test also results from the freedom to stop at any possible time and the uses of two different thresholds $(\lambda_1,\lambda_2)$. If one considers the Bayesian error probability by calculating a weighted sum of three error probabilities,
for $s>1$, the Bayesian exponent is lower bounded by
\begin{align}
\max_{(\lambda_1,\lambda_2)\in\left(0,\left(1-\frac{s}{M-s-1}\right)^2\mathrm{MMD}^2(f_\rmN,f_\rmA)\right):\lambda_1>\lambda_2}
\min\left\{g_1(\lambda_1|f_\rmN,f_\rmA),g_2(\lambda_2|f_\rmN,f_\rmA)\right\},
\end{align}
which is greater than $\frac{\left(\left(1-\frac{s}{M-s-1}\right)^2\mathrm{MMD}^2(f_\rmN,f_\rmA)-\varepsilon\right)^2}{32K_0^2\left(1+\frac{1}{M-T-1}\right)}$ for any $\varepsilon\in\Big(0,\Big(1-\frac{s}{M-s-1}\Big)^2\mathrm{MMD}^2(f_\rmN,f_\rmA)\Big)$ when $\lambda_2=\varepsilon$ and $\lambda_1=\Big(1-\frac{s}{M-s-1}\Big)^2\mathrm{MMD}^2(f_\rmN,f_\rmA)-\varepsilon$.
Since the number of outliers is unknown and might be greater than one, the achievable Bayesian exponent is smaller than the corresponding result for the case of at most one outlier discussed below Theorem \ref{SJMT}. When $T=1$, the Bayesian exponent reduces to the result for the case of at most one outlier. Analogous to the fixed-length test, the Bayesian exponent decreases when the number of outliers $s$ increases.

\subsection{Two-phase test}\label{AFLMTest_unS}

\subsubsection{Test Design and Asymptotic Intuition}

Fix two integers $(K,n)\in\bbN^2$. Similarly to the case of at most one outlier, our two-phase test also has a random stopping time $\tau$ which only take two possible values, either $n$ or $Kn$. Given three positive real numbers $(\lambda_1,\lambda_2,\lambda_3)\in\bbR_+^3$, for any observed sequences $\by^{Kn}=\{y_1^{Kn},\ldots,y_M^{Kn}\}$, the random stopping time satisfies
\begin{equation}\label{Taulength_unS}
\tau:=\left\{
\begin{aligned}
n :& \text{ if }\exists~ t\in[T]:~\forall~ \bar{t}\in[t+1:T] ,~h_{\bar{t}}(\by^n)<\lambda_2,~h_t(\by^n)>\lambda_1,~\text{ or }\forall~ t\in[T],~h_t(\by^n)<\lambda_2,\\
Kn :& \text{ otherwise}.
\end{aligned}
\right.
\end{equation}
Specifically, the test stops at $\tau=n$ if the null hypothesis is believed to be true via the condition that $\forall~ t\in[T]$, $h_t(\by^n)<\lambda_2$ or if a non-null hypothesis with $t$ number of outliers is believed to be true via the condition that $h_t(\by^n)>\lambda_1$ and $\forall~\bar{t}\in[t+1:T]$, $h_{\bar{t}}(\by^n)<\lambda_2$. It both conditions are not satisfied, the test proceeds to collect $(K-1)n$ samples and stops at $\tau=Kn$.

At the stopping time $\tau$, given sequences $\by^\tau$,
our two-phase test operates as follows. When $\tau=n$, our test estimates the number of outliers using \eqref{sdetect} and identifies the set of outliers using the minimal scoring function test $\phi_n(\by^n)$ in \eqref{FLTest_un}. When $\tau=Kn$, we estimate the number of outliers using \eqref{sdetect} with $n$ replaced by $Kn$ and then apply the fixed-length test $\phi_n(\by^n)$ in \eqref{FLTest_un} with $(n,\lambda)$ replaced by $(Kn,\lambda_3)$. In summary, our two-phase test consists of two fixed-length tests using $n$ and $Kn$ samples, respectively. Thus, the asymptotic intuition why the two-phase test works follows from the arguments for the fixed-length test and is omitted. Note that when $\lambda_2>\lambda_1$, we always have $\tau=n$ and thus the two-phase test reduces to the fixed-length test. Thus, to avoid degenerate cases, we require that $\lambda_2<\lambda_1$ as in the sequential test.

\subsubsection{Theoretical Results and Discussions}
Recall the definitions of  $g_1(y|f_\rmN,f_\rmA)$ and $g_2(y|f_\rmN,f_\rmA)$ in \eqref{nota1} and \eqref{nota2}, respectively. The performance of our two-phase test is characterized in the following theorem.
\begin{theorem}\label{E-AFLJMT}
Under any pair of nominal and anomalous distributions $(f_\rmN,f_\rmA)$, given any positive real numbers $(\lambda_1,\lambda_2,\lambda_3)\in\bbR_+^3$ such that $\lambda_1>\lambda_2$, our two-phase test ensures that
\begin{enumerate}
\item when $n$ is sufficiently large and $s>1$, the average stopping time satisfies
\begin{align}
\max_{\calB\in\calC}\bbE_{\bbP_\calB}[\tau]&\leq
\left\{
\begin{array}{ll}
n+1, &\mathrm{if~}\lambda_1<\left(1-\frac{s}{M-s-1}\right)^2\mathrm{MMD}^2(f_\rmN,f_\rmA),\\
Kn&\mathrm{otherwise.}
\end{array}
\right.\\
\bbE_{\bbP_\rmr}[\tau]&\leq n+1.
\end{align}
If $s=1$, the results above still hold by replacing $\left(1-\frac{s}{M-s-1}\right)^2\mathrm{MMD}^2(f_\rmN,f_\rmA)$ with $\mathrm{MMD}^2(f_\rmN,f_\rmA)$.
\item
when $s>1$ and $\max\{\lambda_1,\lambda_3\}<\left(1-\frac{s}{M-s-1}\right)^2\mathrm{MMD}^2(f_\rmN,f_\rmA)$, for each $\calB\in\calC_s$, the misclassification and false reject probabilities satisfy
\begin{align}
\liminf_{n\to\infty}-\frac{1}{\bbE_{\rmP_\calB}[\tau]}\log \beta^\rmtp_{\calB}(\phi_n|f_\rmN,f_\rmA)
&\geq \min\left\{g_1(\lambda_1|f_\rmN,f_\rmA),Kg_1(\lambda_3|f_\rmN,f_\rmA),g_2(\lambda_2|f_\rmN,f_\rmA),Kg_2(\lambda_3|f_\rmN,f_\rmA)\right\},\\
\liminf_{n\to\infty}-\frac{1}{\bbE_{\rmP_\calB}[\tau]}\log \zeta^\rmtp_{\calB}(\phi_n|f_\rmN,f_\rmA)
&\geq \min\left\{g_2(\lambda_2|f_\rmN,f_\rmA),Kg_2(\lambda_3|f_\rmN,f_\rmA)\right\},
\end{align}

If $\lambda_1\geq \left(1-\frac{s}{M-s-1}\right)^2\mathrm{MMD}^2(f_\rmN,f_\rmA)$, the exponents of misclassification and false reject error probabilities are the same as those of fixed-length test with $\lambda_3$ playing the role of $\lambda$.

\item when $s=1$ and $\max\{\lambda_1,\lambda_3\}<\mathrm{MMD}^2(f_\rmN,f_\rmA)$, for each $\calB\in\calC_s$, the misclassification and false reject probabilities satisfy
\begin{align}
\liminf_{n\to\infty}-\frac{1}{\bbE_{\rmP_\calB}[\tau]}\log \beta^\rmtp_{\calB}(\phi_n|f_\rmN,f_\rmA)
&\geq \min\left\{ g_1(\lambda_1|f_\rmN,f_\rmA), Kg_1(\lambda_3|f_\rmN,f_\rmA)\right\},\\
\liminf_{n\to\infty}-\frac{1}{\bbE_{\rmP_\calB}[\tau]}\log \zeta^\rmtp_{\calB}(\phi_n|f_\rmN,f_\rmA)
&\geq
\min\Bigg\{ \frac{\left(\mathrm{MMD}^2(f_\rmN,f_\rmA)-\lambda_2\right)^2}{32K_0^2\left(1+\frac{1}{M-2}\right)},\frac{K\left(\mathrm{MMD}^2(f_\rmN,f_\rmA)-\lambda_3\right)^2}{32K_0^2\left(1+\frac{1}{M-2}\right)}\Bigg\}.
\end{align}
If $\lambda_1\geq \mathrm{MMD}^2(f_\rmN,f_\rmA)$, the exponents of misclassification and false reject error probabilities are the same as those of fixed-length test with $\lambda_3$ playing the role of $\lambda$.
\item the false alarm probability satisfies
\begin{align}
\liminf_{n\to\infty}-\frac{1}{\bbE_{\rmP_\rmr}[\tau]}\log \rmP_\mathrm{FA}^\rmtp(\phi_n|f_\rmN,f_\rmA)
&\geq \min\left\{ g_1(\lambda_1|f_\rmN,f_\rmA), Kg_1(\lambda_3|f_\rmN,f_\rmA)\right\}.
\end{align}
\end{enumerate}
\end{theorem}
The proof of Theorem \ref{E-AFLJMT} generalizes the proof of Theorem \ref{AFLJMT} to the case of unknown number of outliers and is provided in Appendix \ref{proof_of_E-AFLJMT}. 

Analogous to the case of at most one outlier, the performance of the two-phase test bridges over the performance of the fixed-length test in Theorem \ref{FLMT_Un} and the sequential test in Theorem \ref{SJMT_un}. Specifically, if one considers the
Bayesian error probability by calculating a weighted sum of three error probabilities, for $s > 1$, we list the best achievable Bayesian exponent of three tests in table \eqref{Table2}, where $\varepsilon\in\Big(0,\Big(1-\frac{s}{M-s-1}\Big)^2\mathrm{MMD}^2(f_\rmN,f_\rmA)\Big)$ is arbitrary. As observed, by changing the design parameters, the two-phase test bridges over the fixed-length and the sequential tests.
\begin{table}[!h]
\centering
\caption{The best achievable Bayesian exponent of the fixed-length test, the sequential test and the two-phase test with multiple outliers.}
\scalebox{0.8}{
\begin{tabular}{ c| c| c| c}
\hline
Test & Fixed-length &Sequential & Two-phase\\
\hline
\makecell{The best \\achievable\\ Bayesian\\ exponent} & $\frac{\left(1-\frac{s}{M-s-1}\right)^4(\mathrm{MMD}^2(f_\rmN,f_\rmA))^2}{32K_0^2(1+\frac{1}{M-T-1})}$
& $\frac{\left(\left(1-\frac{s}{M-s-1}\right)^2\mathrm{MMD}^2(f_\rmN,f_\rmA)-\varepsilon\right)^2}{32K_0^2(1+\frac{1}{M-T-1})}$
&
$\min\left\{\frac{\left(\left(1-\frac{s}{M-s-1}\right)^2\mathrm{MMD}^2(f_\rmN,f_\rmA)-\varepsilon\right)^2}{32K_0^2\left(1+\frac{1}{M-T-1}\right)},\frac{K\left(1-\frac{s}{M-s-1}\right)^4(\mathrm{MMD}^2(f_\rmN,f_\rmA))^2}{32K_0^2\left(1+\frac{1}{M-T-1}\right)} \right\}
$
\\
\hline
\end{tabular}
}
\label{Table2}
\end{table}

\section{Numerical Results}
\label{simulation}
In this section we simulate the performance of our proposed tests and compare the performance with the tests in~\cite[Eq. (13)]{MMD}. Without loss of generality, assume that the first $s$ sequences of the $M$ observed sequences $\bY^n=(Y_1^n,Y_2^n,...,Y_M^n)$ are outliers.   
The nominal and anomalous distributions are set to be Gaussian distributions: $f_\rmN=\mathcal{N}(0,1)$ and $f_\rmA=\mathcal{N}(1.5,1)$.
In the calculation of the MMD metric~\eqref{MMDcompute}, the Gaussian kernel in \eqref{Gaussiankernel} with $\sigma=1$ is used. Unless otherwise stated, we set $M = 10$.

\begin{figure}[htbp]
\centering
\includegraphics[height=0.4\textwidth]{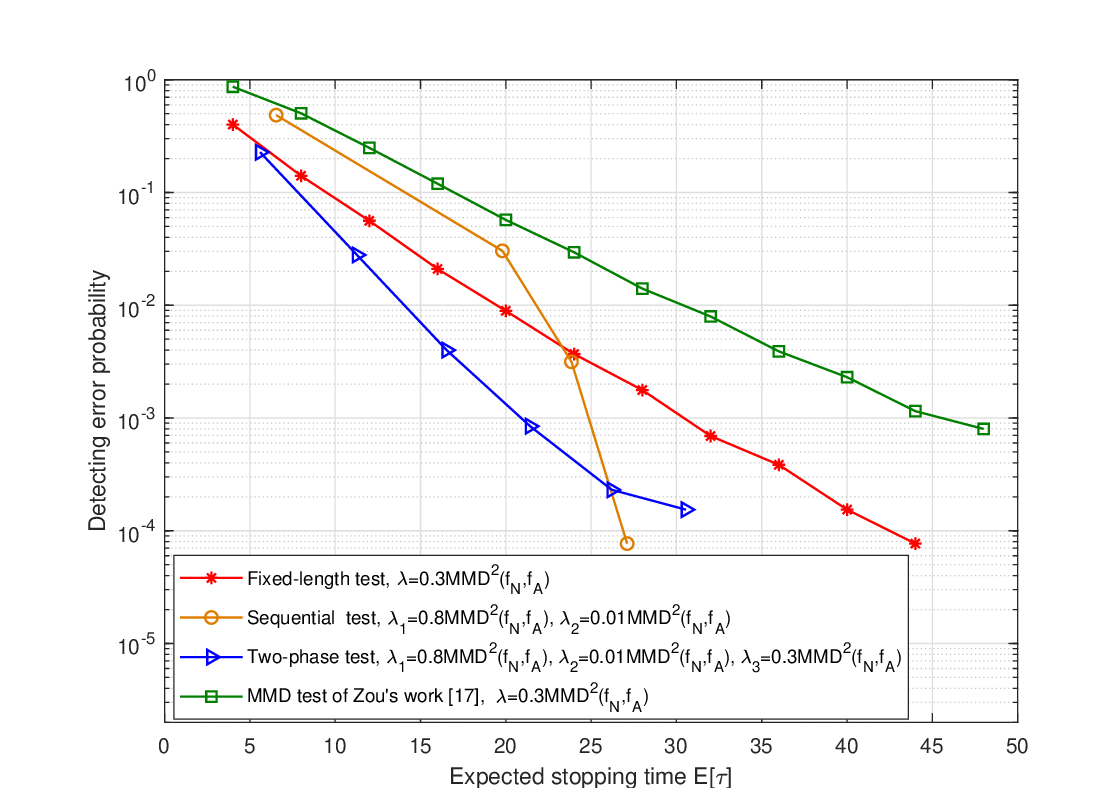}
\caption{Plot of the sum of simulated misclassification and false reject probabilities when there is $s=1$ outlier among $M=10$ observed sequences for our fixed-length test in Section \ref{S-FLMT}, our sequential test in Section \ref{ST}, and our two-phase test in Section \ref{S-AFLJMT} and the fixed-length test of Zou \emph{et al.} in \cite[Eq. (13)]{MMD}. As observed, our fixed-length test outperforms Zou's test. Furthermore, both our two-phase and sequential tests achieve better performance than the fixed-length test. Finally, the sequential test has best performance for 
large expected stopping times while the two-phase test has best performance when $\bbE[\tau]$ is relatively small. The latter occurs since the sequential test tends to stop too early and makes a wrong decision when the key parameter $N$ is small, which leads to a rather small expected stopping time.}\label{detection_error_one}
\end{figure}

In Fig. \ref{detection_error_one}, for the case of at most one outlier, we plot the sum of simulated misclassification and false reject error probabilities of fixed-length, sequential, and two-phase tests in Section \ref{Main} and compare the performance of our tests with the test in  \cite[Eq. (13)]{MMD}. We set $n\in [50]$ as the sample length of the fixed-length test and the first phase of the two-phase test. For the sequential test, we set $N\in[50]$ where $N-1$ is the starting length. The expected stopping length of the fixed-length test is thus $n$ while those of sequential and two-phase tests are obtained by averaging the stopping times over $15000$ independent runs of our tests. As observed from Fig. \ref{detection_error_one}, our proposed fixed-length test has better performance than the test in \cite{MMD}. Furthermore, our two-phase test and sequential test achieve much better performance than the fixed-length test. As the expected stopping time increases, the sequential test achieves best performance while the two-phase test achieves relatively better performance at very small expected stopping times. The latter is because when $N$ is small, the sequential test tends to stop too early and makes wrong decisions. The above numerical results are consistent with our theoretical analyses in Section \ref{Main}.

\begin{figure}[htbp]
\centering
\includegraphics[height=0.4\textwidth]{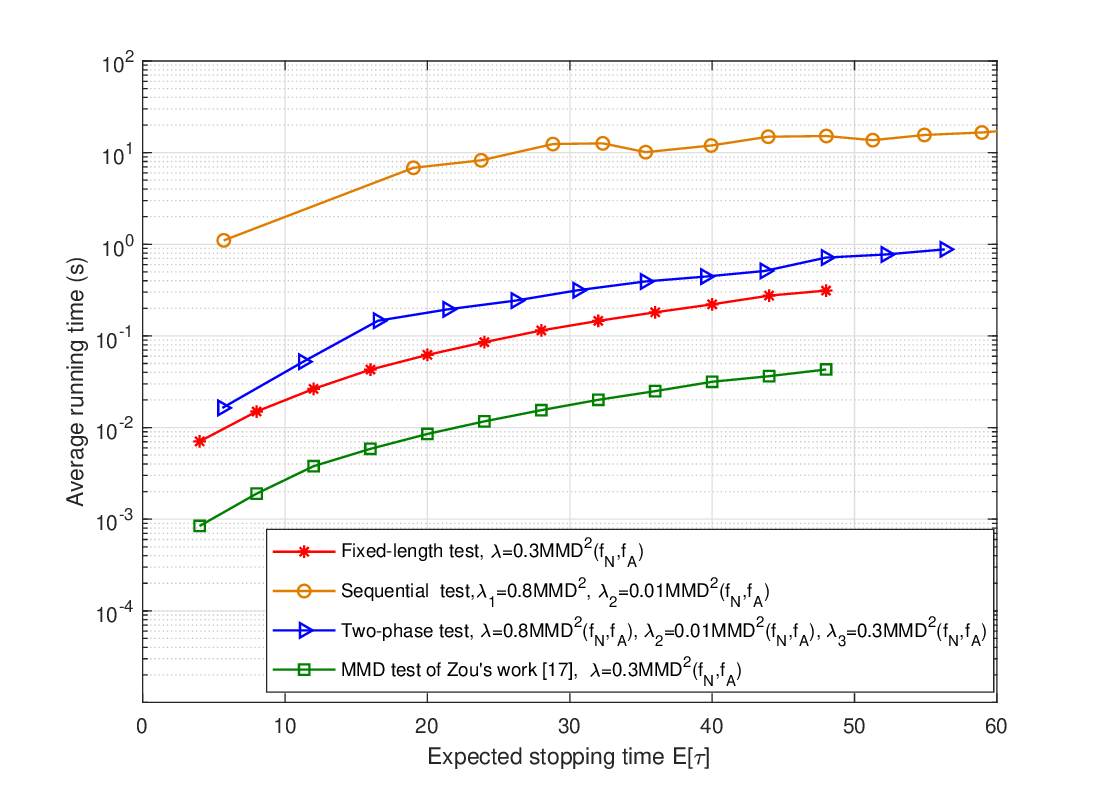}
\caption{Plot of the simulated average running times of our fixed-length, sequential, and two-phase tests in Section \ref{Main} and the fixed-length test of Zou \emph{et al.} in \cite[Eq. (13)]{MMD} as  a function of the expected stopping time when there is $s=1$ outlier among $M=10$ observed sequences. As observed, our fixed-length and two-phase tests have much smaller running times than sequential test. Combining Figs. \ref{detection_error_one} and \ref{compare_FL_ST_AL_computing}, one conclude that our two-phase test strikes a good tradeoff between the performance and the design complexity between our fixed-length and sequential tests.}\label{compare_FL_ST_AL_computing}
\end{figure}

In Fig. \ref{compare_FL_ST_AL_computing}, for the same setting as Fig. \ref{detection_error_one}, we plot the average running times of all four tests as a function of the expected stopping time. As observed, the sequential test is most computationally complicated while the two-phase test and the fixed-length test have roughly the same computational complexity. The above results are consistent with our test design and theoretical findings. Combining Figs. \ref{detection_error_one} and \ref{compare_FL_ST_AL_computing}, one conclude that our two-phase test strikes a good tradeoff between the performance and the design complexity between our fixed-length and sequential tests. One might note that the fixed-length test of Zou \emph{et al.} in \cite{MMD} has the smallest computational complexity. This is because the test of Zou \emph{et al.} checks whether each sequence is an outlier independently by calculating only one MMD metric and comparing the value with a threshold, which has complexity of $O(M)$. In contrast, our fixed-length test needs to calculate $M-1$ MMD values for each sequence, leading to a complexity of $O(M^2)$. However, as shown in Fig. \ref{detection_error_one}, the fixed-length test of Zou \emph{et al.} in \cite{MMD} has the worst performance.

\begin{figure}[htbp]
\centering
\includegraphics[height=0.4\textwidth]{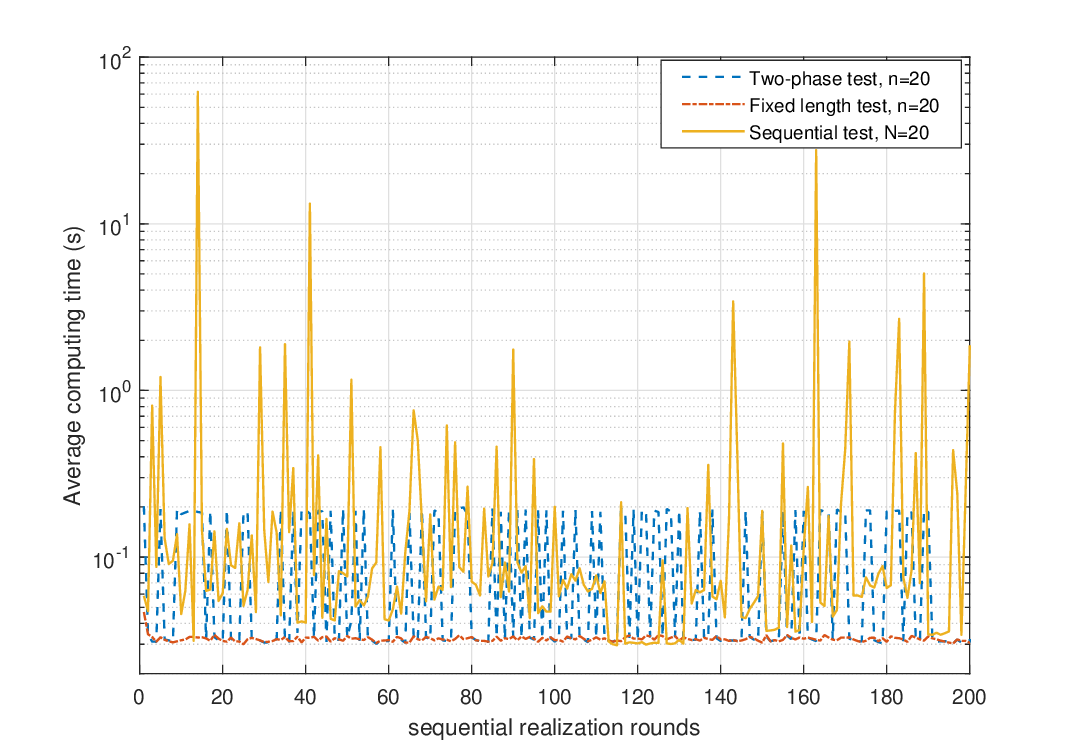}
\caption{Simulated average running times of our fixed-length, sequential, and two-phase tests in Sections \ref{S-FLMT}, \ref{ST}, and \ref{S-AFLJMT} for $300$ independent runs of tests for different realizations of test sequences, when there is $s=1$ outlier among $M=10$ observed sequences. As observed, our two-phase test is much more stable than the sequential test in terms of the random stopping time.}\label{compare_ST_AL_dlength}
\end{figure}

In Fig. \ref{compare_ST_AL_dlength}, again for the same setup as in Fig. \ref{detection_error_one}, we plot running times of the fixed-length, sequential and two-phase tests for $300$ independent runs of tests for different realizations of test sequences. Specifically, we set $n=20$ as sample length for fixed-length test and the sample length in the first phase for the two-phase test, and set $N-1=19$ be the starting sample length of the sequential test. As observed, the sequential test is most sensitive to the variation of test sequences. Specifically, the sequential test tends to have a very large variance for the random stopping time, which leads to a significantly longer running time. In contrast, our two-phase test is much more stable.

\begin{figure}[htbp]
\centering
\includegraphics[height=0.4\textwidth]{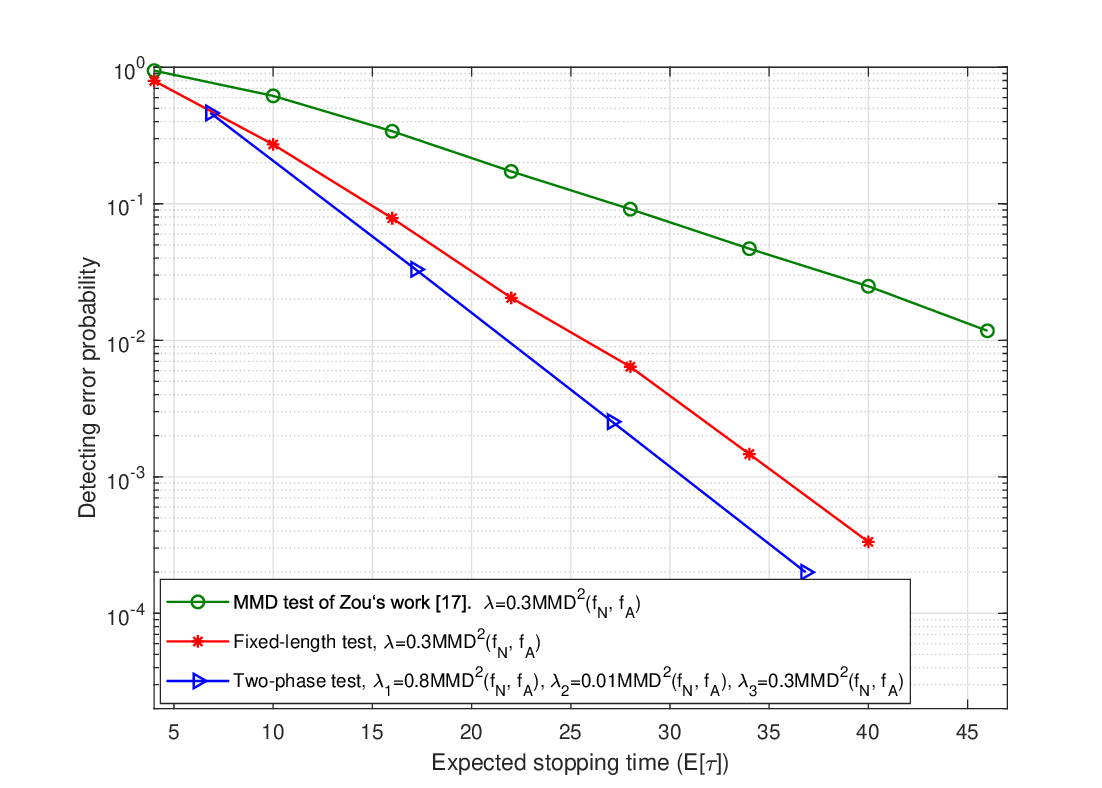}
\caption{Plot for the sum of simulated misclassification and false reject probabilities for our fixed-length test in Section \ref{FLMT_unS} and two-phase test in Section \ref{AFLMTest_unS} and the fixed-length test of Zou \emph{et al.} in \cite[Eq. (13)]{MMD}, when there are $s=2$ outliers among $M=10$ observed sequences . As observed, our fixed-length test also outperforms the test of test of Zou \emph{et al.}. Furthermore, our two-phase test achieves better performance than the fixed-length test as $\bbE[\tau]$ tends to infinity.}\label{detection_error_multi}
\end{figure}

In Fig. \ref{detection_error_multi} we plot the sum of simulated misclassification and false reject error probabilities of our fixed-length and two-phase tests in Sections \ref{FLMT_unS} and \ref{AFLMTest_unS}, respectively. The number of outliers is assumed unknown but upper bounded by $T$ ($T\leq \lceil \frac{M}{2}\rceil-1=4$). Analogously to the case of at most one outlier, we 
compare the performance of our proposed tests with the MMD-based test in \cite[Eq. (13)]{MMD} when there are under multiple outliers and the number of outliers is unknown. Similar conclusions as in the case of at most one outlier hold. In particular, once again, our fixed-length test in Section \ref{FLMT_unS} outperforms the fixed-length test in \cite{MMD}. 

\begin{figure}[htbp]
\centering
\includegraphics[height=0.4\textwidth]{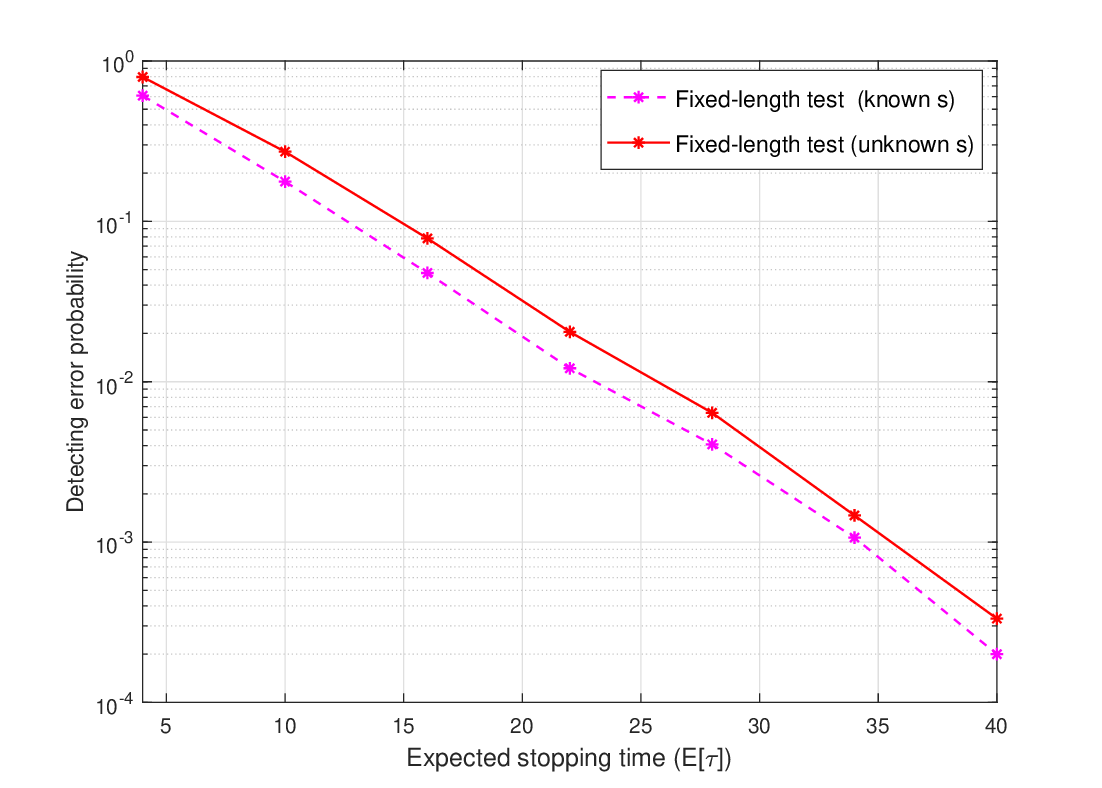}
\caption{Plot of the sum of simulated misclassification and false reject probabilities for our fixed-length test in Section \ref{FLMT_unS} with known and unknown $s$ when there are $s=2$ outliers among $M=10$ observed sequences for  $\lambda=0.3\mathrm{MMD}^2(f_\rmN,f_\rmA)$. As observed, there is a penalty in the performance of not knowing the number of outliers.}\label{detection_error_multi2}
\end{figure}

Finally, in Fig. \ref{detection_error_multi2}, for the same setting as Fig. \ref{detection_error_multi}, we compare the performance of our fixed-length test in Section \ref{FLMT_unS}
when the number of outliers is unknown with the corresponding simpler test when the number of outliers is known. As observed, there is a penalty in the performance of not knowing the number of outliers, which is consistent with our remark below Theorem \ref{FLMT_Un}.

\section{Conclusion}
\label{sec:conc}

We studied outlier hypothesis testing for continuous observed sequences and constructed three kinds of test using MMD: fixed-length, sequential and two-phase tests. We considered both the case of at most one outlier and the case when the number of outliers could be multiple and unknown. For both cases, we analyzed the performance of all three tests by characterizing the achievable exponential decay rates for probabilities of misclassification, false reject and false alarm. In both cases, we showed that the two-phase test bridges over fixed-length and sequential tests by having performance close to the sequential test and having design complexity propositional to the fixed-length test. Our analyses in the case of unknown number of outliers involved a careful analysis of the error event concerning estimating the number of outliers incorrectly and clarified the penalty of not knowing the number of outliers.

We next discuss future directions. Firstly, we studied achievable performance of various tests when the number of outliers is uncertain and the generating distributions are unknown. It would be worthwhile to derive converse results for all settings in this paper to check whether our results are optimal in any regime. Towards this goal, one might modify the generalized Neyman-Pearson criterion used in the discrete case~\cite{second}. Secondly, we were interested in the asymptotic performance and proposed tests that have exponential complexity with respect to the number of outliers. For practical uses, it would be of interest to propose low-complexity tests that could achieve performance close to the benchmarks in this paper. To do so, one might generalize the clustering-based low-complexity test tailored for discrete sequences to continuous sequences~\cite{bu2019linear}. Furthermore, we assumed that each nominal sample follows the same unknown nominal distribution and each outlier follows the same unknown anomalous distribution. In practical applications, the nominal samples and outliers could have different distributions that center around certain distributions. For this case, it would be interesting to generalize the idea in \cite{binaryI-Hsiang} for binary classification of discrete sequences to outlier hypothesis testing of continuous sequences. Finally, it would be of value to generalize the ideas in this paper to other statistical inference problems, e.g., distributed detection \cite{tenney1981detection,tsitsiklis1988decentralized}, quickest change-point detection \cite{poor2008quickest,tartakovsky2014sequential} and clustering \cite{kaufman1990finding,park2009simple}.

\appendix
\subsection{Proof of Theorem \ref{FLMT}}
\label{proof_of_FLMT}
Consistent with \cite{MMD}, we need the following McDiarmid's inequality \cite{mcdiarmid1989method} to bound error probabilities.
\begin{lemma}
\label{McDiarmid}
Let $g:\calX^n\rightarrow \mathbb{R}$ be a
function such that for each $i\in[n]$, there exists a constant $c_i<\infty$ such that
\begin{align}\label{McDiarmid1}
\sup_{x^n\in\calX^n,\tilx\in\calX}\big|g(x_1,\ldots,x_i,\ldots,x_n)-g(x_1,\ldots, x_{i-1},\tilx,x_{i+1},\ldots,x_n)\big|\leq c_i.
\end{align}
Let $X^n$ be generated i.i.d. from a pdf $f\in\calP(\bbR)$. For any $\varepsilon>0$, it follows that
\begin{align}
\label{McDiarmid2}
\mathrm{Pr}\big\{g(X^n)-\bbE_f[g(X^n)]>\varepsilon\big\}< \exp\left\{-\frac{2\varepsilon^2}{\sum_{i=1}^nc_i^2}\right\}.
\end{align}
\end{lemma}

Recall that $\bY^n$ collects all observed sequences $(Y_1^n,\ldots,Y_M^n)$, $\bar{\bY}_{i,j}^n$ collects all sequences except $(Y_i^n,Y_j^n)$ and $\calM=[M]$.  Given above definitions and recalling the definition of the test in \eqref{FLTest}, we can bound three kinds of error probabilities. First consider the misclassification probability. For each $i\in[M]$, the misclassification probability $\beta_i(\phi_n|f_\rmN,f_\rmA)$ satisfies
\begin{align}
\beta_i(\phi_n|f_\rmN,f_\rmA)
&=\bbP_i\{\phi_n(\bY^n)\notin\{\rmH_i,\rmH_\rmr\}\}\label{beta_error_fx0}\\
&=\bbP_i\left\{i^*(\bY^n)\neq i
\text{ and } h(\bY^n)>\lambda\right\}\label{beta_error_fx3}\\
&\leq\bbP_i\left\{\rmG_i(\bY^n)>\lambda\right\}\label{beta_error_fx2}\\
&\leq \bbP_i\left\{\exists~j\in\calM_i, \mathrm{MMD}^2(Y_j^n,\bar{\bY}_{i,j}^n)>\lambda\right\}\label{usedefGi}\\
&\leq \sum_{j\in\calM_i}\bbP_i\left\{\mathrm{MMD}^2(Y_j^n,\bar{\bY}_{i,j}^n)>\lambda\right\}\label{beta_error_fx1}
\end{align}
where \eqref{beta_error_fx2} follows since $\rmG_i(\bY^n)\ge h(\bY^n)$ when $i^*(\bY^n)\neq i$,  \eqref{usedefGi} follows from the definition of $\rmG_i(\cdot)$ in \eqref{GB}.

To further upper bound \eqref{beta_error_fx1}, we need to apply McDiarmid's inequality. To do so, we need  to calculate the expected value of $\mathrm{MMD}^2(Y_j^n,\bar{\bY}_{i,j}^n)$ and the parameters $c_i$ for each $i\in[(M-1)n]$. Note that under hypothesis $\rmH_i$, $Y_i^n$ is the outlier and all other sequences are generated i.i.d. from the nominal distribution. It follows from the definition of the MMD metric in \eqref{MMDcompute} that
\begin{align}
\bbE_{\bbP_i}[\mathrm{MMD}^2(Y_j^n,\bar{\bY}_{i,j}^n)]=0.
\end{align}
To bound the Lipschitz constant $\{c_i\}$, given any observed sequences $\by^n=(y_1^n,\ldots,y_M^n)$, for each $i\in[M]$, define the function $g_{i,j}(\bar{\by}_i^n):=\mathrm{MMD}^2(y_j^n,\bar{\by}_{i,j}^n)$, where $\bar{\by}_i^n$ collects all sequences except $y_i^n$ and $\bar{\by}_{i,j}^n$ collects all sequences except $(y_i^n,y_j^n)$. Note that $g_{i,j}(\bar{\by}_i^n)$ is a function of $(M-1)n$ parameters. For each $k\in[(M-1)n]$, if the $k$-th element of $\bar{\by}_i^n$ is replaced by $\tily$, we use $g_{i,j}(\bar{\by}_i^n,k,\tily)$ to denote the corresponding function value. For each $(j,k)\in\calM_i\times[(M-1)n]$, define
\begin{align}
c_k^{i,j}:=\sup_{\by^n,\tily}|g_{i,j}(\bar{\by}_i^n)-g_{i,j}(\bar{\by}_i^n,k,\tily)|.
\end{align}

Given any $i\in[M]$, for each $j\in\calM_i$ and any $k\in[(M-1)n]$, similarly to [Eg. (27)-(32)]\cite{MMD}, it follows that
\begin{align}
|c_k^{i,j}|&\leq \frac{8K_0}{n},~k\in[n],\\
|c_k^{i,j}|&\leq \frac{8K_0}{(M-2)n},k\in[n+1,(M-1)n].
\end{align}
Thus,
\begin{align}\label{frac}
\sum_{k\in[(M-1)n]} (c_k^{i,j})^2\leq \frac{64K_0^2}{n}\left(1+\frac{1}{M-2}\right).
\end{align}

Using McDiarmid’s inequality, \eqref{beta_error_fx1} and \eqref{frac} leads to
\begin{align}
&\beta_i(\phi_n|f_\rmN,f_\rmA)\leq (M-1)\exp\left\{\frac{-n\lambda^2}{ 32K_0^2\left(1+\frac{1}{M-2}\right)}\right\}.\label{beta_error_fx}
\end{align}

Analogously, the false alarm probability $\rmP_{\mathrm{FA}}$ is upper
bounded as follows
\begin{align}
\rmP_{\mathrm{FA}}(\phi_n|f_\rmN,f_\rmA)
&=\bbP_\rmr\{\phi_n(\bY^n)\neq \rmH_\rmr\}\\
&=\bbP_\rmr\{h(\bY^n)>\lambda\}\\
&=\sum_{i\in[M]} \bbP_\rmr\left\{i^*(\bY^n)=i,\text{ and } h(\bY^n)>\lambda\right\}\\
&\leq \sum_{i\in[M]} \bbP_\rmr\{i^*(\bY^n)=i,\text{ and } \exists~k\in\calM_i:~\rmG_k(\bY^n)>\lambda\}\label{Fa_error_fx2}\\
&\leq \sum_{i\in[M]} \sum_{k\in\calM_i}\bbP_\rmr\{\rmG_k(\bY^n)>\lambda\}\\
&\leq\sum_{i\in[M]}\sum_{k\in\calM_i}\sum_{j\in\calM_k}
\bbP_\rmr\{\mathrm{MMD}^2(Y_j^n,\bar{\bY}_{j,k}^n)>\lambda\}\label{Fa_error_fx1}\\
&\leq M(M-1)^2
\exp\left\{\frac{-n\lambda^2}{32K_0^2\left(1+\frac{1}{M-2}\right)}\right\},\label{Fa_error_fx}
\end{align}
where \eqref{Fa_error_fx2} follows since when $i^*(\bY^n)=i$, $h(\bY^n)=\min_{j\in\calM_i}\rmG_j(\bY^n)$
if a false alarm happens, there must exists $k\in[M]$ that satisfies  $\rmG_k(\bY^n)>\lambda$, \eqref{Fa_error_fx1} follows from the definition of $\rmG_i()$ in \eqref{GB} and the union bound similarly to \eqref{beta_error_fx1}, and \eqref{Fa_error_fx} follows from the McDiarmid's inequality similarly to \eqref{beta_error_fx} since $\bbE_{\bbP_\rmr}[\mathrm{MMD}^2(Y_j^n,\bar{\bY}_{j,k}^n)]=0$ for each $(j,k)\in\calM^2$ under the null hypothesis.

Finally, we bound the false reject probability as follows for each $i\in[M]$:
\begin{align}
\zeta_i(\phi_n|f_\rmN,f_\rmA)
&=\bbP_i(\phi_n(\bY^n)=\rmH_\rmr)\\
&=\bbP_i(h(\bY^n)<\lambda)\\
&\leq\bbP_i\Big\{\exists~(j,k)\in\calM^2:~j\neq k,~\mathrm{and}~\rmG_j(\bY^n)<\lambda,~\rmG_k(\bY^n)<\lambda\Big\}\label{zeta_error_fx00}\\
&\leq \sum_{j\in[M]}\sum_{k\in\calM_j}\bbP_i\{\rmG_j(\by^n)<\lambda\mathrm{~and~}\rmG_k(\bY^n)<\lambda\}\label{zeta_error_fx0}\\
&\leq \sum_{j\in\calM_i}\sum_{k\in\calM_j}\bbP_i\{\rmG_j(\bY^n)<\lambda\}+\sum_{k\in\calM_i}\bbP_i\{\rmG_k(\bY^n)<\lambda\}\label{zeta_error_fx1}\\
&\leq (M-1)\sum_{j\in\calM_i}\bbP_i\{\rmG_j(\bY^n)<\lambda\}+
\sum_{k\in\calM_i}\bbP_i\{\rmG_k(\bY^n)<\lambda\}\\
&\leq M\sum_{j\in\calM_i}\bbP_i\{\rmG_j(\bY^n)<\lambda\}\\
&= M\sum_{j\in \calM_i}
\bbP_i\big\{\max_{k\in\calM_j}\mathrm{MMD}^2(Y_k^n,\bar{\bY}_{j,k}^n)< \lambda\big\}\label{zeta_error_fx2_0}\\
&\leq M\sum_{j\in \calM_i}\bbP_i\{\mathrm{MMD}^2(Y_i^n,\bar{\bY}_{j,i}^n)< \lambda\},\label{zeta_error_fx2}
\end{align}
where \eqref{zeta_error_fx00} follows from the test design in \eqref{FLTest}, \eqref{zeta_error_fx0} follows from the definition of $h(\bY^n)$ in \eqref{secMinG}, \eqref{zeta_error_fx1} follows since $\Pr\{\rmG_j(\bY^n)<\lambda,~\rmG_k(\bY^n)<\lambda\}\leq \min\{\Pr\{\rmG_j(\bY^n)<\lambda\},\Pr\{\rmG_k(\bY^n)<\lambda\}\}$, \eqref{zeta_error_fx2_0} follows from the definition of $\rmG_j(\cdot)$ in \eqref{GB}, \eqref{zeta_error_fx2} follows since when $j\in\calM_i$, $\mathrm{MMD}^2(Y_i^n,\bar{\bY}_{j,i}^n)\leq \max_{k\in\calM_j}\mathrm{MMD}^2(Y_k^n,\bar{\bY}_{j,k}^n)<\lambda$.

It is left to upper bound \eqref{zeta_error_fx2} using the McDiarmid's inequality. To do so, we need to calculate the expected value of $\mathrm{MMD}^2(Y_i^n,\bar{\bY}_{j,i}^n)$ and calculate the Lipschitz continuous parameters. Note that under hypothesis $\rmH_i$, $Y_i^n$ is the outlier and all other sequences are nominal samples, generated i.i.d. from the nominal distribution. It follows from the definition of the MMD metric in \eqref{MMDcompute} that
\begin{align}
\bbE_{\bbP_i}[\mathrm{MMD}^2(Y_i^n,\bar{\bY}_{j,i}^n)]=\mathrm{MMD}^2(f_\rmN,f_\rmA).
\end{align}

To bound the Lipschitz constant $\{c_{i,k}\}$, we follow the idea below \eqref{beta_error_fx1}. Given any observed sequences $\by^n=(y_1^n,\ldots,y_M^n)$, for each $i\in[M]$ and $j\in\calM_i$, define the function $g_{j,i}(\bar{\by}_j^n):=\mathrm{MMD}^2(y_i^n,\bar{\by}_{j,i}^n)$, where $\bar{\by}_j^n$ collects all sequences except $y_j^n$. Note that $g_{j,i}(\bar{\by}_j^n)$ is a function of $(M-1)n$ parameters. For each $k\in[(M-1)n]$, if the $k$-th element of $\bar{\by}_j^n$ is replaced by $\tily$, we use $g_{j,i}(\bar{\by}_j^n,k,\tily)$ to denote the corresponding function value. For each $(j,k)\in\calM_i\times[(M-1)n]$, define
\begin{align}
c_k^{i,j}:=\sup_{\by^n,\tily}|g_{j,i}(\bar{\by}_j^n)-g_{j,i}(\bar{\by}_j^n,k,\tily)|.
\end{align}
Similarly to \eqref{frac}, we obtain
\begin{align}\label{frac2}
\sum_{k\in[(M-1)n]} (c_k^{j,i})^2\leq \frac{64K_0^2}{n}\left(1+\frac{1}{M-2}\right).
\end{align}
Combining \eqref{zeta_error_fx2} and \eqref{frac2}, it follows from McDiarmid's inequality that if $\lambda<\mathrm{MMD}^2(f_\rmN,f_\rmA)$,
\begin{align}
\zeta_i(\phi_n|f_\rmN,f_\rmA)
&\leq M(M-1)\exp\left\{\frac{-n\left(\mathrm{MMD}^2(f_\rmN,f_\rmA)-\lambda\right)^2}{32K_0^2\left(1+\frac{1}{M-2}\right)}\right\},\label{zeta_error_fx}
\end{align}
On the other hand, if $\lambda\geq \mathrm{MMD}^2(f_\rmN,f_\rmA)$, $\zeta_i(\phi_n|f_\rmN,f_\rmA)\leq 1$.

Finally, note that $M$ is finite and taking the sample size $n$ to infity, it follows from \eqref{beta_error_fx}, \eqref{Fa_error_fx}, and \eqref{zeta_error_fx} that the exponent rates of three error probabilities satisfy
\begin{align}
\lim_{n\rightarrow\infty}-\frac{1}{n}\log \beta_i(\phi_n|f_\rmN,f_\rmA)
&\geq\frac{\lambda^2}{ 32K_0^2\left(1+\frac{1}{M-2}\right)},\label{logbeta}\\
\lim_{n\rightarrow\infty}-\frac{1}{n}\log \zeta_i(\phi_n|f_\rmN,f_\rmA)
&\geq\frac{\left(\mathrm{MMD}^2(f_\rmN,f_\rmA)-\lambda\right)^2}{32K_0^2\left(1+\frac{1}{M-2}\right)}\mathbb{I}(\lambda<\mathrm{MMD}^2(f_\rmN,f_\rmA)),\label{logzeta}\\
\lim_{n\rightarrow\infty}-\frac{1}{n}\log \rmP_{\mathrm{FA}}(\phi_n|f_\rmN,f_\rmA)
&\geq\frac{\lambda^2}{ 32K_0^2\left(1+\frac{1}{M-2}\right)}.\label{logFA}
\end{align}
The proof of Theorem \ref{FLMT} is now completed.

\subsection{Proof of Theorem \ref{SJMT}}
\label{proof_of_SJMT}
We first bound the expected stopping time under each non-null and null hypothesis. Subsequently, analogously to the analyses for the fixed-length test, we bound three kinds of error probabilities. Finally, combining the above analyses, we obtain the desired bound on the achievable exponents of the sequential test.

Recall the definition of the stopping time $\tau$ in \eqref{Taulength2}. For each $i\in[M]$, the expected stopping time $\bbE_{\bbP_i}[\tau]$ can be upper bounded as follows:
\begin{align}
\bbE_{\bbP_i}[\tau]
&\leq \sum_{\tau'=1}^\infty \bbP_i\{\tau\geq\tau'\}\\
&\leq N-1+ \sum_{\tau'=N-1}^\infty \bbP_i\{\tau>\tau'\}\label{Etau},
\end{align}
where \eqref{Etau} follows since the random stopping time $\tau\ge N-1$ by the definition in \eqref{Taulength2}.

If $\lambda_1<\mathrm{MMD}^2(f_\rmN,f_\rmA)$, each probability term inside the sum of \eqref{Etau} satisfies
\begin{align}
\bbP_i\{\tau>\tau'\}
&=\bbP_i\{\lambda_2<h(\bY^{\tau'})<\lambda_1\}\\
&\leq\bbP_i\{h(\bY^{\tau'})<\lambda_1\}\\
&\leq M(M-1)\exp\left\{\frac{-\tau'\left(\mathrm{MMD}^2(f_\rmN,f_\rmA)-\lambda_1\right)^2}
{32K_0^2\left(1+\frac{1}{M-2}\right)}\right\},\label{Ptau}
\end{align}
where \eqref{Ptau} follows similarly to \eqref{zeta_error_fx} except that $n$ is replaced with $\tau'$ and $\lambda$ is replaced by $\lambda_1$.
Combining \eqref{Etau} and \eqref{Ptau} leads to
\begin{align}
\bbE_{\bbP_i}[\tau]&\leq N-1+M(M-1)\frac{\exp\left\{\frac{-(N-1)\left(\mathrm{MMD}^2(f_\rmN,f_\rmA)-\lambda_1\right)^2}
{32K_0^2\left(1+\frac{1}{M-2}\right)}\right\}}
{1-\exp\left\{\frac{-\left(\mathrm{MMD}^2(f_\rmN,f_\rmA)-\lambda_1\right)^2}
{32K_0^2\left(1+\frac{1}{M-2}\right)}\right\}},\label{sumPtau}
\end{align}
if $\lambda_1<\mathrm{MMD}^2(f_\rmN,f_\rmA)$. On the other hand, if $\lambda_1\geq\mathrm{MMD}^2(f_\rmN,f_\rmA)$, $\bbP_i\{\tau>\tau'\}\leq 1$ and $\bbE_{\bbP_i}[\tau]\leq \infty$.

Similarly, the expected stopping time under the null hypothesis satisfies:
\begin{align}
\bbE_{\bbP_\rmr}[\tau]
&\leq  N-1+ \sum_{\tau'=N-1}^\infty \bbP_\rmr\{\tau> \tau'\}\\
&= N-1+\sum_{\tau'=N-1}^\infty\bbP_\rmr\{\lambda_2<h(\bY^{\tau'})<\lambda_1\}\\
&\leq N-1+ \sum_{\tau'=N-1}^\infty \bbP_\rmr\{h(\bY^{\tau'})>\lambda_2\}\\
&\leq N-1+M(M-1)^2\frac{\exp\left\{\frac{-(N-1)\lambda_2^2}{ 32K_0^2\left(1+\frac{1}{M-2}\right)} \right\}}
{1-\exp\left\{\frac{-\lambda_2^2}{ 32K_0^2\left(1+\frac{1}{M-2}\right)}\right\}},\label{Etau2}
\end{align}
where \eqref{Etau2} follows similarly to \eqref{Fa_error_fx} except that $n$ is replaced with $\tau'$ and $\lambda$ is replaced by $\lambda_2$.

Note that for $N$ sufficiently large,
\begin{align}
& M(M-1)\frac{\exp\left\{\frac{-(N-1)\left(\mathrm{MMD}^2(f_\rmN,f_\rmA)-\lambda_1\right)^2}{32K_0^2\left(1+\frac{1}{M-2}\right)}\right\}}
{1-\exp\left\{\frac{-\left(\mathrm{MMD}^2(f_\rmN,f_\rmA)-\lambda_1\right)^2}{32K_0^2\left(1+\frac{1}{M-2}\right)}\right\}}\leq 1.\\
&M(M-1)^2\frac{\exp\left\{\frac{-(N-1)\lambda_2^2}{ 32K_0^2\left(1+\frac{1}{M-2}\right)}\right\}}{1-\exp\left\{\frac{-\lambda_2^2}{ 32K_0^2\left(1+\frac{1}{M-2}\right)}\right\}}\leq 1.
\end{align}
It follows from \eqref{sumPtau} and \eqref{Etau2} that $\bbE_{\bbP_i}[\tau]\leq N$ and $\bbE_{\bbP_\rmr}[\tau]\leq N$ when $N$ is sufficiently large and $\lambda_1<\mathrm{MMD}^2(f_\rmN,f_\rmA)$.

For each $i\in[M]$, the misclassification error probability $\beta_i^{\mathrm{seq}}(\phi_\tau|f_\rmN,f_\rmA)$ satisfies
\begin{align}
\beta_i^{\mathrm{seq}}(\phi_\tau|f_\rmN,f_\rmA)
&\leq \sum_{\tau'=N-1}^\infty\bbP_i\{\phi_{\tau'}(\bY^{\tau'})\notin\{\rmH_i,\rmH_\rmr\}\}\\
&\leq \sum_{\tau'=N-1}^\infty\bbP_i\{\rmG_i(\bY^{\tau'})>\lambda_1\}\\
&\leq \sum_{\tau'=N-1}^\infty(M-1)\exp\left\{\frac{-\tau'\lambda_1^2}{ 32K_0^2\left(1+\frac{1}{M-2}\right)}\right\}\label{Pbetas0}\\
&=(M-1)\frac{\exp\left\{\frac{-(N-1)\lambda_1^2}{ 32K_0^2\left(1+\frac{1}{M-2}\right)}\right\}}{1-\exp\left\{\frac{-\lambda_1^2}{ 32K_0^2\left(1+\frac{1}{M-2}\right)}\right\}}, \label{Pbetas}
\end{align}
where \eqref{Pbetas0} from \eqref{beta_error_fx} with $n$ replaced with $\tau'$ and $\lambda$ is replaced by $\lambda_1$.

Similarly, if $\lambda_2<\mathrm{MMD}^2(f_\rmN,f_\rmA)$, the false reject probability $\zeta_i^{\mathrm{seq}}(\phi_\tau|f_\rmN,f_\rmA)$ is upper bounded as follows:
\begin{align}
\zeta_i^{\mathrm{seq}}(\phi_\tau|f_\rmN,f_\rmA)
&\leq
\sum_{\tau'=N-1}^\infty \bbP_i\left\{\phi_{\tau'}(\bY^{\tau'})=\rmH_\rmr\right\}\\
&\leq \sum_{\tau'=N-1}^\infty \bbP_i\left\{h(\bY^{\tau'})<\lambda_2\right\}\\
&\leq \sum_{\tau'=N-1}^\infty M(M-1)\exp\left\{\frac{-\tau'\left(\mathrm{MMD}^2(f_\rmN,f_\rmA)-\lambda_2\right)^2}{32K_0^2\left(1+\frac{1}{M-2}\right)}\right\}\label{Pzetas0}\\
&\leq M(M-1)\frac{\exp\left\{\frac{-(N-1)\left(\mathrm{MMD}^2(f_\rmN,f_\rmA)-\lambda_2\right)^2}
{32K_0^2\left(1+\frac{1}{M-2}\right)}\right\}}
{1-\exp\left\{\frac{-\left(\mathrm{MMD}^2(f_\rmN,f_\rmA)-\lambda_2\right)^2}
{32K_0^2\left(1+\frac{1}{M-2}\right)}\right\}},\label{Pzetas}
\end{align}
where \eqref{Pzetas0} follow the result of \eqref{zeta_error_fx} with $n$ replaced with $\tau'$ and $\lambda$ is replaced by $\lambda_2$.

Finally, the false alarm probability $\rmP_{\mathrm{FA}}^{\mathrm{seq}}(\phi_\tau|f_\rmN,f_\rmA)$ is upper bounded by
\begin{align}
\rmP_{\mathrm{FA}}^{\mathrm{seq}}(\phi_\tau|f_\rmN,f_\rmA)
&\leq \sum_{\tau'=N-1}^\infty
\bbP_\rmr\left\{\phi_{\tau'}(\bY^{\tau'})\neq\rmH_\rmr\right\}\\
&\leq \sum_{\tau'=N-1}^\infty
\bbP_\rmr\left\{h(\bY^{\tau'})>\lambda_1\right\}\\
&\leq \sum_{\tau'=N-1}^\infty M(M-1)^2\exp\left\{\frac{-\tau'\lambda_1^2}
{32K_0^2\left(1+\frac{1}{M-2}\right)}\right\}\label{PFAs0}\\
&\leq M(M-1)^2\frac{\exp\left\{\frac{-(N-1)\lambda_1^2}{32K_0^2\left(1+\frac{1}{M-2}\right)}\right\}}
{1-\exp\left\{\frac{-\lambda_1^2}{32K_0^2\left(1+\frac{1}{M-2}\right)}\right\}}.\label{PFAs}
\end{align}
where \eqref{PFAs0} follow the result of \eqref{Fa_error_fx} with $n$ replaced with $\tau$ and $\lambda$ is replaced by $\lambda_1$.

Using \eqref{Pbetas}, \eqref{Pzetas} and \eqref{PFAs}, we conclude that the achievable exponents of the sequential test satisfy
\begin{enumerate}
\item for each $i\in[M]$, if $0<\lambda_2<\lambda_1<\mathrm{MMD}^2(f_\rmN,f_\rmA)$,
\begin{align}
\lim_{N\rightarrow\infty}-\frac{\log \beta_i^{\mathrm{seq}}(\phi_\tau|f_\rmN,f_\rmA)}{\bbE_{\bbP_i}[\tau]}
&\geq \frac{\lambda_1^2}{32K_0^2\left(1+\frac{1}{M-2}\right)}.\label{logbetas}\\
\lim_{N\rightarrow\infty}-\frac{\log \zeta_i^{\mathrm{seq}}(\phi_\tau|f_\rmN,f_\rmA)}{\bbE_{\bbP_i}[\tau]}
&\geq\frac{\left(\mathrm{MMD}^2(f_\rmN,f_\rmA)-\lambda_2\right)^2}{32K_0^2\left(1+\frac{1}{M-2}\right)}\label{logzetas},
\end{align}
\item for any $\lambda_1>\lambda_2>0$,
\begin{align}
\liminf_{N\to\infty}-\frac{1}{\bbE_{\bbP_\rmr}[\tau]}\log \rmP^{\rmseq}_{\rm{FA}}(\phi_\tau|f_\rmN,f_\rmA)
\geq \frac{\lambda_1^2}{  32K_0^2\left(1+\frac{1}{M-2}\right)}.
\end{align}
\end{enumerate}
The proof of Theorem \ref{SJMT} is now completed.

\subsection{Proof of Theorem \ref{AFLJMT}}\label{proof_of_AFLJMT}

Recall that $0<\lambda_2<\lambda_3<\lambda_1$. We first bound the probability that the two-phase test proceeds in the second phase under each non-null and null hypothesis. For each $i\in[M]$, if $\lambda_1<\mathrm{MMD}^2(f_\rmN,f_\rmA)$, it follows that
\begin{align}
\bbP_i\{\tau=Kn\}
&=\bbP_i\left\{\lambda_2<h(\bY^n)<\lambda_1\right\}\\
&\leq\bbP_i\left\{h(\bY^n)<\lambda_1\right\}\label{P_beta_tau3}\\
&\leq M(M-1)\exp\left\{\frac{-n\left(\mathrm{MMD}^2(f_\rmN,f_\rmA)-\lambda_1\right)^2}{32K_0^2\left(1+\frac{1}{M-2}\right)}\right\},\label{P_tau1}
\end{align}
where \eqref{P_tau1} follows from the same idea to prove \eqref{zeta_error_fx} for the fixed-length test. Therefore, the expected stopping time $\bbE_{\bbP_i}[\tau]$ satisfies that if $\lambda_1<\mathrm{MMD}^2(f_\rmN,f_\rmA)$,
\begin{align}
\bbE_{\bbP_i}[\tau]
&=n\bbP_i\{\tau=n\}+Kn\bbP_i\{\tau=Kn\}\\
& \leq n+Kn\bbP_i\{\tau=Kn\}\\
&\leq  n+nKM(M-1)\exp\left\{\frac{-n\left(\mathrm{MMD}^2(f_\rmN,f_\rmA)-\lambda_1\right)^2}{32K_0^2\left(1+\frac{1}{M-2}\right)}\right\}.\label{Etau1_tp}
\end{align}
On the other hand, if $\lambda_1\geq\mathrm{MMD}^2(f_\rmN,f_\rmA)$, $\bbP_i\{\tau=Kn\}\leq 1$ and thus $\bbE_{\bbP_i}[\tau]\leq Kn$.

Similarly, under the null hypothesis, it follows that
\begin{align}
\bbP_\rmr\{\tau=Kn\}
&=\bbP_\rmr\left\{ \lambda_2<h(\bY^n)<\lambda_1\right\}\\
&\leq\bbP_\rmr\left\{h(\bY^\tau)>\lambda_2\right\}\\
&\leq M(M-1)^2 \exp\left\{\frac{-n\lambda_2^2}{ 32K_0^2\left(1+\frac{1}{M-2}\right)}\right\},\label{P_tau2}
\end{align}
where \eqref{P_tau2} follows from the result in \eqref{Fa_error_fx} for the fixed-length test. The expected stopping time $\bbE_{\bbP_\rmr}[\tau]$ satisfies
\begin{align}
\bbE_{\bbP_\rmr}[\tau]
&=n\bbP_\rmr\{\tau=n\}+Kn\bbP_\rmr\{\tau=Kn\}\nonumber\\
& \leq n+Kn\bbP_\rmr\{\tau=Kn\}\\
&\leq  n+nKM(M-1)^2 \exp\left\{\frac{-n\lambda_2^2}{ 32K_0^2\left(1+\frac{1}{M-2}\right)}\right\}.\label{Etau2_tp}
\end{align}

It follows from \eqref{Etau1_tp} and \eqref{Etau2_tp} that if $\lambda_1<\mathrm{MMD}^2(f_\rmN,f_\rmA)$, for $n$ sufficiently large,
\begin{align}
&n(K-1)M(M-1)\exp\left\{\frac{-n\left(\mathrm{MMD}^2(f_\rmN,f_\rmA)-\lambda_1\right)^2}{32K_0^2\left(1+\frac{1}{M-2}\right)}\right\}\leq1,\\
&n(K-1)M(M-1)^2 \exp\left\{\frac{-n\lambda_2^2}{ 32K_0^2\left(1+\frac{1}{M-2}\right)}\right\}\leq1.
\end{align}
Thus, when $n$ is sufficiently large, the average stopping time satisfies
\begin{align}
\max_{i\in[M]}\bbE_{\bbP_i}[\tau]&\leq
\left\{
\begin{array}{ll}
n+1, &\mathrm{if~}\lambda_1<\mathrm{MMD}^2(f_\rmN,f_\rmA),\\
Kn&\mathrm{otherwise.}
\end{array}
\right.\label{etau:twophase}\\
\bbE_{\bbP_\rmr}[\tau]&\leq n+1\label{etau:null}.
\end{align}

We now upper bound three error probabilities. For each $i\in\calM$, the misclassification probability satisfies
\begin{align}
\beta_i^{\mathrm{tp}}(\phi_n|f_\rmN,f_\rmA)
&=\bbP_i\left\{\tau=n,\phi_n(\bY^n)\notin\{\rmH_i,\rmH_\rmr\}\right\}\\
&+\bbP_i\left\{\tau=Kn,\phi_{Kn}(\bY^{Kn})\notin\{\rmH_i,\rmH_\rmr\}\right\}\\
&\leq \bbP_i\left\{\phi_n(\bY^n)\notin\{\rmH_i,\rmH_\rmr\}\right\}
+\bbP_i\left\{\phi_{Kn}(\bY^{Kn})\notin\{\rmH_i,\rmH_\rmr\}\right\}\\
&\leq  (M-1)\exp\left\{\frac{-n\lambda_1^2}{ 32K_0^2\left(1+\frac{1}{M-2}\right)}\right\}+(M-1)\exp\left\{\frac{-Kn\lambda_3^2}{ 32K_0^2\left(1+\frac{1}{M-2}\right)}\right\},\label{beta_al}
\end{align}
where \eqref{beta_al} follows from the result in \eqref{beta_error_fx} with $(n,\lambda)$ replaced with $(n,\lambda_1)$ and $(Kn,\lambda_3)$, respectively.

If $\lambda_1<\mathrm{MMD}^2(f_\rmN,f_\rmA)$, combining \eqref{etau:twophase} and \eqref{beta_al} leads to
\begin{align}
\lim_{n\rightarrow \infty}-\frac{1}{\bbE_{\bbP_i}[\tau]}\log \beta_i^{\mathrm{tp}}(\phi_n|f_\rmN,f_\rmA)
&\geq\min\left\{\frac{\lambda_1^2}{32K_0^2\left(1+\frac{1}{M-2}\right)},\frac{K\lambda_3^2}{32K_0^2\left(1+\frac{1}{M-2}\right)}\right\}.\label{logP_1}
\end{align}
On the other hand, if $\lambda_1\geq \mathrm{MMD}^2(f_\rmN,f_\rmA)$, asymptotically, the two-phase test reduces to the fixed-length test with the sample size $Kn$. The misclassification exponent thus is the same the result in \eqref{logbeta} for the fixed-length test.

Similarly, if $\max\{\lambda_2,\lambda_3\}<\mathrm{MMD}^2(f_\rmN,f_\rmA)$, for each $i\in[M]$, the false reject probability satisfies
\begin{align}
\zeta_i^{\mathrm{tp}}(\phi_n|f_\rmN,f_\rmA)
&=\bbP_i\left\{\tau=n,\phi_n(\bY^n)=\rmH_\rmr\right\}+\bbP_i\left\{\tau=Kn,\phi_{Kn}(\bY^{Kn})=\rmH_\rmr\right\}\\
&\leq \bbP_i\left\{\phi_n(\bY^n)=\rmH_\rmr\right\}
+\bbP_i\left\{\phi_{Kn}(\bY^{Kn})=\rmH_\rmr\right\}\\
&\leq M(M-1)\exp\left\{\frac{-n\left(\mathrm{MMD}^2(f_\rmN,f_\rmA)-\lambda_2\right)^2}{32K_0^2\left(1+\frac{1}{M-2}\right)}\right\}\nonumber\\
&\qquad+M(M-1)\exp\left\{\frac{-n\left(\mathrm{MMD}^2(f_\rmN,f_\rmA)-\lambda_3\right)^2}{32K_0^2\left(1+\frac{1}{M-2}\right)}\right\}\label{zeta_al},
\end{align}
where \eqref{zeta_al} follows from the result in \eqref{zeta_error_fx} with $(n,\lambda)$ replaced by $(n,\lambda_2)$ and $(Kn,\lambda_3)$, respectively.
In addition, if $\lambda_1<\mathrm{MMD}^2(f_\rmN,f_\rmA)$, combining \eqref{etau:twophase} and \eqref{zeta_al} leads to
\begin{align}
\liminf_{n\to\infty}-\frac{1}{\bbE_{\bbP_i}[\tau]}\log \zeta^{\mathrm{tp}}_i(\phi_\tau|f_\rmN,f_\rmA)
&\geq \min\Bigg\{ \frac{\left(\mathrm{MMD}^2(f_\rmN,f_\rmA)-\lambda_2\right)^2}{32K_0^2\left(1+\frac{1}{M-2}\right)}, \frac{K\left(\mathrm{MMD}^2(f_\rmN,f_\rmA)-\lambda_3\right)^2}{32K_0^2\left(1+\frac{1}{M-2}\right)}\Bigg\}.
\end{align}
Similarly, if $\lambda_1\leq \mathrm{MMD}^2(f_\rmN,f_\rmA)$, asymptotically, the two-phase test achieves the same false reject exponent in \eqref{logzeta} as the fixed-length test.

Finally, we bound the false alarm probability as follows:
\begin{align}
\rmP_{\mathrm{FA}}^{\mathrm{tp}}(\phi_n|f_\rmN,f_\rmA)
&=\bbP_\rmr\left\{\tau=n, \phi_n(\bY^n)\neq\rmH_\rmr
\right\}+\bbP_\rmr\left\{\tau=Kn,\phi_{Kn}(\bY^{Kn})\neq\rmH_\rmr\right\}\\
&\leq \bbP_\rmr\left\{\phi_n(\bY^n)\neq\rmH_\rmr
\right\}
+ \bbP_\rmr\left\{\phi_{Kn}(\bY^{Kn})\neq\rmH_\rmr\right\}\\
&\leq M(M-1)^2 \exp\left\{\frac{-n\lambda_1^2}{  32K_0^2\left(1+\frac{1}{M-2}\right)}\right\}+M(M-1)^2 \exp\left\{\frac{-Kn\lambda_3^2}{ 32K_0^2\left(1+\frac{1}{M-2}\right)}\right\}\label{FA_al},
\end{align}
where \eqref{FA_al} follows from \eqref{Fa_error_fx} with $(n,\lambda)$ replaced by $(n,\lambda_1)$ and $(Kn,\lambda_3)$, respectively.

Combining \eqref{etau:null} and \eqref{FA_al}, the false alarm exponent satisfies
\begin{align}
\liminf_{n\to\infty}-\frac{1}{\bbE_{\bbP_\rmr}[\tau]}\log \rmP_\mathrm{FA}^{\mathrm{tp}}(\phi_n|f_\rmN,f_\rmA)
&\geq \min\left\{\frac{\lambda_1^2}{ 32K_0^2\left(1+\frac{1}{M-2}\right)},\frac{K\lambda_3^2}{ 32K_0^2\left(1+\frac{1}{M-2}\right)}\right\}.
\end{align}

The proof of Theorem \ref{AFLJMT} is now completed.

\subsection{Proof of Theorem \ref{FLMT_Un}}\label{proof_of_FLMT_Un}
Recall that $\bY^n$ denotes the collection of all observed sequences $(Y_1^n,\ldots,Y_M^n)$. For any set $\calD\subset\calM$, recall that we use $\bY_{\calD}^n$ to denote all sequences $Y_j^n$ with $j\in\calD$ and use $\bar{\bY}_{\calD}^n$ to denote all sequences $Y_j^n$ with $j\in\calM_\calD$. Furthermore, given any $j\in\calM_\calD$, recall that we use $\bar{\bY}_{\calD,j}^n$ to denote all sequences in $\bar{\bY}_{\calD}^n$ except $Y_j^n$.

Fix a positive integer $s\in[T]$ and fix $\calB\in\calC_s$. Under hypothesis $\rmH_\calB$, the indices of the outliers are denoted by the set $\calB$ and the number of outliers is $|\calB|=s$. Given any integers $(t,k)\in[T]^2$ such that $k\leq t$, define the following set
\begin{align}
\calC_{t,k}^{\calB}:=\big\{\calD\in\calC_t~:|\calD\cap\calB|=k,~|\calD\cap\calM_\calB|=t-k\big\},
\label{def:calCtk}
\end{align}
where $\calM_\calB$ denotes the set $\calM\setminus\calB$. Note that $\calC_{t,k}^{\calB}$ collect the set of indices of sequences that have exactly $k$ outliers and $t-k$ nominal samples under hypothesis $\rmH_\calB$.

Under hypothesis $\rmH_\calB$, the misclassification probability is upper bounded as follows:
\begin{align}
\beta_{\calB}(\phi_n|f_\rmN,f_\rmA)
&=\bbP_{\calB}\{\phi_n(\bY^n)\notin\{\rmH_\calB,\rmH_\rmr\}\}\\
&\leq \bbP_{\calB}\{\hatS=s,~\phi_n(\bY^n)\neq\rmH_\calB\}+\bbP_{\calB}\{\hatS\neq s\}\label{beta_UNs_fx0}\\
&=\bbP_{\calB}\{\hatS=s,~\phi_n(\bY^n)\neq\rmH_\calB\}+\bbP_{\calB}\{\hatS>s\}+\bbP_{\calB}\{0<\hatS<s\}\\
\nn&=\bbP_\calB\left\{\hatS=s,~\calI_s^*(\bY^n)\neq\calB, h_s(\bY^n)>\lambda\right\}+ \bbP_\calB\left\{\exists ~t\in [s+1:T], h_t(\bY^n)>\lambda\right\}\\*
&\qquad+\bbP_\calB\left\{h_s(\bY^n)<\lambda\right\}\mathbb{I}(s>1)\label{beta_UNs_fx1},
\end{align}
where \eqref{beta_UNs_fx0} follows since a misclassification error occurs if the number of outliers is estimated wrongly or if the set of outliers is identified wrongly when the number of outliers is estimated correctly, and \eqref{beta_UNs_fx1} follows from the definition of the fixed-length test specified in \eqref{sdetect} and \eqref{FLTest_un}. Specifically, the third term in \eqref{beta_UNs_fx1} appears since the event $0<\hatS<s$ occurs if $s>1$ and we use the fact that $\bigcup_{t\in[s]}\{\by^n:~\max_{t'\in[t:s]}h_{t'}(\by^n)<\lambda\}\subseteq\{\by^n:~h_s(\by^n)<\lambda\}$.

Analogously to \eqref{beta_error_fx2} to \eqref{beta_error_fx1}, the first term in \eqref{beta_UNs_fx1} is upper bounded as follows:
\begin{align}
\bbP_\calB\big\{\hatS=s,~\calI_s^*(\bY^n)\neq\calB, h_s(\bY^n)>\lambda\big\}
&\leq \bbP_\calB\left\{\rmG_{\calB}(\bY^n)>\lambda\right\}  \\
&= \bbP_\calB\Big\{\max_{j\in\calM_\calB}\mathrm{MMD}^2(\bY_j^n,\bar{\bY}^n_{\calB,j})>\lambda \Big\}\\
&\leq\bbP_\calB\big\{\exists~j\in\calM_\calB,~\mathrm{MMD}^2(\bY_j^n,\bar{\bY}^n_{\calB,j})>\lambda\big\}\\
&\leq\sum_{j\in\calM_\calB} \bbP_\calB\left\{\mathrm{MMD}^2(\bY_j^n,\bar{\bY}^n_{\calB,j})>\lambda\right\}\label{PGD1}.
\end{align}

Using the definition of the set $\calC_{t,k}^\calB$, the second item in \eqref{beta_UNs_fx1} is upper bounded as follows:
\begin{align}
\nn &\bbP_\calB\left\{\exists ~t\in [s+1:T], h_t(\bY^n)>\lambda\right\}\\*
&\leq \sum_{t\in[s+1:T]}\bbP_\calB\{h_t(\bY^n)>\lambda\}\\
&=\sum_{t\in[s+1:T]}\sum_{\calV\in\calC_t}\bbP_\calB\{\calI_t^*(\bY^n)=\calV,h_t(\bY^n)>\lambda\}\\
&=\sum_{t\in[s+1:T]}\sum_{\calV\in\calC_{t,s}^{\calB}}\bbP_\calB\left\{\calI_t^*(\bY^n)=\calV,h_t(\bY^n)>\lambda\right\}+\sum_{t\in[s+1:T]}\sum_{\calV\in\calC_t\setminus\calC_{t,s}^{\calB}}\bbP_\calB\left\{\calI_t^*(\bY^n)=\calV,h_t(\bY^n)>\lambda\right\}\label{PGD2-0}\\
&\leq\sum_{t\in[s+1:T]}\sum_{\calV\in\calC_{t,s}^{\calB}}\bbP_\calB\left\{\forall~\calD\in\calC_{t,s}^{\calB}\setminus \{\calV\},~ \rmG_{\calD}(\bY^n)>\lambda\right\}+\sum_{t\in[s+1:T]}\sum_{\calV\in\calC_t\setminus\calC_{t,s}^{\calB}}\bbP_\calB\left\{\forall~\calD\in\calC_{t,s}^{\calB},~\rmG_{\calD}(\bY^n)>\lambda\right\}\label{PGD2-1}\\
&\leq \sum_{t\in[s+1:T]}\sum_{\calV\in\calC_{t,s}^{\calB}}\min_{\calD\in\calC_{t,s}^{\calB}\setminus\{\calV\}}\bbP_\calB\left\{\rmG_{\calD}(\bY^n)>\lambda\right\}+\sum_{t\in[s+1:T]}\sum_{\calV\in\calC_t\setminus\calC_{t,s}^{\calB}}\min_{\calD\in\calC_{t,s}^{\calB}}\bbP_\calB\left\{\rmG_{\calD}(\bY^n)>\lambda\right\}\label{PGD2-2}\\
\nn &\leq \sum_{t\in[s+1:T]}\sum_{\calV\in\calC_{t,s}^{\calB}}\min_{\calD\in\calC_{t,s}^{\calB}\setminus\{\calV\}}\bbP_\calB\left\{\exists~j\in\calM_\calD,~\mathrm{MMD}^2(\bY_j^n,\bar{\bY}^n_{\calD,j})>\lambda\right\}\\
&\qquad+\sum_{t\in[s+1:T]}(|\calC_t|-|\calC_{t,s}^{\calB}|)\min_{\calD\in\calC_{t,s}^{\calB}}
\bbP_\calB\left\{\exists~j\in\calM_\calD,~\mathrm{MMD}^2(\bY_j^n,\bar{\bY}^n_{\calD,j})>\lambda\right\}\label{PGD2-3}\\
\nn&\leq \sum_{t\in[s+1:T]}\sum_{\calV\in\calC_{t,s}^{\calB}}\min_{\calD\in\calC_{t,s}^{\calB}\setminus\{\calV\}}\sum_{j\in\calM_\calD}\bbP_\calB\left\{\mathrm{MMD}^2(\bY_j^n,\bar{\bY}^n_{\calD,j})>\lambda\right\}\\
&\qquad+\sum_{t\in[s+1:T]}(|\calC_t|-|\calC_{t,s}^{\calB}|)\min_{\calD\in\calC_{t,s}^{\calB}}\sum_{j\in\calM_\calD}
\bbP_\calB\left\{\mathrm{MMD}^2(\bY_j^n,\bar{\bY}^n_{\calD,j})>\lambda\right\}\label{PGD2},
\end{align}
where \eqref{PGD2-0} follows by decomposing the set $\calC_t$ as $\calC_{t,s}^{\calB}$ and $\calC_t\setminus\calC_{t,s}^{\calB}$, \eqref{PGD2-1} follows from the definition of $h_t(\by^n)$ in \eqref{secMinG2}, \eqref{PGD2-2} follows from the bound that $\Pr\{\calA_1\mathrm{~and~}\calA_2\}\leq \min\{\Pr\{\calA_1\},\Pr\{\calA_2\}\}$ for any events $\calA_1$ and $\calA_2$ under any probability measure, and \eqref{PGD2-3} follows from the definition of the scoring function $\rmG_{\calD}(\cdot)$ in \eqref{GB2}.

When $s>1$, the third term in \eqref{beta_UNs_fx1} can be further upper bounded as follows:
\begin{align}
\bbP_\calB\left\{h_s(\bY^n)<\lambda\right\}
&\leq \bbP_\calB\Big\{\exists~(\calD,\calV)\in\calC_s^2:~\calD\neq \calV,~\mathrm{and}~\rmG_\calD(\bY^n)<\lambda,~\rmG_\calV(\bY^n)<\lambda\Big\}\label{PGB_1}\\
&\leq\sum_{\calD\in\calC_s\setminus\{\calB\}}\sum_{\calV\in\calC_s\setminus\{\calD\}}\bbP_\calB\{\rmG_\calD(\bY^n)\}<\lambda\}+\sum_{\calV\in\calC_s\setminus\{\calB\}}\bbP_\calB\{\rmG_\calV(\bY^n)<\lambda\}\label{PGB_2}\\
&\leq (|\calC_s|-1)\sum_{\calD\in\calC_s\setminus\{\calB\}}\bbP_\calB\{\rmG_\calD(\bY^n)<\lambda\}+\sum_{\calV\in\calC_s\setminus\{\calB\}}\bbP_\calB\{\rmG_\calV(\bY^n)<\lambda\}\\
&=|\calC_s|\sum_{\calD\in\calC_s\setminus\{\calB\}}\bbP_\calB\{\rmG_\calD(\bY^n)<\lambda\}\\
&=|\calC_s|\sum_{\calD\in\calC_s\setminus\{\calB\}}\bbP_\calB\left\{\max_{j\in\calM_\calD}\mathrm{MMD}^2(\bY_j^n,\bar{\bY}^n_{\calD,j})<\lambda\right\}\label{PGB_3}\\
&\leq  |\calC_s|\sum_{\calD\in\calC_s\setminus\{\calB\}}\bbP_\calB\left\{\max_{j\in\calM_\calD\cap\calB}\mathrm{MMD}^2(\bY_j^n,\bar{\bY}^n_{\calD,j})<\lambda\right\}\label{PGB}\\
&\leq |\calC_s|\sum_{\calD\in\calC_s\setminus\{\calB\}}\min_{j\in\calM_\calD\cap\calB}\bbP_\calB\{\mathrm{MMD}^2(\bY_j^n,\bar{\bY}^n_{\calD,j})<\lambda\}\label{PGD3},
\end{align}
where \eqref{PGB_1} follows from the definition of $h_s(\bY^n)$ in \eqref{secMinG2}, \eqref{PGB_2} follows since $\Pr\{\rmG_\calD(\bY^n)<\lambda,~\rmG_\calV(\bY^n)<\lambda\}\leq \min\{\Pr\{\rmG_\calD(\bY^n)<\lambda\},\Pr\{\rmG_\calV(\bY^n)<\lambda\}\}$, \eqref{PGB_3} follows from the definition of $\rmG_\calD(\cdot)$ in \eqref{GB2}, \eqref{PGB} follows because $\calM_\calD\cap\calB$ is a non-empty set when $\calD\neq\calB$ and $\max_{j\in\calM_\calD\cap\calB}\mathrm{MMD}^2(Y_j^n,\bar{\bY}_{\calD,j}^n)\leq \max_{k\in\calM_\calD}\mathrm{MMD}^2(Y_k^n,\bar{\bY}_{\calD,k}^n)$, and \eqref{PGD3} follows similarly to \eqref{PGD2-2}.

The results in \eqref{PGD1}, \eqref{PGD2} and \eqref{PGD3} can be further upper bounded by applying the McDiarmid's inequality in Lemma \ref{McDiarmid}, analogously to the case of at most one outlier. Consider any set $\calD\in\calC$ such that $\calB\subseteq\calD$. Under hypothesis $\rmH_\calB$, for each $j\in\calM_\calD$, each sequence $\bY_j^n$ is a nominal sample and $\bar{\bY}^n_{\calD,j}$ collect nominal samples. It follows from the definition of the MMD metric in \eqref{MMDcompute} that
\begin{align}\label{EMMD_PB}
\bbE_{\rmP_\calB}[\mathrm{MMD}^2(\bY_j^n,\bar{\bY}^n_{\calD,j})]=0.
\end{align}
To bound the Lipschitz constant, given any observed sequences $\by^n=(y_1^n,\ldots,y_M^n)$, for each $t\in[s:T]$, given $\calD\in\calC_t$ such that $\calB\subseteq\calD$ define the function $g_{\calD,j}(\bar{\by}_\calD^n):=\mathrm{MMD}^2(y_j^n,\bar{\by}_{\calD,j}^n)$. Note that $g_{\calD,j}(\bar{\by}_\calD^n)$ is a function of $(M-t)n$ parameters. For each $k\in[(M-t)n]$, if the $k$-th element of $\bar{\by}_\calD^n$ is replaced by $\tily$, we use $g_{\calD,j}(\bar{\by}_\calD^n,k,\tily)$ to denote the corresponding function value. For each $(j,k)\in\calM_\calD\times[(M-t)n]$, define
\begin{align}
\label{lips:uk}
c_k^{\calD,j}:=\sup_{\by^n,\tily}|g_{\calD,j}(\bar{\by}_\calD^n)-g_{\calD,j}(\bar{\by}_\calD^n,k,\tily)|.
\end{align}
Similarly to \eqref{frac}, we obtain
\begin{equation}\label{E-frac}
\sum_{k\in[(M-t)n]}(c_k^{\calD,j})^2\leq \frac{64K_0^2}{n}\left(1+\frac{1}{M-t-1}\right).
\end{equation}

Applying the McDiarmid's inequality in Lemma \ref{McDiarmid} and invoking \eqref{E-frac} with $\calD=\calB$ such that $t=s$, it follows from \eqref{PGD1} that
\begin{align}
\bbP_\calB\left\{\hatS=s,~\calI_s^*(\bY^n)\neq\calB, h_s(\bY^n)>\lambda\right\}
% &\leq \sum_{j\in\calM_\calB} \bbP_\calB\left\{\mathrm{MMD}^2(\bY_j^n,\bar{\bY}^n_{\calB,j})>\lambda\right\}\\
&\leq (M-s) \exp\left\{\frac{-n\lambda^2}{32K_0^2\left(1+\frac{1}{M-s-1}\right)}\right\}\leq (M-s)g_1(\lambda|f_\rmN,f_\rmA)\label{PGD_1}.
\end{align}
Similarly, it follows from \eqref{PGD2} that
\begin{align}
\nn&\bbP_\calB\left\{\exists ~t\in [s+1:T], h_t(\bY^n)>\lambda\right\}\\
&\leq \sum_{t\in[s+1:T]}|\calC_{t,s}^{\calB}|(M-t)\exp\left\{\frac{-n\lambda^2}{32K_0^2\left(1+\frac{1}{M-t-1}\right)}\right\}+
\sum_{t\in[s+1:T]}(|\calC_t|-|\calC_{t,s}^{\calB}|)(M-t)\exp\left\{\frac{-n\lambda^2}{32K_0^2\left(1+\frac{1}{M-t-1}\right)}\right\}\label{PGD_2-0}\\
&=\sum_{t\in[s+1:T]}|\calC_t|(M-t)\exp\left\{\frac{-n\lambda^2}{32K_0^2\left(1+\frac{1}{M-t-1}\right)}\right\}\\
&\leq \sum_{t\in[s+1:T]}|\calC_t|(M-t)\exp\left\{-ng_1(\lambda|f_\rmN,f_\rmA)\right\}\label{PGD_2}.
\end{align}
where $g_1(\lambda|f_\rmN,f_\rmA)$ was defined in \eqref{nota1}.

We now upper bound \eqref{PGD3} when $s>1$. Given any $\calD\in\calC_s\setminus\{\calB\}$, for each $j\in\calM_\calD\cap\calB$, $Y_j^n$ is an outlier while $\bar{\bY}_{\calD,j}$ collects both outliers and nominal samples when $s>1$ and collects nominal samples when $s=1$. It follows that
\begin{align}
\bbE_{\rmP_\calB}[\mathrm{MMD}^2[\bY_j^n,\bar{\bY}^n_{\calD,j}]
&=\frac{(M-s-(s-t_1))((M-s-(s-t_1))n-1)\mathrm{MMD}^2(f_\rmN,f_\rmA)}{(M-s-1)((M-s-1)n-1)}\label{EGB_PB2-0},
\end{align}
where $t_1=|\calM_\calD\cap\calB|<s$ denotes the number of outliers in $\bar{\bY}^n_{\calD}$. When $s>1$, it follows that for $n$ sufficiently large,
\begin{align}
\bbE_{\rmP_\calB}[\mathrm{MMD}^2[\bY_j^n,\bar{\bY}^n_{\calD,j}]
&\geq \left(1-\frac{s-t_1}{M-s-1}\right)^2\mathrm{MMD}^2(f_\rmN,f_\rmA)\label{EGB_PB2-1}\\
&\geq\left(1-\frac{s}{M-s-1}\right)^2\mathrm{MMD}^2(f_\rmN,f_\rmA)\label{EGB_PB2}.
\end{align}
We now calculate the Lipschitz constant. Recall the definitions of $g_{\calD,j}(\bar{\by}_\calD^n)$ and  $c_k^{\calD,j}$ around \eqref{lips:uk}. Since $\calD\in\calC_s$, $g_{\calD,j}(\bar{\by}_\calD^n)$ is a function of $(M-s)n$ parameters. Similarly to \eqref{frac}, we have
\begin{equation}\label{E-frac2}
\sum_{k\in[(M-s)n]}(c_k^{\calD,j})^2\leq \frac{64K_0^2}{n}\left(1+\frac{1}{M-s-1}\right).
\end{equation}
Combing \eqref{EGB_PB2}, \eqref{E-frac2}, when $s>1$ and $\lambda<\left(1-\frac{s}{M-s-1}\right)^2\mathrm{MMD}^2(f_\rmN,f_\rmA)$, it follows from the McDiarmid's inequality in Lemma \ref{McDiarmid} that
\begin{align}
\bbP_\calB\left\{h_s(\bY^n)<\lambda\right\}
&\leq  |\calC_s|(|\calC_{s}|-1)\exp\left\{-ng_2(\lambda|f_\rmN,f_\rmA)\right\},\label{PGB2}
\end{align}
where $g_2(\lambda|f_\rmN,f_\rmA)$ was defined in \eqref{nota2}.

Combining \eqref{beta_UNs_fx1}, \eqref{PGD_1}, \eqref{PGD_2} and \eqref{PGB2}, we conclude that when $s>1$ and $\lambda<\left(1-\frac{s}{M-s-1}\right)^2\mathrm{MMD}^2(f_\rmN,f_\rmA)$, for each $\calB\in\calC_s$,
\begin{align}
\lim_{n\to\infty}-\frac{1}{n}\log \beta_\calB(\phi_n|f_\rmN,f_\rmA)
&\geq
\min\left\{ g_1(\lambda|f_\rmN,f_\rmA), g_2(\lambda|f_\rmN,f_\rmA)\right\}.\label{beta_UNs_fx}
%& \calB\in\calC_s,~s=1.
\end{align}
and when $s=1$ and $\lambda<\mathrm{MMD}^2(f_\rmN,f_\rmA)$, for each $\calB\in\calC_s$,
\begin{align}
\lim_{n\to\infty}-\frac{1}{n}\log \beta_\calB(\phi_n|f_\rmN,f_\rmA)
&\geq g_1(\lambda|f_\rmN,f_\rmA).
\end{align}

We next analyze the false reject probability. When $s>1$, for each $\calB\in\calC_s$, if $\lambda<\left(1-\frac{s}{M-s-1}\right)^2\mathrm{MMD}^2(f_\rmN,f_\rmA)$, it follows that
\begin{align}
\zeta(\phi_n|f_\rmN,f_\rmA)
&=\bbP_\calB\{\phi_n(\bY^n)=\rmH_\rmr\}\\
&=\bbP_\calB\{\forall~ t\in[T], h_t(\bY^n)<\lambda\}\\
&\leq\bbP_\calB\{h_s(\bY^n)<\lambda\}\\
&\leq |\calC_{s}|(|\calC_{s}|-1)\exp\left\{-ng_2(\lambda|f_\rmN,f_\rmA)\right\},\label{zeta_UNs_fx}
\end{align}
where \eqref{zeta_UNs_fx} follows from the result in \eqref{PGB2}. When $s=1$, similarly to the analyses leading to \eqref{zeta_error_fx}, if $\lambda<\mathrm{MMD}^2(f_\rmN,f_\rmA)$, we have
\begin{align}
\zeta(\phi_n|f_\rmN,f_\rmA)
&=\bbP_\calB\{\forall~ t\in[T], h_t(\bY^n)<\lambda\}\\
&\leq\bbP_\calB\{h_1(\bY^n)<\lambda\}\\
&=M(M-1)\exp\left\{\frac{-n\left(\mathrm{MMD}^2(f_\rmN,f_\rmA)-\lambda\right)^2}
{32K_0^2\left(1+\frac{1}{M-2}\right)}\right\}\label{zeta_UNs_fx-s1}.
\end{align}
Combining \eqref{zeta_UNs_fx} and \eqref{zeta_UNs_fx-s1}, we conclude that when  $s>1$ and $\lambda<\left(1-\frac{s}{M-s-1}\right)^2\mathrm{MMD}^2(f_\rmN,f_\rmA)$, for each $\calB\in\calC_s$,
\begin{align}
\lim_{n\rightarrow\infty}-\frac{1}{n}\log \zeta_\calB (\phi_n|f_\rmN,f_\rmA)  \geq
g_2(\lambda|f_\rmN,f_\rmA),
\end{align}
and when $s=1$ and $\lambda<\mathrm{MMD}^2(f_\rmN,f_\rmA)$, for each $\calB\in\calC_s$,

\begin{align}
\lim_{n\rightarrow\infty}-\frac{1}{n}\log \zeta_\calB (\phi_n|f_\rmN,f_\rmA)  \geq
\frac{\left(\mathrm{MMD}^2(f_\rmN,f_\rmA)-\lambda\right)^2}
{32K_0^2\left(1+\frac{1}{M-2}\right)}.
\end{align}

Finally, we upper bound the false alarm probability as follows:
\begin{align}
\rmP_{\mathrm{FA}}(\phi_n|f_\rmN,f_\rmA)
&=\bbP_\rmr\{\phi_n(\bY^n)\neq\rmH_\rmr\}\\
&=\bbP_\rmr\left\{\exists~ t\in[T], h_t(\bY^n)>\lambda\right\}\\
&\leq \sum_{t\in[T]}\sum_{\calV\in\calC_t}\sum_{\calD\in\calC_t\setminus\{\calV\}}
\bbP_\rmr\{\rmG_\calD(\bY^n)>\lambda\}\\
&\leq \sum_{t\in[T]}|\calC_{t}|^2(M-t)\exp\left\{\frac{-n\lambda^2}{ 32K_0^2\left(1+\frac{1}{M-t-1}\right)}\right\}\label{FA_UNs_fx:step0}\\
&\leq\sum_{t\in[T]}|\calC_{t}|^2(M-t)\exp\left\{-ng_1(\lambda|f_\rmN,f_\rmA)\right\},\label{FA_UNs_fx}
\end{align}
where \eqref{FA_UNs_fx:step0} follows from the McDiarmid's inequality in Lemma \ref{McDiarmid} similarly to \eqref{PGD2} and noting that $\bbE_{\bbP_\rmr}[\rmG_\calD(\bY^n)]=0$. Thus,
\begin{align}
&\lim_{n\rightarrow\infty}-\frac{1}{n}\log \rmP_{\mathrm{FA}}(\phi_n|f_\rmN,f_\rmA)
\geq g_1(\lambda|f_\rmN,f_\rmA).\label{logFA_un}
\end{align}
The proof of Theorem \ref{FLMT_Un} is now completed.

\subsection{Proof of Theorem \ref{SJMT_un}}
\label{proof_of_SJMT_un}
The proof of Theorem \ref{SJMT_un} is similar to the proof of Theorems \ref{SJMT} and \ref{FLMT_Un}. Thus, we only emphasize the differences here.  

Recall the definitions of exponent functions $g_1(\cdot)$ in \eqref{nota1} and $g_2(\cdot)$ in \eqref{nota2}. Recall the definition of the stopping time $\tau$ in \eqref{Taulength2_unS}. For each non-empty set $\calB\in\calC$, the expected stopping time under hypothesis $\rmH_\calB$ is upper bounded as follows:
\begin{align}
\bbE_{\bbP_\calB}[\tau]
&\leq \sum_{\tau'=1}^\infty \bbP_\calB\{\tau\geq\tau'\}\\
&\leq N-1+ \sum_{\tau'=N-1}^\infty \bbP_\calB\{\tau>\tau'\}\label{Etau_unS-0}\\
&\leq N-1+\sum_{\tau'=N-1}^\infty
\bbP_\calB\left\{\exists ~t\in [T],~\lambda_2<h_t(\bY^{\tau'})<\lambda_1\right\}\label{Etau_unS-1}\\
&\leq N-1+ \sum_{\tau'=N-1}^\infty\bigg(\bbP_\calB\left\{\exists~t\in[s+1:T],~h_t(\bY^{\tau'})>\lambda_2\right\}+\bbP_\calB\left\{\exists~t\in [s],~h_t(\bY^{\tau'})<\lambda_1\right\}\bigg)
\label{Etau_unS},
\end{align}
where \eqref{Etau_unS-0} and \eqref{Etau_unS-1} follow from the definition of the random stopping time $\tau$ in \eqref{Taulength2_unS}.

Recall the definitions of $g_1(y|f_\rmN,f_\rmA)$ and $g_2(y|f_\rmN,f_\rmA)$ in \eqref{nota1} and \eqref{nota2}, respectively. Using result in \eqref{PGD_2} by replacing ($n,\lambda$) by ($\tau',\lambda_2$), the first term inside the sum of \eqref{Etau_unS} satisfies
\begin{align}
\bbP_\calB\left\{\exists~t\in [s+1:T],~h_t(\bY^{\tau'})>\lambda_2\right\}
&\leq \sum_{t\in[s+1:T]}|\calC_t|(M-t)\exp\left\{-\tau'g_1(\lambda_2|f_\rmN,f_\rmA)\right\}.\label{PB_second1}
\end{align}
Similarly to the derivations leading to \eqref{PGD3}, the second term inside the sum of \eqref{Etau_unS} satisfies
\begin{align}
\nn&\bbP_\calB\left\{\exists~t\in [s],~h_t(\bY^{\tau'})<\lambda_1\right\}\\*
&\leq \sum_{t\in[s-1]}\bbP_\calB\{h_t(\bY^{\tau'})<\lambda_1\}\mathbb{I}(s>1)+\bbP_\calB\{h_s(\bY^{\tau'})<\lambda_1\}\\
&\leq \sum_{t\in[s-1]}(|\calC_t|-1)\sum_{\calD\in\calC_t}\min_{j\in\calM_\calD\cap\calB}\bbP_\calB\big\{\mathrm{MMD}^2(\bY_j^{\tau'},\bar{\bY}^{\tau'}_{\calD,j})<\lambda_1\big\}\mathbb{I}(s>1) +\bbP_\calB\{h_s(\bY^{\tau'})<\lambda_1\}\label{PB_second2}.
\end{align}
Using \eqref{PGB2} with $(n,\lambda)$ replaced by $(\tau',\lambda_2)$, the second term of \eqref{PB_second2} satisfies
\begin{align}
\bbP_\calB\{h_s(\bY^{\tau'})<\lambda_1\}\leq (M-1)M\exp\{-\tau'g_2(\lambda_1|f_\rmN,f_\rmA)\}\label{PB_second2-2}.
\end{align}

It now suffices to bound the first term in \eqref{PB_second2} using McDiarmid's inequality in Lemma \ref{McDiarmid}. Recall the definition of $\calC_{t,k}^\calB$ in \eqref{def:calCtk}. Fix $t\in[s-1]$. Given any set $\calD\in\calC_t$, let $t_1=|\calM_\calD\cap\calB|$ the number of outliers in $\bar{\bY}^{\tau'}_{\calD}$. When $s>1$, for any $n\in\bbN$, $t_1\leq t$ and $\calD\in\calC_{t,t_1}^\calB$, analogously to \eqref{EGB_PB2-0}, we have
\begin{align}
\bbE_{\bbP_\calB}[\mathrm{MMD}^2(\bY_j^n,\bar{\bY}^n_{\calD,j})]
&=\frac{(M-t-(t-t_1))((M-t-(t-t_1))n-1)\mathrm{MMD}^2(f_\rmN,f_\rmA)}{(M-t-1)((M-t-1)n-1)}\\
&\geq \left(1-\frac{t-t_1}{M-t-1}\right)^2\mathrm{MMD}^2(f_\rmN,f_\rmA)\\
&\geq \left(1-\frac{t}{M-t-1}\right)^2\mathrm{MMD}^2(f_\rmN,f_\rmA)\\
&>\left(1-\frac{s}{M-s-1}\right)^2\mathrm{MMD}^2(f_\rmN,f_\rmA).\label{E_uk}
\end{align}
When $s=1$, $\bbE_{\bbP_\calB}[\mathrm{MMD}^2(\bY_j^n,\bar{\bY}^n_{\calD,j})]=\mathrm{MMD}^2(f_\rmN,f_\rmA)$.
Similarly to \eqref{frac}, the Lipschitz continuous constants $c_k^{\calD,j}$ defined around \eqref{lips:uk} satisfy
\begin{align}
\sum_{k\in[(M-t)n]}(c_k^{\calD,j})^2
&\leq \frac{64K_0^2}{n}\left(1+\frac{1}{M-t-1}\right)<\frac{64K_0^2}{n}\left(1+\frac{1}{M-s-1}\right).\label{frac_uk}
\end{align}

Combining \eqref{PB_second2}, \eqref{PB_second2-2}, \eqref{E_uk} and \eqref{frac_uk}, it follows that
\begin{align}
\bbP_\calB\left\{\exists~t\in [s],~h_t(\bY^{\tau'})<\lambda_1\right\}
\nn&\leq\sum_{t\in[1:s]}(|\calC_t|-1)|\calC_t|\exp\{-\tau'g_2(\lambda_1|f_\rmN,f_\rmA)\}\mathbb{I}(s>1)\\*
&\qquad+(M-1)M\exp\left\{\frac{-\tau'\left(\mathrm{MMD}^2(f_\rmN,f_\rmA)-\lambda_1\right)^2}
{32K_0^2\left(1+\frac{1}{M-s-1}\right)}\right\}\mathbb{I}(s=1).
\label{PB_second3}
\end{align}

It follows from \eqref{Etau_unS}, \eqref{PB_second1} and \eqref{PB_second3} that, for $s>1$, and $\lambda_1<\left(1-\frac{s}{M-s-1}\right)^2\mathrm{MMD}^2(f_\rmN,f_\rmA)$, the expected stopping time $\bbE_{\rmP_\calB}[\tau]$ satisfies
\begin{align}
&\bbE_{\rmP_\calB}[\tau]
\leq N-1+ \sum_{\tau'=N-1}^\infty \bbP_\calB\{\tau>\tau'\}\\
&\leq N-1+ \left(\sum_{t\in[s+1:T]}|\calC_t|(M-t)+\sum_{t\in[1:s]}(|\calC_t|-1)|\calC_t|\right)\sum_{\tau'=N-1}^\infty \exp\Big\{
-\tau'\min\big\{g_1(\lambda_2|f_\rmN,f_\rmA),g_2(\lambda_1|f_\rmN,f_\rmA)\big\}\Big\} \\
&= N-1+\left(\sum_{t\in[s+1:T]}|\calC_t|(M-t)+\sum_{t\in[1:s]}(|\calC_t|-1)|\calC_t|\right)
\frac{\exp\Big\{-(N-1)\min\big\{g_1(\lambda_2|f_\rmN,f_\rmA),g_2(\lambda_1|f_\rmN,f_\rmA)\big\}\Big\}}
{1-\exp\Big\{\min\big\{g_1(\lambda_2|f_\rmN,f_\rmA),g_2(\lambda_1|f_\rmN,f_\rmA)\big\}\Big\}}
.\label{EBtau_un}
\end{align}
Similarly, when $s=1$, given any $\lambda_1<\mathrm{MMD}^2(f_\rmN,f_\rmA)$, the expected stopping time $\bbE_{\rmP_\calB}[\tau]$ satisfies
\begin{align}
&\bbE_{\rmP_\calB}[\tau]
\leq
N-1+\sum_{t\in[T]}|\calC_t|(M-t) \frac{\exp\left\{-(N-1)\min\left\{g_1(\lambda_2|f_\rmN,f_\rmA),\frac{\left(\mathrm{MMD}^2(f_\rmN,f_\rmA)-\lambda_1\right)^2}
{32K_0^2\left(1+\frac{1}{M-s-1}\right)}\right\}\right\}}
{1-\exp\Big\{\min\big\{g_1(\lambda_2|f_\rmN,f_\rmA),\frac{\left(\mathrm{MMD}^2(f_\rmN,f_\rmA)-\lambda_1\right)^2}
{32K_0^2\left(1+\frac{1}{M-s-1}\right)}\big\}\Big\}}
.\label{EBtau_un-s1}
\end{align}
Thus, if $N$ is sufficiently large, $\bbE_{\bbP_\calB}[\tau]\leq N$ if $\lambda_1<\left(1-\frac{s}{M-s-1}\right)^2\mathrm{MMD}^2(f_\rmN,f_\rmA)$ when $s>1$ and $\lambda_1<\mathrm{MMD}^2(f_\rmN,f_\rmA)$ when $s=1$.

We next upper bound the average stopping time under the null hypothesis. It follows from the definition of $\tau$ in \eqref{Taulength2_unS} that
\begin{align}
\bbE_{\rmP_\rmr}[\tau]
&\leq N-1+\sum_{\tau'=N-1}^\infty\bbP_\rmr\{\tau>\tau'\}\\
&\leq N-1+\sum_{\tau'=N-1}^\infty\bbP_\rmr\left\{\exists~ t\in[T], ~\lambda_2<h_t(\bY^{\tau'})<\lambda_1 \right\}\\
&\leq  N-1+\sum_{\tau'=N-1}^\infty\bbP_\rmr\left\{\exists~ t\in[T],~h_t(\bY^{\tau'}) >\lambda_2\right\}\\
& \leq N-1+\sum_{t\in[T]}|\calC_{t}|^2\sum_{\tau'=N-1}^\infty\exp\left\{-\tau'g_1(\lambda_2|f_\rmN,f_\rmA)\right\}\label{PR_second}\\
&\leq N-1+\sum_{t\in[T]}|\calC_{t}|^2 \frac{\exp\Big\{-(N-1)g_1(\lambda_2|f_\rmN,f_\rmA)\Big\}}{1-\exp\Big\{-g_1(\lambda_2|f_\rmN,f_\rmA)\Big\}}\label{Ertau_un}
\end{align}
where \eqref{PR_second} follows from \eqref{FA_UNs_fx} with $(n,\lambda)$ replaced by $(\tau',\lambda_2)$. If $N$ is sufficiently large, $\bbE_{\bbP_\rmr}[\tau]\leq N$.

In the remaining part of the proof, we upper bound three kinds of error probabilities. We first bound the misclassification and false reject probabilities by considering $\calB\in\calC_s$ when $s>1$ and  $\lambda_1<\left(1-\frac{s}{M-s-1}\right)^2\mathrm{MMD}^2(f_\rmN,f_\rmA)$. The misclassification probability satisfies
\begin{align}
\beta_\calB^{\mathrm{seq}}(\phi_\tau|f_\rmN,f_\rmA)
&\leq \sum_{\tau'=N-1}^\infty\bbP_\calB\{\phi_{\tau'}(\bY^{\tau'})\notin\{\rmH_\calB,\rmH_\rmr\}\}\\
&\leq \sum_{\tau'=N-1}^\infty\bbP_{\calB}\{\hatS=s,~\phi_{\tau'}(\bY^{\tau'})\neq\rmH_\calB\}+\sum_{\tau'=N-1}^\infty\bbP_{\calB}\{\hatS\neq s\}\\
\nn&\leq
\Bigg((M-s)+\sum_{t\in[s+1:T]}|\calC_t|(M-t)\Bigg)\sum_{\tau'=N-1}^\infty \exp\Big\{-\tau'g_1(\lambda_1|f_\rmN,f_\rmA)\Big\}\\
&\qquad+|\calC_s|(|\calC_{s}|-1)\sum_{\tau'=N-1}^\infty\exp\Big\{-\tau'g_2(\lambda_2|f_\rmN,f_\rmA)\Big\},
\label{beta_error_ST_un}
\end{align}
where \eqref{beta_error_ST_un} follows from \eqref{PGD_1}, \eqref{PGD_2}, and \eqref{PGB2}  with $(n,\lambda)$ replaced with $(\tau',\lambda_1)$ and $(\tau',\lambda_2)$, respectively. 
The false reject probability satisfies
\begin{align}
\zeta_\calB^{\mathrm{seq}}(\phi_\tau|f_\rmN,f_\rmA)
&\leq
\sum_{\tau'=N-1}^\infty \bbP_\calB\left\{\phi_{\tau'}(\bY^{\tau'})=\rmH_\rmr\right\}\\
&\leq  |\calC_{s}|(|\calC_{s}|-1)\sum_{\tau'=N-1}^\infty\exp\Big\{-\tau'g_2(\lambda_2|f_\rmN,f_\rmA)\Big\},\label{zeta_uNs_ST}
\end{align}
where \eqref{zeta_uNs_ST} follows from \eqref{zeta_UNs_fx} with $(n,\lambda)$ replaced by $(\tau',\lambda_2)$.

Combining \eqref{EBtau_un}, \eqref{beta_error_ST_un} and \eqref{zeta_uNs_ST}, when $s>1$ and $\lambda_2<\left(1-\frac{s}{M-s-1}\right)^2$, for each $\calB\in\calC_s$, it follows that
\begin{align}
\lim_{N\to\infty}-\frac{1}{\bbE_{\bbP_\calB}[\tau]}\log \beta_\calB^{\mathrm{seq}}(\phi_n|f_\rmN,f_\rmA)
&\geq
\min\left\{g_1(\lambda_1|f_\rmN,f_\rmA),g_2(\lambda_2|f_\rmN,f_\rmA)\right\}.\label{logbeta_UNs_ST}\\
\lim_{N\rightarrow\infty}-\frac{1}{\bbE_{\bbP_\calB}[\tau]}\log \zeta_\calB^{\mathrm{seq}} (\phi_n|f_\rmN,f_\rmA) 
&\geq g_2(\lambda_2|f_\rmN,f_\rmA)\label{logzeta_uNs_ST}.
\end{align}

We next bound the misclassification and false reject probabilities by considering $\calB\in\calC_s$ when $s=1$ and $\lambda_1<\mathrm{MMD}^2(f_\rmN,f_\rmA)$. Using \eqref{PGD_2} with ($n,\lambda$) replaced by ($\tau',\lambda_1$), the misclassification error probability satisfies
\begin{align}
\beta_\calB^{\mathrm{seq}}(\phi_\tau|f_\rmN,f_\rmA)
\leq
\sum_{t\in[1:T]}|\calC_t|(M-t)\sum_{\tau'=N-1}^\infty \exp\left\{-\tau'g_1(\lambda_1|f_\rmN,f_\rmA)\right\}.
\label{beta_error_ST_un-s1}
\end{align}
Similarly, using \eqref{zeta_UNs_fx-s1}, the false reject probability satisfies
\begin{align}
\zeta_\calB^{\mathrm{seq}}(\phi_\tau|f_\rmN,f_\rmA)
&\leq  M(M-1)\sum_{\tau'=N-1}^\infty\exp\left\{\frac{-\tau'\left(\mathrm{MMD}^2(f_\rmN,f_\rmA)-\lambda_2\right)^2}
{32K_0^2\left(1+\frac{1}{M-2}\right)}\right\}\label{zeta_uNs_ST-s1}.
\end{align}
Thus, Combining \eqref{EBtau_un-s1}, \eqref{EBtau_un-s1} and \eqref{zeta_uNs_ST-s1}, when $s=1$ and $\lambda_2<\mathrm{MMD}^2(f_\rmN,f_\rmA)$, it follows that for each $\calB\in\calC_s$,
\begin{align}
\lim_{N\to\infty}-\frac{1}{\bbE_{\bbP_\calB}[\tau]}\log \beta_\calB^{\mathrm{seq}}(\phi_n|f_\rmN,f_\rmA)
&\geq g_1(\lambda_1|f_\rmN,f_\rmA).\label{logbeta_UNs_ST-s1},\\
\lim_{N\rightarrow\infty}-\frac{1}{\bbE_{\bbP_\calB}[\tau]}\log \zeta_\calB^{\mathrm{seq}}(\phi_n|f_\rmN,f_\rmA)  
&\geq
\frac{(\mathrm{MMD}^2(f_\rmN,f_\rmA)-\lambda_2)^2}{32K_0^2\left(1+\frac{1}{M-2}\right)}. \label{logzeta_uNs_ST-s1}
\end{align}

Finally, the false alarm probability is upper bounded by
\begin{align}
\rmP_{\mathrm{FA}}^{\mathrm{seq}}(\phi_\tau|f_\rmN,f_\rmA)
&\leq \sum_{\tau'=N-1}^\infty
\bbP_\rmr\left\{\phi_{\tau'}(\bY^{\tau'})\neq\rmH_\rmr\right\}\\
&\leq \sum_{\tau'=N-1}^\infty
\bbP_\rmr\left\{\exists~ t\in[T], h_t(\bY^{\tau'})>\lambda_1\right\}\\
&\leq \sum_{t\in[T]}|\calC_{t}|^2(M-t)\sum_{\tau'=N-1}^\infty\exp\left\{-\tau'g_1(\lambda_2|f_\rmN,f_\rmA)\right\}\label{FA_UNs_ST}
\end{align}
where \eqref{FA_UNs_ST} follows from \eqref{FA_UNs_fx} with $(n,\lambda)$ replaced by $(\tau',\lambda_1)$. Combining \eqref{Ertau_un} and \eqref{FA_UNs_ST}, it follows that
\begin{align}
\lim_{N\rightarrow\infty}-\frac{1}{\bbE_{\bbP_\rmr}[\tau]}\log \rmP_{\mathrm{FA}}^{\mathrm{seq}}(\phi_\tau|f_\rmN,f_\rmA)  \geq
g_1(\lambda_1|f_\rmN,f_\rmA).\label{logFA_UNs_ST}
\end{align}

The proof of Theorem \ref{SJMT_un} is now completed.

\subsection{Proof of Theorem \ref{E-AFLJMT}}
\label{proof_of_E-AFLJMT}

The proof of Theorem \ref{E-AFLJMT} is similar to the proof of Theorems \ref{AFLJMT}, \ref{FLMT_Un} and \ref{SJMT_un}. Thus, we only emphasize the differences here.  

Recall the definitions of exponent functions $g_1(\cdot)$ in \eqref{nota1} and $g_2(\cdot)$ in \eqref{nota2}. We first bound the probability that the two-phase test proceeds to the second phase under each hypothesis so that the expected stopping time under each hypothesis is well bounded. Subsequently, we upper bound three kinds of error probabilities using the results for the fixed-length test in \ref{FLMT_Un} and obtain the desired exponential lower bounds. 

Fix a non-empty set $\calB\in\calC_s$ for some $s>1$.  When $\lambda_1<\left(1-\frac{s}{M-s-1}\right)^2\mathrm{MMD}^2(f_\rmN,f_\rmA)$, it follows that
\begin{align}
\bbP_\calB\{\tau=Kn\}
&\leq
\bbP_\calB\left\{\exists ~t\in [T],~\lambda_2<h_t(\bY^n)<\lambda_1\right\}\\
&\leq\bbP_\calB\left\{\exists~t\in[s+1:T],~h_t(\bY^n)>\lambda_2\right\}+\bbP_\calB\left\{\exists~t\in [s],~h_t(\bY^n)<\lambda_1\right\}.\\
&\leq \sum_{t\in[s+1:T]}|\calC_t|(M-t)\exp\left\{-ng_1(\lambda_2|f_\rmN,f_\rmA)\right\}+\sum_{t\in[1:s]}(|\calC_t|-1)|\calC_t|\exp\left\{-ng_2(\lambda_1|f_\rmN,f_\rmA)\right\}, \label{PB_second0}
\end{align}
where \eqref{PB_second0} follows from the results of \eqref{PB_second1} and \eqref{PB_second3} with $\tau'$ replaced by $n$. Similarly,
when $s=1$ and $\lambda_1<\mathrm{MMD}^2(f_\rmN,f_\rmA)$, we have
\begin{align}
\bbP_\calB\{\tau=Kn\}
&\leq \sum_{t\in[s+1:T]}|\calC_t|(M-t)\exp\left\{-ng_1(\lambda_2|f_\rmN,f_\rmA)\right\}\\
&\qquad +M(M-1)\exp\left\{\frac{-n\left(\mathrm{MMD}^2(f_\rmN,f_\rmA)-\lambda_1\right)^2}
{32K_0^2\left(1+\frac{1}{M-2}\right)}\right\}.\label{PB_second0-s1}
\end{align}
Thus,  if  $\lambda_1<\left(1-\frac{s}{M-s-1}\right)^2\mathrm{MMD}^2(f_\rmN,f_\rmA)$ when $s>1$ or if  $\lambda_1<\mathrm{MMD}^2(f_\rmN,f_\rmA)$ when $s=1$, when $n$ is sufficient large, the expected stopping time satisfies that for each $\calB\in\calC_s$,
\begin{align}
\bbE_{\rmP_\calB}[\tau]
&=n +(K-1)n\bbP_\calB\{\tau=Kn\}\\
&\leq n+1, \label{EBtau_un_tp}
\end{align}
where \eqref{EBtau_un_tp} follows by using \eqref{PB_second0} and \eqref{PB_second0-s1}, analogously to the case of sequential test.

Similarly, under the null hypothesis, it follows that
\begin{align}
\bbP_\rmr\{\tau =Kn\}
&\leq\bbP_\rmr\left\{\exists~ t\in[T], ~\lambda_2<h_t(\bY^n)<\lambda_1 \right\}\\
&\leq \bbP_\rmr\left\{\exists~ t\in[T],~h_t(\bY^n) >\lambda_2\right\}\\
& \leq\sum_{t\in[T]}|\calC_{t}|^2\exp\left\{-ng_1(\lambda_2|f_\rmN,f_\rmA)\right\}.\label{PR_second_tp}
\end{align}
where \eqref{PR_second_tp} follows from \eqref{FA_UNs_fx}. Thus, when $n$ is sufficiently large,
\begin{align}
\bbE_{\rmP_\rmr}[\tau]
&= n +(K-1)n\bbP_\rmr\{\tau=Kn\}\\
&\leq n+1\label{Ertau_un_tp}.
\end{align}

In subsequent analyses, we upper bound each of the three error probabilities. First consider the misclassification and false reject probability for $\calB\in\calC_s$ when $s>1$. When $\max\{\lambda_1,\lambda_3\}<\left(1-\frac{s}{M-s-1}\right)^2\mathrm{MMD}^2(f_\rmN,f_\rmA)$, it follows that
\begin{align}
\beta_\calB^{\rmtp}(\phi^n|f_\rmN,f_\rmA)
&\leq\bbP_{\calB}\{\phi_n(\bY^n)\notin\{\rmH_\calB,\rmH_\rmr\}\}+
\bbP_{\calB}\{\phi_{Kn}(\bY^{Kn})\notin\{\rmH_\calB,\rmH_\rmr\}\}\\
\nn&\leq \sum_{t\in[s:T]}|\calC_{t}|(M-t)\exp\left\{-ng_1(\lambda_1|f_\rmN,f_\rmA)\right\}+|\calC_s|(|\calC_{s}|-1)\exp\left\{-ng_2(\lambda_2|f_\rmN,f_\rmA)\right\}
\\
&+\sum_{t\in[s:T]}|\calC_{t}|(M-t)\exp\left\{-ng_1(\lambda_3|f_\rmN,f_\rmA)\right\}+|\calC_s|(|\calC_{s}|-1)\exp\left\{-ng_2(\lambda_3|f_\rmN,f_\rmA)\right\}\label{beta_al_UnS_1}\\
\nn&\leq\left(\sum_{t\in[s:T]}|\calC_{t}|(M-t)+|\calC_s|(|\calC_{s}|-1)\right)\\
&\times\Bigg(\exp\Big\{-n\min\big\{g_1(\lambda_1|f_\rmN,f_\rmA),g_2(\lambda_2|f_\rmN,f_\rmA)\big\}\Big\}+\exp\Big\{-n\min\big\{Kg_1(\lambda_3|f_\rmN,f_\rmA),Kg_2(\lambda_3|f_\rmN,f_\rmA)\big\}\Big\}\Bigg)\label{beta_al_UnS},
\end{align}
where \eqref{beta_al_UnS_1} follows by combining the results in \eqref{beta_UNs_fx1}, \eqref{PGD_1}, \eqref{PGD_2} and \eqref{PGB2} by replacing $\lambda$ with $\lambda_1$ and $\lambda_2$ for $\tau=n$, and replacing $(n,\lambda)$ with $(Kn,\lambda_3)$ for $\tau=Kn$.
Thus, combining \eqref{EBtau_un_tp} and \eqref{beta_al_UnS} leads to
\begin{align}
\lim_{n\rightarrow \infty}-\frac{1}{\bbE_{\rmP_\calB}[\tau]}\log \beta_\calB^{\rmtp}(\phi^n|f_\rmN,f_\rmA)
&\geq \min\Big\{g_1(\lambda_1|f_\rmN,f_\rmA),Kg_1(\lambda_3|f_\rmN,f_\rmA),g_2(\lambda_2|f_\rmN,f_\rmA),Kg_2(\lambda_3|f_\rmN,f_\rmA)\Big\}.
\label{logbeta_al_UnS}
\end{align}
The false reject probability satisfies
\begin{align}
\zeta_{\calB}^{\rmtp}(\phi^n|f_\rmN,f_\rmA)
&\leq \bbP_{\calB}\{\phi_n(\bY^n)=\rmH_\rmr\}+
\bbP_{\calB}\{\phi_{Kn}(\bY^{Kn})=\rmH_\rmr\}\\
&\leq  |\calC_{s}|(|\calC_{s}|-1)\Big(\exp\left\{-ng_2(\lambda_2|f_\rmN,f_\rmA)\right\}+\exp\left\{-Kng_2(\lambda_3|f_\rmN,f_\rmA)\right\}\Big),\label{zeta_al_UnS}
\end{align}
where \eqref{zeta_al_UnS} follows from \eqref{zeta_UNs_fx} where $\lambda$ is replaced with $\lambda_2$ and $\lambda_3$. Thus, using \eqref{EBtau_un_tp}, it follows that
\begin{align}
\lim_{n\rightarrow \infty}-\frac{1}{\bbE_{\rmP_\calB}[\tau]}\log \zeta_\calB^{\rmtp}(\phi^n|f_\rmN,f_\rmA)
&\geq\min\left\{g_2(\lambda_2|f_\rmN,f_\rmA),Kg_2(\lambda_3|f_\rmN,f_\rmA)\right\}.\label{logzeta_al_UnS}
\end{align}
If $\lambda_1\geq\left(1-\frac{s}{M-s-1}\right)^2\mathrm{MMD}^2(f_\rmN,f_\rmA)$, $\bbE_{\rmP_\calB}[\tau]=Kn$ and the two-phase test reduces to the fixed-length one. Therefore, the exponent is the same as that of the fixed-length one.

Now consider the case of $s=1$. When $\max\{\lambda_1,\lambda_3\}<\mathrm{MMD}^2(f_\rmN,f_\rmA)$, it follows that
\begin{align}
\nn\beta_\calB^{\rmtp}(\phi^n|f_\rmN,f_\rmA)
&\leq (M-1)\exp\left\{-ng_1(\lambda_1|f_\rmN,f_\rmA)\right\}+\sum_{t\in[2:T]}|\calC_{t}|(M-t)\exp\left\{-ng_1(\lambda_1|f_\rmN,f_\rmA)\right\}\\*
&\qquad+(M-1)\exp\left\{-Kng_1(\lambda_3|f_\rmN,f_\rmA)\right\}+\sum_{t\in[2:T]}|\calC_{t}|(M-t)\exp\left\{-Kng_1(\lambda_3|f_\rmN,f_\rmA)\right\}\label{beta_al_UnS_1-s1}\\
&\leq\sum_{t\in[T]}|\calC_{t}|(M-t)\left(\exp\left\{-ng_1(\lambda_1|f_\rmN,f_\rmA)\right\}+\exp\left\{-Kng_1(\lambda_3|f_\rmN,f_\rmA)\right\}\right)\\
&\leq 2\sum_{t\in[T]}|\calC_{t}|(M-t)\exp\left\{-n\min\left\{g_1(\lambda_1|f_\rmN,f_\rmA),Kg_1(\lambda_3|f_\rmN,f_\rmA)\right\}\right\}\label{beta_al_UnS-s1}
\end{align}
where \eqref{beta_al_UnS_1-s1} follows by using the results of \eqref{PGD_1} and \eqref{PGD_2} with $s=1$ and by replacing $(n,\lambda)$ with $(n,\lambda_1)$ and $(Kn,\lambda_3)$, respectively. Thus, using \eqref{EBtau_un_tp}, it follows that
\begin{align}
\lim_{n\rightarrow \infty}-\frac{1}{\bbE_{\rmP_\calB}[\tau]}\log \beta_\calB^{\rmtp}(\phi^n|f_\rmN,f_\rmA)
&\geq \min\left\{g_1(\lambda_1|f_\rmN,f_\rmA),Kg_1(\lambda_3|f_\rmN,f_\rmA)\right\}.
\label{logbeta_al_UnS-s1}
\end{align}
Similarly, we have
\begin{align}
&\zeta_{\calB}^{\rmtp}(\phi^n|f_\rmN,f_\rmA)
\leq  M(M-1)\left(\exp\left\{\frac{-n\left(\mathrm{MMD}^2(f_\rmN,f_\rmA)-\lambda_2\right)^2}
{32K_0^2\left(1+\frac{1}{M-2}\right)}\right\}+\exp\left\{\frac{-Kn\left(\mathrm{MMD}^2(f_\rmN,f_\rmA)-\lambda_3\right)^2}
{32K_0^2\left(1+\frac{1}{M-2}\right)}\right\}\right)\label{zeta_al_UnS-s1}
\end{align}
and
\begin{align}
\lim_{n\rightarrow \infty}-\frac{1}{\bbE_{\rmP_\calB}[\tau]}\log \zeta_\calB^{\rmtp}(\phi^n|f_\rmN,f_\rmA)
&\geq\min\Bigg\{\frac{\left(\mathrm{MMD}^2(f_\rmN,f_\rmA)-\lambda_2\right)^2}
{32K_0^2\left(1+\frac{1}{M-2}\right)},\frac{K\left(\mathrm{MMD}^2(f_\rmN,f_\rmA)-\lambda_3\right)^2}
{32K_0^2\left(1+\frac{1}{M-2}\right)}\Bigg\},\label{logzeta_al_UnS-s1}
\end{align}
where \eqref{zeta_al_UnS-s1} follows from the result of \eqref{zeta_UNs_fx-s1}. If $\lambda_1\geq\mathrm{MMD}^2(f_\rmN,f_\rmA)$, the exponent is the same as that of the fixed-length one.

Finally, the false alarm probability satisfies
\begin{align}
\rmP_{\mathrm{FA}}^{\rmtp}(\phi^n|f_\rmN,f_\rmA)
&\leq\bbP_\rmr\{\phi_n(\bY^n)\neq\rmH_\rmr\}+
\bbP_\rmr\{\phi_{Kn}(\bY^{Kn})\neq\rmH_\rmr\}\\ &\leq\sum_{t\in[T]}|\calC_{t}|^2(M-t)\exp\left\{-ng_1(\lambda_1|f_\rmN,f_\rmA)\right\}+\sum_{t\in[T]}|\calC_{t}|^2\exp\left\{-Kng_1(\lambda_3|f_\rmN,f_\rmA)\right\}\label{FA_al_UnS},
\end{align}
where \eqref{FA_al_UnS} follows from the result in \eqref{FA_UNs_fx} by replacing $(n,\lambda)$ with $(n,\lambda_1)$ and $(n,\lambda_3)$.

Thus, using \eqref{Ertau_un_tp}, it follows that
\begin{align}
\lim_{n\to \infty}-\frac{1}{\bbE_{\rmP_\rmr}[\tau]}\log \rmP_{\mathrm{FA}}^{\rmtp}(\phi^n|f_\rmN,f_\rmA)
&\geq
\min\left\{g_1(\lambda_1|f_\rmN,f_\rmA),Kg_1(\lambda_3|f_\rmN,f_\rmA)\right\}.\label{logFA_al_UnS}
\end{align}

The proof of Theorem \ref{E-AFLJMT} is now completed.

\bibliographystyle{IEEEtran}
\bibliography{BiB}

\begin{thebibliography}{10}
\providecommand{\url}[1]{#1}
\csname url@samestyle\endcsname
\providecommand{\newblock}{\relax}
\providecommand{\bibinfo}[2]{#2}
\providecommand{\BIBentrySTDinterwordspacing}{\spaceskip=0pt\relax}
\providecommand{\BIBentryALTinterwordstretchfactor}{4}
\providecommand{\BIBentryALTinterwordspacing}{\spaceskip=\fontdimen2\font plus
\BIBentryALTinterwordstretchfactor\fontdimen3\font minus
  \fontdimen4\font\relax}
\providecommand{\BIBforeignlanguage}[2]{{%
\expandafter\ifx\csname l@#1\endcsname\relax
\typeout{** WARNING: IEEEtran.bst: No hyphenation pattern has been}%
\typeout{** loaded for the language `#1'. Using the pattern for}%
\typeout{** the default language instead.}%
\else
\language=\csname l@#1\endcsname
\fi
#2}}
\providecommand{\BIBdecl}{\relax}
\BIBdecl

\bibitem{kumar2005parallel}
V.~Kumar, ``Parallel and distributed computing for cybersecurity,'' \emph{IEEE
  Distributed Systems Online}, vol.~6, no.~10, 2005.

\bibitem{spence2001detection}
C.~Spence, L.~Parra, and P.~Sajda, ``Detection, synthesis and compression in
  mammographic image analysis with a hierarchical image probability model,'' in
  \emph{IEEE MMBIA}.\hskip 1em plus 0.5em minus 0.4em\relax IEEE, 2001, pp.
  3--10.

\bibitem{Chernoff1952AMO}
H.~Chernoff, ``A measure of asymptotic efficiency for tests of a hypothesis
  based on the sum of observations,'' \emph{Ann. Math. Stat.}, vol.~23, pp.
  493--507, 1952.

\bibitem{Blahut1974HypothesisTA}
R.~E. Blahut, ``Hypothesis testing and information theory,'' \emph{IEEE Trans.
  Inf. Theory}, vol.~20, pp. 405--417, 1974.

\bibitem{KL}
S.~Kullback, ``Kullback-leibler divergence,'' 1951.

\bibitem{wald1948optimum}
A.~Wald and J.~Wolfowitz, ``Optimum character of the sequential probability
  ratio test,'' \emph{Ann. Math. Stat.}, pp. 326--339, 1948.

\bibitem{AFL}
A.~Lalitha and T.~Javidi, ``On error exponents of almost-fixed-length channel
  codes and hypothesis tests,'' \emph{arXiv preprint arXiv:2012.00077}, 2020.

\bibitem{Asymptotically_optimal_classification}
M.~Gutman, ``Asymptotically optimal classification for multiple tests with
  empirically observed statistics,'' \emph{IEEE Trans. Inf. Theory}, vol.~35,
  no.~2, pp. 401--408, 1989.

\bibitem{second}
L.~Zhou, V.~Y. Tan, and M.~Motani, ``Second-order asymptotically optimal
  statistical classification,'' \emph{Information and Inference: A Journal of
  the IMA}, vol.~9, no.~1, pp. 81--111, 2020.

\bibitem{Sequential_classification}
M.~Haghifam, V.~Y. Tan, and A.~Khisti, ``Sequential classification with
  empirically observed statistics,'' \emph{IEEE Trans. Inf. Theory}, vol.~67,
  no.~5, pp. 3095--3113, 2021.

\bibitem{hsu2022universal}
C.-Y. Hsu, C.-F. Li, and I.-H. Wang, ``On universal sequential classification
  from sequentially observed empirical statistics,'' in \emph{IEEE ITW}.\hskip
  1em plus 0.5em minus 0.4em\relax IEEE, 2022, pp. 642--647.

\bibitem{bai2022achievable}
L.~Bai, J.~Diao, and L.~Zhou, ``Achievable error exponents for almost
  fixed-length binary classification,'' in \emph{IEEE ISIT}.\hskip 1em plus
  0.5em minus 0.4em\relax IEEE, 2022, pp. 1336--1341.

\bibitem{diao2023achievable}
J.~Diao, L.~Zhou, and L.~Bai, ``Achievable error exponents for almost
  fixed-length m-ary classification,'' in \emph{IEEE ISIT}.\hskip 1em plus
  0.5em minus 0.4em\relax IEEE, 2023, pp. 1568--1573.

\bibitem{Universal_Outlier_Hypothesis}
Y.~Li, S.~Nitinawarat, and V.~V. Veeravalli, ``Universal outlier hypothesis
  testing,'' \emph{IEEE Trans. Inf. Theory}, vol.~60, no.~7, pp. 4066--4082,
  2014.

\bibitem{Joint}
L.~Zhou, Y.~Wei, and A.~O. Hero, ``Second-order asymptotically optimal outlier
  hypothesis testing,'' \emph{IEEE Trans. Inf. Theory}, vol.~68, no.~6, pp.
  3585--3607, 2022.

\bibitem{Universal_sequential_Hypothesis}
Y.~Li, S.~Nitinawarat, and V.~V. Veeravalli, ``Universal sequential outlier
  hypothesis testing,'' \emph{Sequential Analysis}, vol.~36, no.~3, pp.
  309--344, 2017.

\bibitem{MMD}
S.~Zou, Y.~Liang, H.~V. Poor, and X.~Shi, ``Nonparametric detection of
  anomalous data streams,'' \emph{IEEE Trans. Signal Process.}, vol.~65,
  no.~21, pp. 5785--5797, 2017.

\bibitem{Hilbert}
A.~Berlinet and C.~Thomas-Agnan, \emph{Reproducing kernel Hilbert spaces in
  probability and statistics}.\hskip 1em plus 0.5em minus 0.4em\relax Springer
  Science \& Business Media, 2011.

\bibitem{Moderate_Deviations}
I.~Sason, ``Moderate deviations analysis of binary hypothesis testing,'' in
  \emph{IEEE ISIT}, 2012, pp. 821--825.

\bibitem{generalized_likelihoodratio_test}
O.~Zeitouni, J.~Ziv, and N.~Merhav, ``When is the generalized likelihood ratio
  test optimal?'' \emph{IEEE Trans. Inf. Theory}, vol.~38, no.~5, pp.
  1597--1602, 1992.

\bibitem{Merhav1991ABA}
M.~Neri and Z.~Jacob, ``A bayesian approach for classification of markov
  sources,'' \emph{IEEE Trans. Inf. Theory}, vol.~37, pp. 1067--1071, 1991.

\bibitem{Hsu2020OnBS}
H.~Hung-Wei and W.~I-Hsiang, ``On binary statistical classification from
  mismatched empirically observed statistics,'' \emph{IEEE ISIT}, pp.
  2533--2538, 2020.

\bibitem{bu2019linear}
Y.~Bu, S.~Zou, and V.~V. Veeravalli, ``Linear-complexity
  exponentially-consistent tests for universal outlying sequence detection,''
  \emph{IEEE Trans. Signal Process.}, vol.~67, no.~8, pp. 2115--2128, 2019.

\bibitem{gretton2012kernel}
A.~Gretton, K.~M. Borgwardt, M.~J. Rasch, B.~Sch{\"o}lkopf, and A.~Smola, ``A
  kernel two-sample test,'' \emph{J. Mach. Learn. Res.}, vol.~13, no.~1, pp.
  723--773, 2012.

\bibitem{sun2023kernel}
Z.~Sun and S.~Zou, ``Kernel robust hypothesis testing,'' \emph{IEEE Trans. Inf.
  Theory}, 2023.

\bibitem{mcdiarmid1989method}
C.~McDiarmid \emph{et~al.}, ``On the method of bounded differences,''
  \emph{Surveys in combinatorics}, vol. 141, no.~1, pp. 148--188, 1989.

\bibitem{binaryI-Hsiang}
H.-W. Hsu and I.-H. Wang, ``On binary statistical classification from
  mismatched empirically observed statistics,'' in \emph{IEEE ISIT}, 2020, pp.
  2533--2538.

\bibitem{tenney1981detection}
R.~R. Tenney and N.~R. Sandell, ``Detection with distributed sensors,''
  \emph{IEEE Trans. Aerosp. Electron. Syst.s}, no.~4, pp. 501--510, 1981.

\bibitem{tsitsiklis1988decentralized}
J.~N. Tsitsiklis, ``Decentralized detection by a large number of sensors,''
  \emph{Math Control Signal}, vol.~1, no.~2, pp. 167--182, 1988.

\bibitem{poor2008quickest}
H.~V. Poor and O.~Hadjiliadis, \emph{Quickest detection}.\hskip 1em plus 0.5em
  minus 0.4em\relax Cambridge University Press, 2008.

\bibitem{tartakovsky2014sequential}
A.~Tartakovsky, I.~Nikiforov, and M.~Basseville, \emph{Sequential analysis:
  Hypothesis testing and changepoint detection}.\hskip 1em plus 0.5em minus
  0.4em\relax CRC press, 2014.

\bibitem{kaufman1990finding}
L.~Kaufman and P.~J. Rousseeuw, ``Finding groups in data. an introduction to
  cluster analysis,'' \emph{Wiley Series in Probability and Mathematical
  Statistics. Applied Probability and Statistics}, 1990.

\bibitem{park2009simple}
H.-S. Park and C.-H. Jun, ``A simple and fast algorithm for k-medoids
  clustering,'' \emph{Expert Syst. Appl.}, vol.~36, no.~2, pp. 3336--3341,
  2009.

\end{thebibliography}
%\bibliography{IEEEfull}
\end{document}